\documentclass[aps,pra,reprint,amsmath,amssymb,groupedaddress,showpacs]{revtex4-1}
\usepackage{graphicx}
\usepackage{float}
\usepackage{dcolumn} 
\usepackage{bm}      
\usepackage{epsdice} 
\usepackage[bookmarks=false]{hyperref} 
\usepackage[font={small}]{caption}     
\hypersetup{
    unicode=false,          
    pdftoolbar=true,        
    pdfmenubar=true,        
    pdffitwindow=true,      
    pdfstartview={},        
    pdftitle={Noise-Resistant Quantum Teleportation, Ansibles, and the No-Projector Theorem},                  
    pdfauthor={Samuel R. Hedemann},     
    pdfsubject={Quantum Communication}, 
    pdfcreator={},                      
    pdfproducer={},                     
    pdfkeywords={Ansible,} {Pseudo-Ansible,} {No-Projector Theorem,} {Quantum Teleportation,} {Noise-Resistant Quantum Teleportation,} {Unassisted Quantum Error Correction,} {No-Cloning Theorem,} {Superluminal,} {Quantum Operations,} {Noise},                    
    pdfnewwindow=true,      
    colorlinks=true,        
    linkcolor=black,        
    citecolor=black,        
    filecolor=black,        
    urlcolor=black          
}
\newcommand{\Sec}[1]{\hyperref[sec:#1]{Sec.{\kern 2pt}\ref*{sec:#1}}}
\newcommand{\Section}[1]{\hyperref[sec:#1]{Section~\ref*{sec:#1}}}
\newcommand{\Fig}[2][]{\hyperref[fig:#2]{Fig.{\kern 2pt}\ref*{fig:#2}#1}}
\newcommand{\Figure}[2][]{\hyperref[fig:#2]{Figure~\ref*{fig:#2}#1}}
\newcommand{\App}[1]{\hyperref[sec:#1]{App.{\kern 2pt}\ref*{sec:#1}}}
\newcommand{\Appendix}[1]{\hyperref[sec:#1]{Appendix~\ref*{sec:#1}}}
\begin{document}
\title{Noise-Resistant Quantum Teleportation, Ansibles, and the No-Projector Theorem}
\author{Samuel R. Hedemann}
\affiliation{The Johns Hopkins University Applied Physics Laboratory, Laurel, MD 20723, USA}
\date{\today}
\begin{abstract}
A method is presented for achieving entanglement-free teleportation of a quantum state subject to any quantum noise. We apply this as a light-speed noise-resistant communicator, but also treat the possibility of a quantum ansible, a device for effectively superluminal communication and quantum broadcasting. The results suggest a ``no-projector theorem'' analogous to the no-cloning theorem. We then show how to build a pseudo-ansible for connection-free light-speed communication.
\end{abstract}
\maketitle
\section{\label{sec:I}Introduction}
Since the birth of special relativity in 1905 \cite[]{EiSR,EiLo}, many have tried to break the classical light-speed limit to achieve superluminal communication.  The most popular attempts use entanglement and quantum cloning \cite[]{Her1,Her2}, but such methods have been proven impossible \cite[]{Buss,WeSc,BDMS,SiBG}.

Here, we propose a \textit{new} communication method that pairs two big ideas: entanglement-free quantum teleportation to send information as a state, and quantum error correction (QEC) to protect the state during teleportation.  We call this \textit{noise-resistant quantum teleportation} (NRQT). In one realization, it is light-speed limited and requires a physical connection.  However, the theory also suggests that if true projectors are possible, then information could be sent faster than light, \textit{without violating special relativity}.  We can call any device that achieves this an \textit{ansible}, a term coined by science-fiction author Ursula K. Le Guin for the idea proposed by earlier authors like Isaac Asimov, inspired by Einstein. This suggests a ``\hyperlink{NoProjThm}{no-projector theorem},'' but also indicates that a \textit{connection-free light-speed pseudo-ansible is possible}.
\begin{figure}[H]
\centering
\includegraphics[width=0.99\linewidth]{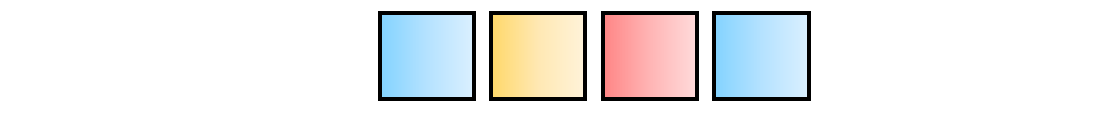}%
\vspace{-12pt}
\setlength{\unitlength}{0.01\linewidth}
\begin{picture}(100,0)
\put(10,3.8){\footnotesize Alice}
\put(20,3.8){\footnotesize $\rho\to\,\sum_j$}
\put(37,3.8){\footnotesize $L_j$}
\put(45.9,3.8){\footnotesize $\mathcal{E}(\cdot)$}
\put(55.5,3.8){\footnotesize $\mathcal{R}(\cdot)$}
\put(66.7,3.8){\footnotesize $L_{j}^{\dag}$}
\put(75,3.8){\footnotesize $\to\rho$}
\put(83.6,3.8){\footnotesize Bob}
\end{picture}
\caption[]{(color online) Unassisted QEC (UQEC) combined with teleportation to achieve NRQT. \smash{$\mathcal{E}$} is \textit{any} error channel, \smash{$L_j$} and \smash{$L_j^{\dag}$} are synchronized encoding/decoding operations, and $\mathcal{R}$ is a recovery channel, \textit{all in the space of} $\rho$. Alice can send any message to Bob, regardless of noise.}
\label{fig:1}
\end{figure}
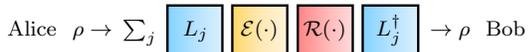
The protection against quantum noise comes from a new form of QEC that is capable of correcting \textit{all} errors, \textit{without help} from ancillas, called \textit{unassisted quantum error correction} (UQEC), depicted in \Fig{1}.  UQEC both protects against noise and enables the teleportation, but is impractical for quantum computation (see \App{App.A}).

To present ideas organically, \Sec{II} reviews UQEC and its role in NRQT, showing how the existence of projectors leads to effective superluminality.  Then, \Sec{III} shows how to build an ansible if true projectors could be realized, motivating a ``\hyperlink{NoProjThm}{no-projector theorem}'' in \Sec{IV}.  Finally, \Sec{V} shows how to build a \textit{pseudo-ansible} for connection-free light-speed communication.
\section{\label{sec:II}UQEC Applied to Noise-Resistant Quantum Teleportation}
Unassisted quantum error correction (UQEC) in its full form can protect \textit{any} quantum state from \textit{any} noise channel, as shown in \App{App.A}.  However, for conceptual simplicity we begin by reviewing a special case called \textit{limited-direct} UQEC, which only protects a limited family of states from all noise. We then discuss how to realize the process and examine the crucial role of projectors.
\subsection{\label{sec:II.A}Overview of Unassisted Quantum Error Correction}
The process of limited-direct UQEC starts by only considering as input \textit{families} of states with the form
\begin{equation}
\rho\equiv \rho _{(\bm{\theta} )}  \equiv \sum\limits_{j = 1}^n {\lambda _j (\bm{\theta} )|e_j \rangle \langle e_j |},
\label{eq:1}
\end{equation}
where $\{ |e_j \rangle \}$ is a set of $n$ orthonormal states that are the defining constant of the family $\rho _{(\bm{\theta} )}$, and where $\lambda _j  \equiv \lambda _j (\bm{\theta} ) \in [0,1]$ are eigenvalues where $\sum\nolimits_{j = 1}^n {\lambda _j }  = 1$, so the family's free parameters are $\bm{\theta} \equiv \{ \theta _1 , \ldots ,\theta _{n - 1} \}  \in [0,{\textstyle{\pi  \over 2}}]$.

Given input $\rho _{(\bm{\theta} )}$, we can get \textit{limited-direct} UQEC as
\begin{equation}
\rho _{(\bm{\theta} )}  = \sum\limits_{j = 1}^n {\sum\limits_{q = 1}^n {\sum\limits_{k} {L_j^{\dag}  R_q E_k L_j \rho _{(\bm{\theta} )} L_j^{\dag}  E_k^{\dag}  R_q^{\dag}  L_j } } },
\label{eq:2}
\end{equation}
where we define \textit{encoding} and \textit{recovery} operators as
\begin{equation}
L_j  \equiv |\psi \rangle \langle e_j |\;\;\text{and}\;\;R_q  \equiv |\psi \rangle \langle \phi_{q}|,
\label{eq:3}
\end{equation}
where $|\psi \rangle$ is the \textit{reference state}, which is any pure state, $\{|\phi_{q}\rangle\}$ is any orthonormal complete basis, $\{|e_j \rangle\}$ is the constant set of orthonormal eigenstates defining $\rho _{(\bm{\theta} )}$, and $\{E_k\}$ are Kraus operators of noise channel $\mathcal{E}$ where \smash{$\sum\nolimits_k \! E_k^{\dag}E_k =I$}.   Operators like those in (\ref{eq:3}) can be decomposed as a unitary operation and rank-$1$ projector, as shown in \Sec{II.D.3}.  See \App{App.A} for a derivation of (\ref{eq:2}).

Next, to see how (\ref{eq:2}) enables us to protect any family of states $\rho _{(\bm{\theta} )}$ in the form of (\ref{eq:1}) from any quantum noise channel, we will look at a single-qubit example.
\subsection{\label{sec:II.B}Single-Qubit Example of UQEC}
Here, we do a step-by-step example showing that all possible input states of a certain family can be recovered from all possible error channels acting on one qubit ($n=2$).  For simplicity, let \smash{$\{ |u\rangle \}$} be computational basis kets generically labeled starting on $1$ as \smash{$\{ |1\rangle , \ldots ,|n\rangle \}$}.

First, define the state family of constant eigenstates by picking \smash{$\{ |e_{j}\rangle \}\equiv\{|1\rangle,|2\rangle\}$}, so that
\begin{equation}
\rho\equiv\rho_{(\theta)}  \equiv \sum\limits_{j = 1}^2 {\lambda _j |e_j \rangle \langle e_j |}= c_{\theta}^2|1\rangle\langle 1| +s_{\theta}^2|2\rangle\langle 2|,
\label{eq:4}
\end{equation}
where $c_{\theta}\equiv\cos(\theta)$ and $s_{\theta}\equiv\sin(\theta)$.  Then, choosing $|\psi \rangle \! =\! |1\rangle$ and \smash{$\{|\phi_q\rangle\}\equiv\{|1\rangle,|2\rangle\}$}, where $|1\rangle\!\equiv\!\binom{1}{0}$ and $|2\rangle\!\equiv\!\binom{0}{1}$, (\ref{eq:3}) gives both the encoding and recovery operators as
\begin{equation}
L_1 = R_1 = \left(\! {\begin{array}{*{20}c}
   1 & 0  \\
   0 & 0  \\
\end{array}}\! \right)\;\;\text{and}\;\;\,L_2 = R_2 = \left(\! {\begin{array}{*{20}c}
   0 & 1  \\
   0 & 0  \\
\end{array}}\! \right)\!,
\label{eq:5}
\end{equation}
where \Sec{II.D.3} explains their physical implementation.

The first step in the protection procedure is to randomly choose and apply one of the $L_j$ to $\rho$, keeping in mind that later, we will also have to apply its adjoint \smash{$L_j^{\dag}$}.  A physical process for this is given in \Sec{II.D.2}, but for now, the possible results of the encoding step are
\begin{equation}
L_j \rho L_j^{\dag} = {\lambda_j}\!\left(\! {\begin{array}{*{20}c}
   1 & 0  \\
   0 & 0  \\
\end{array}}\! \right)\!,
\label{eq:6}
\end{equation}
where the coefficients $\lambda_{j}\equiv\langle e_{j}|\rho|e_{j}\rangle$ are given by
\begin{equation}
\lambda _1  = \rho _{1,1} =c_{\theta}^2\;\;\;\text{and}\;\;\;\lambda _2  = \rho _{2,2} =s_{\theta}^2 ,
\label{eq:7}
\end{equation}
where $\rho _{a,b}\equiv\langle a|\rho|b\rangle$. Then one of the $E_k$ acts, producing
\begin{equation}
E_k L_j \rho L_j^{\dag}  E_k^{\dag}   = \lambda_j \left(\! {\begin{array}{*{20}c}
   {|(E_k )_{1,1} |^2 } & {(E_k )_{1,1} (E_k )_{2,1} ^* }  \\
   {(E_k )_{2,1} (E_{k})_{1,1} ^* } & {|(E_k )_{2,1} |^2 }  \\
\end{array}}\! \right)\!,
\label{eq:8}
\end{equation}
where again, $(E_k)_{a,b}\equiv\langle a|E_{k}|b\rangle$.  Next we randomly apply one of the recovery operators, producing
\begin{equation}
R_q E_k L_j \rho L_j^{\dag}  E_k^{\dag}  R_q^{\dag} = \lambda_j |(E_k )_{q,1} |^2 \left( {\begin{array}{*{20}c}
   1 & 0  \\
   0 & 0  \\
\end{array}} \right)\!.
\label{eq:9}
\end{equation}
Then, applying the matched decoder $L_j^{\dag}=|e_{j}\rangle\langle 1|$ yields
\begin{equation}
L_j^{\dag}  R_q E_k L_j \rho L_j^ {\dag}  E_k^{\dag}  R_q^{\dag}  L_j  = \lambda_j |(E_k )_{q,1} |^2 |e_j \rangle \langle e_j |.
\label{eq:10}
\end{equation}

Now, since we don't know \textit{which} \smash{$E_k$} happened, or \textit{which} \smash{$R_q$} was applied, or \textit{which pair} \smash{$\{L_j ,L_j^{\dag}\}$} was used, then before we look at the result, the potentially unnormalized state is the sum over all possible results, as
\begin{equation}
\sum\limits_{j = 1}^2 \sum\limits_{q = 1}^2 {\sum\limits_k {L_j^{\dag}  R_q E_k L_j \rho L_j^ {\dag}  E_k^{\dag}  R_q^{\dag}  L_j } }  = C\sum\limits_{j = 1}^2 \lambda_j |e_j \rangle \langle e_j |,
\label{eq:11}
\end{equation}
where $C$ is an ``error-dependent'' scalar, defined as
\begin{equation}
C \equiv \sum\limits_{q = 1}^n {\sum\limits_k {|(E_k )_{q,1} |^2 } }. 
\label{eq:12}
\end{equation}
\Appendix{App.A.1} proves that $C=1$ for all channels, but here we keep it.  Now we see that the matrix to the right of $C$ in (\ref{eq:11}) only contains information about $\rho$, and putting (\ref{eq:4}) into (\ref{eq:11}) and normalizing (unnecessarily) gives
\begin{equation}
{\textstyle{{1} \over {\text{tr}(C\rho)}}}\sum\limits_{j = 1}^2 {\sum\limits_{q = 1}^2 {\sum\limits_k \!{L_j^{\dag}  R_q E_k L_j \rho L_j^{\dag}  E_k^{\dag}  R_q^{\dag}  L_j } } }=\rho,
\label{eq:13}
\end{equation}
which is identical to the input state, as desired.  Notice that no ancilla systems are needed, and this same method works for all $n$-level systems, even with \textit{nonlocal} noise.
\subsection{\label{sec:II.C}Numerical Demonstration of UQEC}
\begin{figure}[H]
\centering
\includegraphics[width=0.99\linewidth]{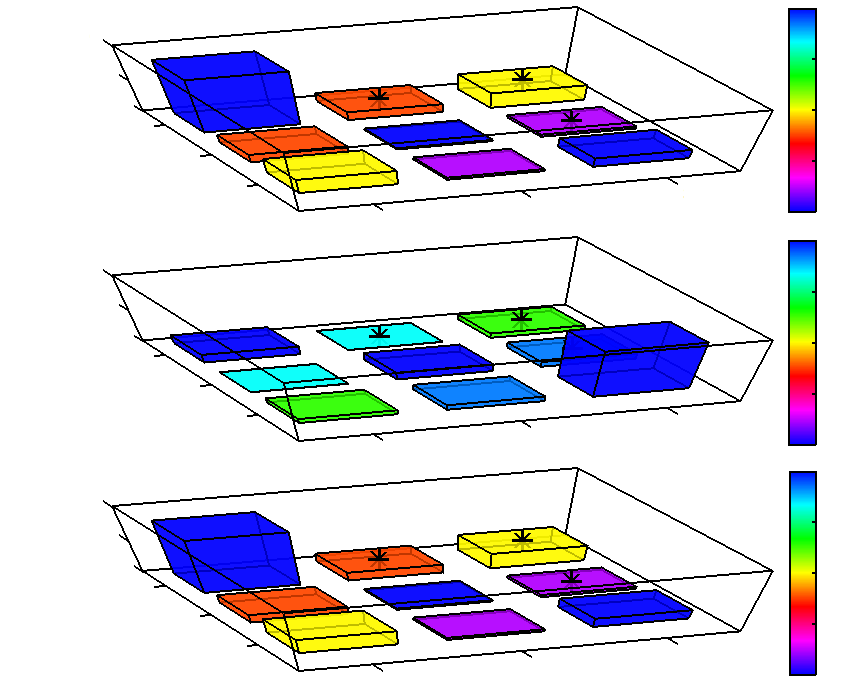}%
\vspace{-12pt}
\setlength{\unitlength}{0.01\linewidth}
\begin{picture}(100,0)
\put(0.7,75){\small (a)}
\put(0.7,48.5){\small (b)}
\put(0.7,22){\small (c)}
\put(8.0,61.2){\small $\rho$}
\put(8.0,34.5){\small $\rho_E$}
\put(8.0,7){\small $\rho_R$}
\put(88.4,74.5){\footnotesize $\phi$}
\put(96,77.0){\footnotesize $2\pi$}
\put(95.7,71.6){\footnotesize $\frac{3\pi}{2}$}
\put(96.7,65.7){\footnotesize $\pi$}
\put(96.3,59.8){\footnotesize $\frac{\pi}{2}$}
\put(96.8,53.95){\footnotesize $0$}
\put(88.4,47.6){\footnotesize $\phi$}
\put(96,50.1){\footnotesize $2\pi$}
\put(95.7,44.7){\footnotesize $\frac{3\pi}{2}$}
\put(96.7,38.8){\footnotesize $\pi$}
\put(96.3,32.9){\footnotesize $\frac{\pi}{2}$}
\put(96.8,27.05){\footnotesize $0$}
\put(88.4,21.0){\footnotesize $\phi$}
\put(96,23.5){\footnotesize $2\pi$}
\put(95.7,18.1){\footnotesize $\frac{3\pi}{2}$}
\put(96.7,12.2){\footnotesize $\pi$}
\put(96.3,6.3){\footnotesize $\frac{\pi}{2}$}
\put(96.8,0.45){\footnotesize $0$}
%
\put(7.7,74.1){\scriptsize $1.0$}
\put(9.4,70.3){\scriptsize $0.5$}
\put(10.9,66.5){\scriptsize $0.0$}
\put(7.7,47.5){\scriptsize $1.0$}
\put(9.4,43.7){\scriptsize $0.5$}
\put(10.9,39.9){\scriptsize $0.0$}
\put(7.7,20.9){\scriptsize $1.0$}
\put(9.4,17.1){\scriptsize $0.5$}
\put(10.9,13.3){\scriptsize $0.0$}
%
\put(16.2,63.4){\scriptsize $1$}
\put(21.2,59.9){\scriptsize $2$}
\put(26.8,56.3){\scriptsize $3$}
\put(16.2,36.8){\scriptsize $1$}
\put(21.2,33.3){\scriptsize $2$}
\put(26.8,29.7){\scriptsize $3$}
\put(16.2,10.2){\scriptsize $1$}
\put(21.2,6.7){\scriptsize $2$}
\put(26.8,3.1){\scriptsize $3$}
%
\put(44.8,52.8){\scriptsize $1$}
\put(61.9,54.2){\scriptsize $2$}
\put(78.8,55.6){\scriptsize $3$}
\put(44.8,26.2){\scriptsize $1$}
\put(61.9,27.6){\scriptsize $2$}
\put(78.8,29.0){\scriptsize $3$}
\put(44.8,-0.4){\scriptsize $1$}
\put(61.9,1.0){\scriptsize $2$}
\put(78.8,2.4){\scriptsize $3$}
\end{picture}
\caption[]{(color online) Limited-direct UQEC on an arbitrary qutrit in an arbitrary noise channel. These plots depict density-matrix representations of the states. For each matrix element, labeled by row indices on the bottom left and column indices on the bottom right, the height of the box gives the magnitude, the color represents the phase-angle magnitude, and an asterisk on top indicates a negative phase angle.  (a) shows input $\rho$, (b) shows damaged state $\rho_{E}$ resulting from no protection, and (c) shows recovered state $\rho_{R}$ resulting from full protection.  Note that $\rho_{R}=\rho$ exactly.}
\label{fig:2}
\end{figure}
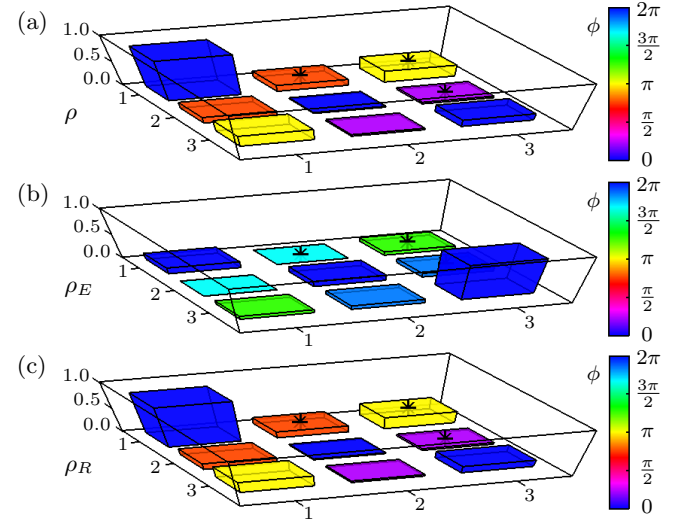

Here, we look at a numerical test to see that UQEC works, but see \App{App.A} for a \textit{proof} that it \textit{always} works.  \Figure{2} subjects an arbitrary qutrit $\rho$ to an arbitrary noise channel $\mathcal{E}(\sigma)\equiv \sum\nolimits_{k}E_{k}\sigma E_{k}^{\dag}$.\hspace{\stretch{1}}This qutrit is treated%
{\noindent}as part of a constant family \smash{$\rho_{(\theta_{1},\theta_{2})}$} by using its eigenstates to construct \smash{$L_j$}.  \Figure{2} then compares \smash{$\rho$} to the error-damaged state \smash{$\rho_{E}\equiv\mathcal{E}(\rho)$}, and to the recovered state \smash{$\rho_{R}\equiv \sum\nolimits_{j = 1}^n \sum\nolimits_{q = 1}^n {\sum\nolimits_k {L_j^{\dag}  R_q E_k L_j \rho L_j^ {\dag}  E_k^{\dag}  R_q^{\dag}  L_j } } $}, which includes the effects of \smash{$\mathcal{E}$}.
\subsection{\label{sec:II.D}How to Realize the Quantum Operations for Noise-Resistance}
Here we discuss the fine points of the UQEC procedure, focusing on practical methods of implementation, to see how it applies to noise-resistant quantum teleportation (NRQT), and to identify the problem with projectors.
\subsubsection{\label{sec:II.D.1}Classical Generation of Quantum Operations}
To prepare for later discussions, here we briefly review quantum operations \cite[]{HKr1,HKr2,Choi,Kra1}. A dimension-preserving quantum operation $\mathcal{F}(\rho)$ maps states in the space of $\rho$ to other states $\rho'\equiv\mathcal{F}(\rho)$ in the same space, as
\begin{equation}
\mathcal{F}(\rho ) = \sum\limits_{k = 1}^{n_F } {F_k \rho F_k^{\dag}  },
\label{eq:14}
\end{equation}
where $\{F_k\}$ is a set of $n_F$ Kraus operators in the space of $\rho$, obeying Kraus completeness,
\begin{equation}
\sum\limits_{k = 1}^{n_F } {F_k^{\dag} F_k }=I,
\label{eq:15}
\end{equation}
where the identity $I$ is also in the space of $\rho$.

There are many ways to physically interpret quantum operations, but the one of interest for NQRT is as follows.

\hypertarget{int:2.1}{\textbf{State-Free Interpretation:~~}}Apply all $F_k$ with equal probabilities $p_k^{\prime} \equiv {\textstyle{1 \over {n_F }}}$, and then normalize the result as
\begin{equation}
\mathcal{F}(\rho ) = \frac{\sum\nolimits_{k = 1}^{n_F } {{\textstyle{1 \over {n_F }}}F_k \rho F_k^{\dag}  }}{\text{tr}(\sum\nolimits_{k = 1}^{n_F } {{\textstyle{1 \over {n_F }}}F_k \rho F_k^{\dag}  } )}= n_F \sum\limits_{k = 1}^{n_F } {{\textstyle{1 \over {n_F }}}F_k \rho F_k^{\dag}  }.
\label{eq:16}
\end{equation}
The normalization is state-independent because of (\ref{eq:15}) and the fact that $\text{tr}(\rho)=1$, which together enable \smash{$\text{tr}(\sum\nolimits_{k = 1}^{n_F } {{\textstyle{1 \over {n_F }}}F_k \rho F_k^{\dag}  } ) = {\textstyle{1 \over {n_F }}}\sum\nolimits_{k = 1}^{n_F } {\text{tr}(\rho F_k^{\dag}  F_k )}  = {\textstyle{1 \over {n_F }}}$}.

Thus, we will use the \hyperlink{int:2.1}{State-Free Interpretation} to treat all quantum operations as applications of Kraus operators with equal probability, followed by normalization.

Another ingredient useful for classical implementation of UQEC is a \textit{method for probabilistic application of operations}.  This is simpler than it sounds; just use a random or true random number generator (TRNG).

\begin{figure}[H]
\centering
\includegraphics[width=0.99\linewidth]{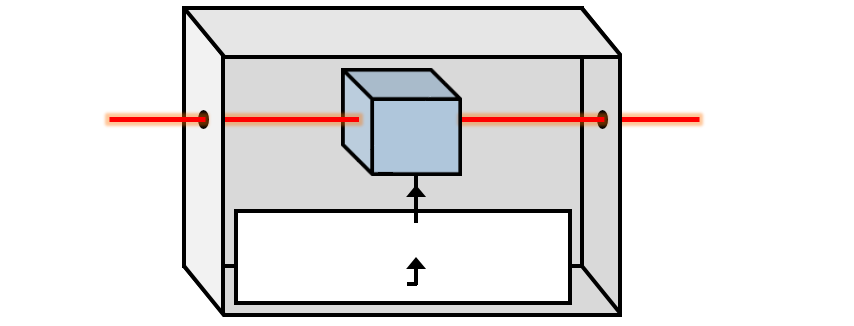}%
\vspace{-12pt}
\setlength{\unitlength}{0.01\linewidth}
\begin{picture}(100,0)
\put(16.6,18.6){\footnotesize $\rho$}
\put(74,17.6){\footnotesize $\mathcal{F}(\rho)\!=\!\!\sum\limits_{k = 1}^{n_F }\!\! {F_k \rho F_k^{\dag}}$}
\put(56,18.6){\footnotesize $F_k \rho F_k^{\dag}$}
\put(46.9,20.5){\footnotesize $F_{k}$}
\put(39.5,8.4){\footnotesize $\{F_{1},\ldots,F_{n_{F}\!}\}$}
\put(29.0,3.25){\footnotesize TRNG}
\put(39.2,3.7){\footnotesize :}
\put(44.9,3.3){\footnotesize $k$}
\put(40.6,2.7){\large \epsdice{5}}
\end{picture}
\caption[]{(color online) Schematic of classical implementation of the \hyperlink{int:2.1}{State-Free Interpretation} of a quantum operation \smash{$\mathcal{F}(\rho ) = \sum\nolimits_{k = 1}^{n_F } {F_k \rho F_k^{\dag}  }$}. A closed box contains a true random number generator (TRNG) that makes random number $k$, selects the $k$th operation $F_k$, and applies it to input state $\rho$.  Since each $F_k$ has equal probability of being applied, and since observers outside the box do not know \textit{which} $F_k$ was applied, the output state is $\mathcal{F}(\rho )$, the normalized sum of all possible results \smash{$\{F_k \rho F_k^{\dag}\}$}.}
\label{fig:3}
\end{figure}
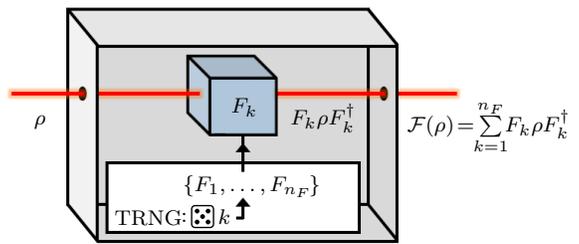

The scenario goes like this: As shown in \Fig{3}, a device containing a TRNG internally produces a number $k$ between $1$ and $n_F$, where the TRNG produces any of these numbers with equal probability.  The device applies the operation $F_{k}$ to the system in the input state $\rho$, but no information about which operation was applied is allowed to leave the device.  Then, after the operation was applied, since we know any one of the $F_k$ could have been applied, but we do not know \textit{which}, then the state of the output system is $\mathcal{F}(\rho ) = \sum\nolimits_{k = 1}^{n_F } {F_k \rho F_k^{\dag}  }$, thus classically realizing some intended quantum operation $\mathcal{F}(\rho )$.
Note that this involves only \textit{classical probability}, and thus it is  our \textit{ignorance} of which operation was applied that ``creates'' the classical mixture of all the operations acting on the input state.  However, this step-function method only works perfectly if the Kraus operators can be applied alone deterministically, as \Sec{IV} explains.

Alternatively, we could make quantum operations by reducing from larger systems, but this classical method is what lets us \textit{synchronize} the encoders and decoders.
\subsubsection{\label{sec:II.D.2}Implementation of Index-Linked Operations}
\begin{figure}[H]
\centering
\includegraphics[width=0.99\linewidth]{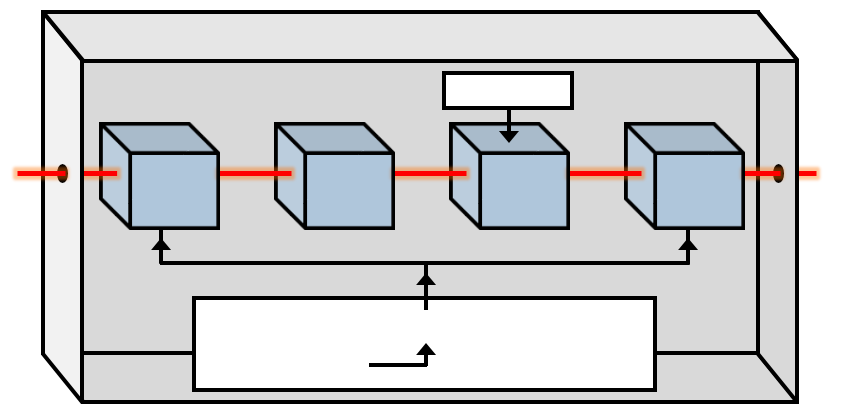}%
\vspace{-12pt}
\setlength{\unitlength}{0.01\linewidth}
\begin{picture}(100,0)
\put(2,24){\footnotesize $\rho$}
\put(95,24){\footnotesize $\rho$}
\put(52.6,36.05){\footnotesize TRNG}
\put(62.8,36.6){\footnotesize :}
\put(64.1,36.8){\footnotesize $q$}
\put(59,25){\footnotesize $R_{q}$}
\put(32.1,9){\footnotesize $\{{\{L_{1},\! L_{1}^{\dag}\},\ldots,\{L_{n},\! L_{n}^{\dag}\}}\}$}
\put(24.1,4.35){\footnotesize TRNG}
\put(34.3,4.8){\footnotesize :}
\put(40.9,4.8){\footnotesize $j$}
\put(36.1,3.8){\large \epsdice{5}}
\put(18.5,25){\footnotesize $L_{j}$}
\put(79.2,25){\footnotesize $L_{j}^{\dag}$}
\put(38.8,25){\footnotesize $E_{k}$}
\end{picture}
\caption[]{(color online) Schematic of an implementation of limited-direct UQEC.  A closed box contains a true random number generator (TRNG) that makes random number $j$, selects the $j$th encoding/decoding pair $\{L_{j},L_{j}^{\dag}\}$, and first applies $L_{j}$ to input state $\rho$.  Then, error channel $\mathcal{E}$ randomly applies one of its errors $E_k$.  Then, another TRNG selects and applies the $q$th recovery operator $R_q$.  Finally, the other member $L_{j}^{\dag}$ of the \smash{$\{L_{j},L_{j}^{\dag}\}$} pair is applied, and since information about \textit{which} operators were applied inside the box is not learned by observers outside, the output state is $\rho$, from (\ref{eq:2}).}
\label{fig:4}
\end{figure}
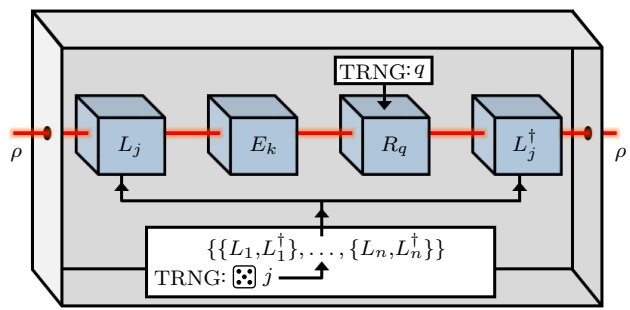

The peculiar thing about (\ref{eq:2}) is its index-linked operators, the $j$-indexed pairs in \smash{$L_{j}^{\dag}R_{q}E_{k}L_{j}$}.  While \App{App.A.2} proves the existence of physical index-linked operations, here we simply show one way to realize them, in \Fig{4}.
\subsubsection{\label{sec:II.D.3}How to ``Realize'' the Individual Kraus Operators}
Recall from (\ref{eq:3}) that we need operators of the form $|\psi \rangle \langle \phi |$, which are in general both nonHermitian and nonunitary.  To ``realize'' these, if we use the \textit{descending eigenvalue convention}, such that eigenvalues of all states are labeled as $\lambda _1  \ge  \cdots  \ge \lambda _n  \ge 0$ where $\sum\nolimits_{k = 1}^n {\lambda _k }  = 1$, then we can express $L_j \equiv |\psi \rangle \langle e_j |$ and $R_q \equiv |\psi \rangle \langle \phi_{q}|$ as
\begin{equation}
L_j  = P_{|\psi \rangle}\epsilon _{|\psi\rangle}  \epsilon _{|e_j \rangle }^{\dag} \;\;\; \text{and}\;\;\; R_q  = P_{|\psi \rangle}\epsilon _{|\psi\rangle}  \epsilon _{|\phi_q\rangle}^{\dag},
\label{eq:17}
\end{equation}
where \smash{$P_{|\psi \rangle}\equiv|\psi \rangle \langle \psi |$} is a projector, and $\epsilon_{|A\rangle}$ is the unitary descending-order eigenvector matrix of $|A\rangle\langle A|$, for which columns of $\epsilon _{|A\rangle}$ are eigenstates of $|A\rangle\langle A|$, and \smash{$\epsilon _{|A\rangle}^{\dag}  A\epsilon _{|A\rangle}  = \Lambda _{|A\rangle} \equiv\text{diag}\{\lambda_1 ,\ldots,\lambda_n\}$} is the diagonal eigenvalue matrix of $n$-level matrix $|A\rangle\langle A|$.  Thus, each $L_j$ and $R_q$ can be ``realized'' by a unitary gate and projective filter.  Note that $|e_{j}\rangle$ are eigenstates of $\rho$ and that $\epsilon _{|e_j \rangle }$ is the descending-order eigenvector matrix of $|e_{j}\rangle\langle e_{j}|$.

If we choose $|\psi \rangle$ as the first computational basis state $|1\rangle$, then we can set \smash{$\epsilon _{|\psi\rangle}   = \epsilon _{|1\rangle}  = I$}, and (\ref{eq:17}) simplifies to
\begin{equation}
L_j  = P_{|1 \rangle} \epsilon _{|e_j \rangle }^{\dag} \;\;\; \text{and}\;\;\; R_q  = P_{|1 \rangle}  \epsilon _{|\phi_q\rangle}^{\dag}.
\label{eq:18}
\end{equation}

As \Sec{II.D.4} explains, \textit{true} projectors may be difficult (or impossible) to achieve, since \textit{absorptive} projective filters such as polarizers \textit{do not produce scalar zero from orthogonal input}, but instead produce the vacuum state.  However, for light-speed limited (light-limited) communication, absorptive projectors will be adequate.
\subsubsection{\label{sec:II.D.4}True Projectors Cannot be Implemented with Absorptive Projective Filters}
Consider a state in an orthonormal basis $\{|\widetilde{1}\rangle,|\widetilde{2}\rangle\}$ as
\begin{equation}
|\psi \rangle  = d|\widetilde{1}\rangle  + e|\widetilde{2}\rangle,
\label{eq:19}
\end{equation}
where $|d|^2 +|e|^2=1$, and the tilde is used since this basis is not necessarily composed of Fock states, but our results here will involve the Fock vacuum \smash{$|0\rangle$}.

Now, a \textit{true projector}, such as \smash{$P_{|\widetilde{1}\rangle}\equiv|\widetilde{1}\rangle\langle\widetilde{1}|$} causes
\begin{equation}
P_{|\widetilde{1}\rangle } |\psi \rangle\langle\psi | P_{|\widetilde{1}\rangle } = \left\{ {\begin{array}{*{20}l}
   {|d|^2 |\widetilde{1}\rangle \langle \widetilde{1}| ;} & {|\psi \rangle  \ne |\widetilde{2}\rangle }  \\
   {\;\;\;\;\;\;\;\;\,{\kern 0.4pt}0;} & {|\psi \rangle  = |\widetilde{2}\rangle, }  \\
\end{array}} \right.
\label{eq:20}
\end{equation}
before normalization, where the scalar zero in the second line is due to the fact that the overlap of the pass-state \smash{$|\widetilde{1}\rangle$} with an orthogonal state \smash{$|\widetilde{2}\rangle$} is zero; \smash{$\langle\widetilde{1}|\widetilde{2}\rangle=0$}. (We write separate cases to make a point.)

However, as shown in \App{App.B}, \textit{absorptive} projective filters (APFs) such as linear polarizers do \textit{not} produce the scalar zero, and instead produce the vacuum state, as
\begin{equation}
P_{|\widetilde{1}\rangle }^{(\text{APF})} (|\psi \rangle\langle\psi| ) =\! \left\{\! {\begin{array}{*{20}l}
   {|d|^2 |\widetilde{1}\rangle \langle \widetilde{1}| + |e|^2 |0\rangle \langle 0| ;} & {|\psi \rangle  \ne |\widetilde{2}\rangle }  \\
   {\;\;\;\;\;\;\;\;\;\;\;{\kern 2.6pt}|0\rangle\langle 0|;} & {|\psi \rangle  = |\widetilde{2}\rangle, }  \\
\end{array}} \right.
\label{eq:21}
\end{equation}
where the superscript $\text{APF}$ reminds us that \smash{$P_{|\widetilde{1}\rangle }^{(\text{APF})}$} is really a selective absorption channel, which is why it merely produces the vacuum state in the orthogonal-input case.

Furthermore, the ``pass-input'' case of the APF is a \textit{mixture} with the vacuum, whereas the corresponding case in (\ref{eq:20}) is a \textit{pure} state.

As we will see in \Sec{II.E}, the ability of true projectors to produce scalar zero in the case of limited-direct UQEC is the key ingredient for the possibility of superluminal information transfer.  Without it, one must use \textit{postselection}, which we will discuss in \Sec{II.F}.
\subsection{\label{sec:II.E}How to Achieve Noise-Resistant Quantum Teleportation with UQEC}
Before using UQEC within a communication device, we need to incorporate quantum teleportation (QT), which we define as \textit{transfer of a quantum state from one spatial mode to another spatial mode}, where \textit{mode} simply means subsystem of the total Hilbert space.  Since communication involves transfer of information from one place to another, then our noise-resistant quantum teleportation (NRQT) scheme must be able to maintain its UQEC action while achieving QT.

Traditionally, QT involves entanglement \cite[]{BBCJ,BPM1,BPM2}, but actually that is unnecessary.  For example, an ideal (lossless) phase shifter acts as a kind of teleporter to flying states, at the cost of some phase factors, since it transmits the state from the spatial mode at its input port to the spatial mode at its output port (see \App{App.C}).

Here we will show that the UQEC process still works if we apply the recovery and decoding stages in a \textit{different} location than the encoding.  To start, suppose that our input is $\rho  \otimes \upsilon$, and recalling that $L_j  \equiv |\psi \rangle \langle e_j |$, the encoded state results are
\begin{equation}
\begin{array}{*{20}l}
   {L_j \rho L_j ^{\dag}  \otimes \upsilon} &\!\! { = |\psi \rangle \langle e_j |\rho |e_j \rangle \langle \psi | \otimes \upsilon}  \\
   {} &\!\! { = \lambda _j |\psi \rangle \langle \psi | \otimes \upsilon},  \\
\end{array}
\label{eq:22}
\end{equation}
where $\lambda _j  \equiv \langle e_j |\rho |e_j \rangle$, and $|e_j \rangle$ are eigenstates of $\rho$.  Now suppose that some noise channel $\mathcal{E}$ acts, possibly on \textit{both} spatial modes, as
\begin{equation}
\mathcal{E}(L_j \rho L_j ^{\dag}   \otimes \upsilon) = \lambda _j \mathcal{E}(\varsigma _{(\psi ,\upsilon)} ),
\label{eq:23}
\end{equation}
where $\varsigma _{(\psi ,\upsilon)}  \equiv |\psi \rangle \langle \psi | \otimes \upsilon$.  Then, recalling that $R_q  \equiv |\psi \rangle \langle \phi _q |$, if we apply the $R_q$ in the \textit{second spatial mode}, and sum over all of them since we do not know which is applied at any time, we get
\begin{equation}
\mathcal{R}^{(2)} (\lambda _j \mathcal{E}(\varsigma _{(\psi ,\upsilon)} )) = \lambda _j \text{tr}_{2} (\mathcal{E}(\varsigma _{(\psi ,\upsilon)} )) \otimes |\psi \rangle \langle \psi |,
\label{eq:24}
\end{equation}
where $\mathcal{R}^{(2)} (\sigma ) \equiv \sum\nolimits_{q = 1}^n {(I^{(1)}  \otimes R_q )\sigma (I^{(1)}  \otimes R_q ^{\dag}  } )$, and parenthetical superscripts are subsystem labels.

Finally, letting $\eta_{j}\equiv\mathcal{R}^{(2)} (\lambda _j \mathcal{E}(\varsigma _{(\psi ,\upsilon)} ))$ and recalling that $L_j  \equiv |\psi \rangle \langle e_j |$, decoding in subsystem 2 gives results
\begin{equation}
(I^{(1)}  \otimes L_j ^{\dag}  )\eta_{j}(I^{(1)}  \otimes L_j ) = \text{tr}_{2} (E(\varsigma _{(\psi ,\upsilon)} )) \otimes \lambda _j |e_j \rangle \langle e_j |,
\label{eq:25}
\end{equation}
where we used the fact that $\lambda_j$ is a scalar to shift it to subsystem 2.  Finally, summing over all the results gives
\begin{equation}
\sum\limits_{j = 1}^n {(I^{(1)}  \otimes L_j ^{\dag}  )\eta _j (I^{(1)}  \otimes L_j )}  = \text{tr}_{2} (\mathcal{E}(\varsigma _{(\psi ,\upsilon)} )) \otimes \rho ,
\label{eq:26}
\end{equation}
which we can also express as
\begin{equation}
\sum\limits_{j = 1}^n {\sum\limits_{q = 1}^n {\sum\limits_k \!{\Omega _{(j,q,k)} (\rho  \otimes \upsilon)\Omega _{(j,q,k)}^{\dag}  } } }\!  = \text{tr}_{2} (\mathcal{E}(\varsigma _{(\psi ,\upsilon)} )) \otimes \rho ,
\label{eq:27}
\end{equation}
where $\Omega _{(j,q,k)}  \equiv (I \otimes L_j ^{\dag}  R_q )E_k (L_j  \otimes I)$, and $ \text{tr}_{2} (\mathcal{E}(\varsigma _{(\psi ,\upsilon)} ))$ is some ``garbage state'' about which we do not care.  Thus, we have shown that we can indeed successfully teleport $\rho$, the family of states of constant eigenstates $|e_{j}\rangle$, in the presence of any noise channel $\mathcal{E}$, \textit{even nonlocal noise over both modes of the teleportation}.

We have just \textit{mathematically} described \textit{superluminal communication}\ldots however, we will now see that \textit{how} we realize the operators can cause major limitations.
\subsection{\label{sec:II.F}NRQT via Passive Absorptive Projectors is Light-Speed Limited and Needs Lossless Connection}
Notice that the crucial step in (\ref{eq:25}) was that the scalar $\lambda_j$ could slide from the sender's mode to the receiver's mode, thus enabling proper reconstruction of the input state in a different location.

Focusing on the encoding, the ideal action should be
\begin{equation}
\begin{array}{*{20}l}
   {L_j \rho L_j ^{\dag}  } &\!\! { = |e_{1} \rangle \langle e_j |(\sum\limits_{k = 1}^n {\lambda _k |e_k \rangle \langle e_k |} )|e_j \rangle \langle e_{1} |}  \\
   {} &\!\! { = |e_{1} \rangle \langle e_{1} |\sum\limits_{k = 1}^n {\lambda _k \delta _{j,k} }  = \lambda _j |e_{1} \rangle \langle e_{1} |,}  \\
\end{array}
\label{eq:28}
\end{equation}
where we have assumed the eigenbasis to be the computational basis, and the reference state is the first basis state.  From this, we see that there should be an appearance of a scalar Kronecker delta to properly sift $\lambda_j$.

However, if we use \textit{absorptive} projective filters, then as in \App{App.B}, letting \smash{$\rho_{L_j}^{(\text{APF})}\equiv P_{|e_1 \rangle }^{(\text{APF})} (L_j \rho L_j ^{\dag})$} where \smash{$P_{|e_1 \rangle }^{(\text{APF})} (\rho ) \equiv \text{tr}_2 ((I \otimes A_0 ( \cdot ))\rho_{\text{DR}})$} is a selective absorption channel where \smash{$\rho_{\text{DR}} \equiv U_{\text{PBS}} (\rho  \otimes |0\rangle \langle 0|)U_{\text{PBS}} ^{\dag}$} is a dual-rail version of input $\rho$ such that the pass-state is kept in mode $1$, \smash{$U_{\text{PBS}}$} is a polarizing beam-splitter unitary, and \smash{$\mathcal{A}_0(\rho)=|0\rangle\langle 0|\;\forall\rho$} is the absorption channel (see \Sec{III.A}), we get
\begin{equation}
\begin{array}{*{20}l}
   {\rho_{L_j}^{(\text{APF})}  } &\!\! { = P_{|e_1 \rangle }^{(\text{APF})}\!\left({ \epsilon _{|e_j \rangle }^{\dag}  (\sum\limits_{k = 1}^n {\lambda _k |e_k \rangle \langle e_k |} )\epsilon _{|e_j \rangle } }\right) }  \\
   {} &\!\! { = P_{|e_1 \rangle }^{(\text{APF})}\! \left({\lambda _j |e_1 \rangle \langle e_1 | +\!\! \sum\limits_{k = 1 \ne j}^n \!\!\!{\lambda _k \epsilon _{|e_j \rangle }^{\dag}|e_k \rangle \langle e_k |\epsilon _{|e_j \rangle }} }\right) }  \\
   {} &\!\! { = \lambda _j |e_1 \rangle \langle e_1 | + (1 - \lambda _j )|0\rangle \langle 0|,}  \\
\end{array}
\label{eq:29}
\end{equation}
which shows that \textit{the only way to tell if the correct scalar is present is to wait for a nonvacuum event}.

Thus, the use of absorptive projective filters (APFs) causes the communication of the scalar information $\lambda_j$ to be \textit{event-based} and \textit{comparative to vacuum}.  Because the nonvacuum state must belong to a particle, this means that \textit{the use of APFs restricts us to light-speed communication}, because we must wait for the particle to reach the receiver and thus communicate the nonvacuum event.

If we use \textit{passive} APFs, such as polarizers, the need to wait for a nonvacuum event has another consequence; we must not allow the encoded state to experience further absorption, or else the whole state will be vacuum with unit probability as
\begin{equation}
\mathcal{A}_0 (\rho_{L_j}^{(\text{APF})} ) = \lambda _j |0\rangle \langle 0| + (1 - \lambda _j )|0\rangle \langle 0| = |0\rangle \langle 0|,
\label{eq:30}
\end{equation}
which conveys no scalar information.

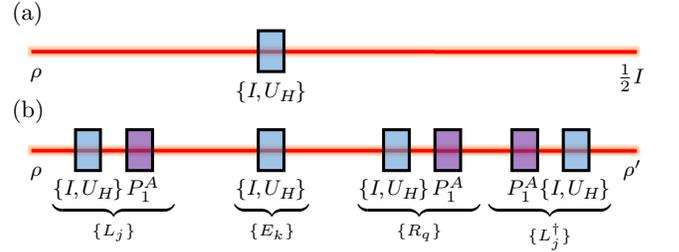
\begin{figure}[H]
\centering
\includegraphics[width=0.99\linewidth]{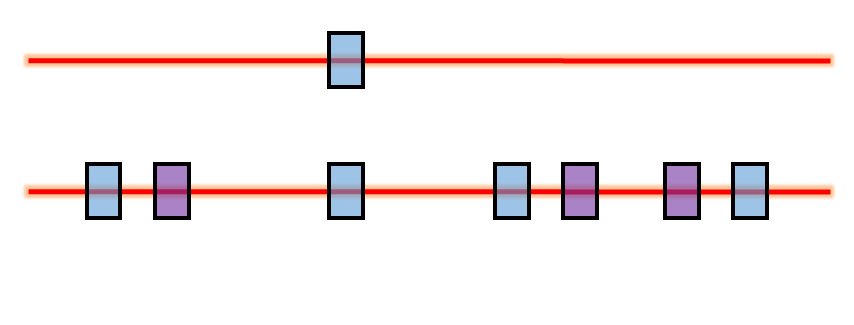}%
\vspace{-12pt}
\setlength{\unitlength}{0.01\linewidth}
\begin{picture}(100,0)
\put(0.7,35){\small (a)}
\put(0.7,20){\small (b)}
\put(3.5,26){\footnotesize $\rho$}
\put(3.5,10.9){\footnotesize $\rho$}
\put(93.8,25.2){\footnotesize $\frac{1}{2}I$}
\put(94.7,10.9){\footnotesize $\rho'$}
\put(35.1,23){\footnotesize $\{{\kern 0.0pt}I{\kern -1.0pt},{\kern -0.5pt}U_{H}{\kern -1.0pt}\}$}
\put(35.1,8){\footnotesize $\{{\kern 0.0pt}I{\kern -1.0pt},{\kern -0.5pt}U_{H}{\kern -1.0pt}\}$}
\put(35,7){\footnotesize $\underbrace{\kern 28pt}_{\{ E_k \}}$}
\put(7,8){\footnotesize $\{{\kern 0.0pt}I{\kern -1.0pt},{\kern -0.5pt}U_{H}{\kern -1.0pt}\}$}
\put(18.3,8){\footnotesize $P_{1}^{A}$}
\put(7,7){\footnotesize $\underbrace{\kern 46pt}_{\{ L_j \}}$}
\put(54,8){\footnotesize $\{{\kern 0.0pt}I{\kern -1.0pt},{\kern -0.5pt}U_{H}{\kern -1.0pt}\}$}
\put(65.3,8){\footnotesize $P_{1}^{A}$}
\put(54.0,7){\footnotesize $\underbrace{\kern 46pt}_{\{ R_q \}}$}
\put(76.8,8){\footnotesize $P_{1}^{A}$}
\put(81.8,8){\footnotesize $\{{\kern 0.0pt}I{\kern -1.0pt},{\kern -0.5pt}U_{H}{\kern -1.0pt}\}$}
\put(74.1,7){\footnotesize $\underbrace{\kern 46pt}_{\{ L_j^{\dag} \}}$}
\end{picture}
\caption[]{(color online) Example of light-limited noise-resistant quantum teleportation of \smash{$\rho=c_{\theta}^{2}|\widetilde{1}\rangle\langle \widetilde{1}|+s_{\theta}^{2}|\widetilde{2}\rangle\langle \widetilde{2}|$} as in (\ref{eq:4}), where \smash{$\{|\widetilde{1}\rangle,|\widetilde{2}\rangle\}$} are orthonormal single-photon polarization states. (a) Input $\rho$ is subjected to an emulated bit-flip channel in which 50\% of the trials a half-wave plate (HWP) \smash{$U_{H}$} is placed, and 50\% of the trials get nothing, yielding an output (if information about the HWP is ignored) of \smash{$\frac{1}{2}I$}, the maximally mixed state.  (b) The same emulated noise but now within the UQEC process using absorptive polarizers \smash{$P_{1}^{A}\equiv P_{|1\rangle}^{(\text{APF})}(\cdot)$}, producing \smash{$\rho'\equiv{\textstyle{1 \over 4}}\rho  + {\textstyle{3 \over 4}}|0\rangle \langle 0|$}.  Postselecting for nonvacuum results, tomography on $\rho'$ yields the message state \smash{$\rho$} in spite of having the same noise as (a).}
\label{fig:5}
\end{figure}

As \Fig{5} shows, light-limited NRQT is possible.  The restrictions of light-limiting and lossless transmission are honored by maintaining a line-of-sight connection and by restricting the error channel to the basis of the qubit.  In this kind of setup, ``all errors'' means ``all errors within the space of the input state.''
Thus, using passive APFs rather than true projectors limits the teleportation to light speed, and imposes the further requirement that the sender and receiver be connected by a lossless transmission line.

Later we will see the particularly interesting alternative that if we use an \textit{active} APF in the form of a detector at the sender, we can achieve ``connection-free'' postselected light-speed communication, where there still needs to be a path to send the classical postselection results, but the quantum part can and does get completely destroyed and therefore does not require a connection.
\subsection{\label{sec:II.G}A Way to Achieve True Projectors}
\textit{Mathematically}, a joint-system two-qubit unitary that yields orthogonal projectors is the CNOT gate,
\begin{equation}
U_{\text{CNOT}}  = \left( {\begin{array}{*{20}c}
   1 & 0 & 0 & 0  \\
   0 & 1 & 0 & 0  \\
   0 & 0 & 0 & 1  \\
   0 & 0 & 1 & 0  \\
\end{array}} \right)\!.
\label{eq:31}
\end{equation}
Preparing the qubits as \smash{$\rho\otimes|\widetilde{1}\rangle\langle\widetilde{1}|$} so the target qubit is in the first computational basis state, then applying \smash{$U_{\text{CNOT}}$} and tracing over the target yields Kraus operators,
\begin{equation}
P_{|\widetilde{r}\rangle}  = (I \otimes \langle \widetilde{r}|)U_{\text{CNOT}} (I \otimes |\widetilde{1}\rangle ) = \{ P_{|\widetilde{1}\rangle } ,P_{|\widetilde{2}\rangle } \} ,
\label{eq:32}
\end{equation}
where \smash{$P_{|\widetilde{r}\rangle } \equiv |\widetilde{r}\rangle \langle \widetilde{r}|$}.  \textit{Physically}, although we only want \textit{one} projector, we need \textit{both} projectors in (\ref{eq:32}) for Kraus completeness.  However, we do not want both simultaneously, because then both sides of the communicator could not be synchronized.  Therefore, suppose these projectors \textit{happen in different time bins of a time-bin qubit}.

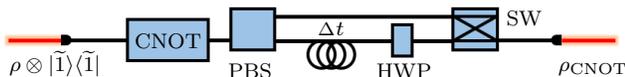
\begin{figure}[H]
\centering
\includegraphics[width=0.99\linewidth]{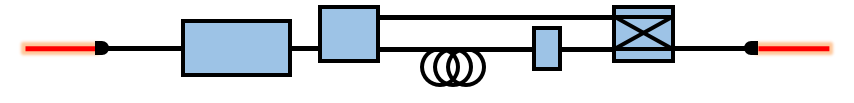}%
\vspace{-12pt}
\setlength{\unitlength}{0.01\linewidth}
\begin{picture}(100,0)
\put(3.5,1.1){\footnotesize $\rho\otimes|\widetilde{1}\rangle\langle\widetilde{1}|$}
\put(88,1.1){\footnotesize $\rho_{\text{CNOT}}$}
\put(22.8,4.6){\footnotesize CNOT}
\put(37.2,0){\footnotesize PBS}
\put(60,0){\footnotesize HWP}
\put(51,6.55){\scriptsize $\Delta$}
\put(53.8,6.55){\footnotesize $t$}
\put(80,8){\footnotesize SW}
\end{picture}
\caption[]{(color online) Example of a time-separated true-projector channel (see \App{App.D} for details). A photon's internal polarization and momentum qubits enter a deterministic CNOT \cite[]{FiWo}, then time-bin conversion \cite[]{BSBL} and reduction yield \smash{$\rho_{\text{CNOT}}=\rho_{1,1}|\widetilde{1}_{\tau_1},0_{\tau_2}\rangle\langle \widetilde{1}_{\tau_1},0_{\tau_2}|\!+\!\rho_{2,2}|0_{\tau_1},\widetilde{1}_{\tau_2}\rangle\langle 0_{\tau_1},\widetilde{1}_{\tau_2}|$}.}
\label{fig:6}
\end{figure}

\Figure{6} shows how to make a projector channel where the same \textit{true} polarization projector appears to act at different times, so the only difference is the time-bin 2 scalar, easily dealt with by treating the input of time-bin 2 as if it were a bit-flipped version of that of time-bin 1.
In \Sec{V}, we will see that while (\ref{eq:32}) \textit{does} achieve true projectors, it still leads to ansible failure.
\subsection{\label{sec:II.H}Insulating UQEC Enables Locally Generated Superposition Scalars}
Another possible problem for an ansible is the meaning of the scalars based on how the input state was made.  

If we create $\rho$ as a step-function of pure states, then scalars $\lambda_j$ are just \textit{statistical} representations of classical events.  However, if we instead obtain $\rho$ as the \textit{reduced state} from the \textit{pure} state of a larger system, then scalars $\lambda_j$ are directly inherited from pure-state superposition scalars, and are products of wave-function overlaps.

However, since the momentum ancilla is for the projectors, and since deterministic preparation of two-qubit photonic states is difficult, it would be nice if we could just use \textit{local pure states} as input, instead of reductions.  That way we could harvest the superposition scalars directly.  The problem is that limited-direct UQEC can only convey information with mixed-state input.

Fortunately, the method of \textit{insulating} UQEC from \App{App.E} can handle \textit{all} input states, not just those of known eigenstates. Therefore, by restricting ourselves to pure input states and insulating UQEC, we can guarantee locally-prepared superposition scalars in our input.

\Appendix{App.E} gives the details of insulating UQEC, but the only procedural differences for one qubit are that we now need \textit{six} encoding/decoding pairs, and the output state $\rho_D$ is \textit{isomorphic} to the input $\rho$, meaning that an extra numerical computation called \textit{extraction} must be done to obtain $\rho$ from $\rho_D$.

The important thing is that with insulating UQEC, $\rho$ can be \textit{pure}, such as $\rho=|\xi\rangle\langle\xi|$, where
\begin{equation}
|\xi\rangle  = c_\theta  |\widetilde{1}\rangle + s_\theta  e^{i\phi } |\widetilde{2}\rangle,
\label{eq:33}
\end{equation}
where $c_{\theta}\equiv \cos(\theta)$ and $s_{\theta}\equiv \sin(\theta)$. Then, using the overcomplete basis given in \App{App.E}, the information-carrying scalars are (instead of $\lambda_j$),
\begin{equation}
\begin{array}{*{20}l}
   {d_1  = {\textstyle{1 \over 2}}(1 + s_{2\theta } c_\phi  ),} & {d_3  = {\textstyle{1 \over 2}}(1 + s_{2\theta } s_\phi  ),} & {d_5  = c_\theta^2 ,}  \\
   {d_2  = {\textstyle{1 \over 2}}(1 - s_{2\theta } c_\phi  ),} & {d_4  = {\textstyle{1 \over 2}}(1 - s_{2\theta } s_\phi  ),} & {d_6  = s_\theta^2 ,}  \\
\end{array}
\label{eq:34}
\end{equation}
which all obey $d_j\geq 0$, so the quantum operation is physical, and most importantly, these scalars are \textit{directly inherited from the pure superposition scalars of (\ref{eq:33})}, provided that we use true projectors.

Thus we now have a strategy for locally producing pure input states that ensure our UQEC operation only uses \textit{superposition scalars}, which may avoid the light-speed limits of event-based statistical scalars.

For simplicity, in everything that follows, we will use the limited-direct UQEC variables, since technically we can get superposition scalars in mixed states by reducing from a larger pure multiqubit system, but keep in mind the advantages of using insulating UQEC instead.
\subsection{\label{sec:II.I}Vacuum/Single-Photon Basis Conversion}
For single-photon (SP) qubits, the all-mode vacuum is not part of their Hilbert space, yet general noise channels can easily take the state \textit{outside} the SP subspace.

One way to handle this is to \textit{force} vacuum involvement by converting encoded results to vacuum.  If we use the methods of \Sec{II.H} and \Sec{II.G}, our encoded results are all proportional to the same reference state, such as
\begin{equation}
L_j \rho L_j ^{\dag}   = \lambda _j |1\rangle \langle 1|\;\;\,\text{or}\;\;D_j \rho D_j ^{\dag}   = d_j |1\rangle \langle 1|,
\label{eq:35}
\end{equation}
where the scalars are true overlaps of pure states and are \textit{not} event-based.  Then, it is easy to see that conversion to vacuum via total absorption $\mathcal{A}_0$ preserves the scalars,
\begin{equation}
\mathcal{A}_{0}(D_j \rho D_j ^{\dag}) = d_j \mathcal{A}_{0}(|1\rangle \langle 1|)=d_j|0\rangle \langle 0|.
\label{eq:36}
\end{equation}
Later, we will see an example showing what happens at the receiver; the important point here is that (\ref{eq:36}) shows that (if projectors exist) no classical signal needs to leave the encoding half of the device, vacuum will suffice!
\section{\label{sec:III}How to Build an Ansible if True Lone Projectors were Possible}
Supposing that true lone projectors could be realized, here we organically show how to build an ansible, a device for effectively superluminal communication.
\subsection{\label{sec:III.A}Using NRQT for Communication in the Context of the Quantum Vacuum}
For the purposes of communication by quantum states, suppose that we have some source of flying qubits capable of making a steady stream of identically prepared states.  Also suppose that these states can be accurately measured by tomography in reasonable amounts of time.  We will further suppose that our qubits are optical, and therefore the first thing we must confront is how we can achieve NRQT in the context of the quantum vacuum.

To begin, we define the \textit{attenuation channel} as a variable-transmittivity beam splitter \smash{$B_\theta \equiv e^{\theta(a_{1}^{\dag}a_{2}-a_{1}a_{2}^{\dag})}$} of transmittivity $T$, where $a_1$ and $a_2$ are annihilation operators for modes $1$ and $2$, and \smash{$\cos(\theta)\equiv\sqrt{\,\!T^{\,^{\,}}\!\!}$}, with vacuum in its auxiliary input (mode $2$), and reduced as
\begin{equation}
\mathcal{A}_{T}(\rho)\equiv\mathcal{A}_{\theta(T)}(\rho)\equiv\text{tr}_{2}(B_\theta(\rho\otimes|0^{(2)}\rangle\langle 0^{(2)}|) B_\theta^{\dag}).
\label{eq:37}
\end{equation}
This model of loss acknowledges that total absorption never yields just ``$0$'', but instead outputs the vacuum state $|0\rangle$, since absorption is just part of a unitary process on a larger Hilbert space. 

Now suppose we must send a state $\rho$ over a lossy transmission line modeled as an ideal unitary phase shifter $U$ of any length, preceded by an attenuation channel $\mathcal{A}_T$ of any value $T$, as in \Fig{7}. Then the output is
\begin{equation}
\rho ' = U\mathcal{A}_T (\rho )U^{\dag} .
\label{eq:38}
\end{equation}
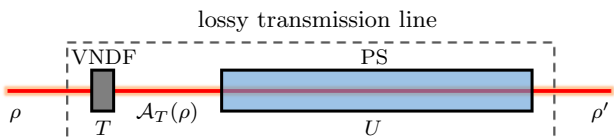
\begin{figure}[H]
\centering
\includegraphics[width=0.99\linewidth]{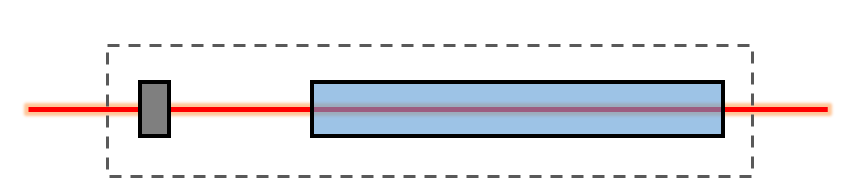}%
\vspace{-12pt}
\setlength{\unitlength}{0.01\linewidth}
\begin{picture}(100,0)
\put(4,5){\footnotesize $\rho$}
\put(93.5,5){\footnotesize $\rho'$}
\put(33,19){\small lossy transmission line}
\put(13.5,13.3){\footnotesize VNDF}
\put(58,13.3){\footnotesize PS}
\put(17.2,2.35){\footnotesize $T$}
\put(23.5,5){\footnotesize $\mathcal{A}_{T}(\rho)$}
\put(58.6,2.35){\footnotesize $U$}
\end{picture}
\caption[]{(color online) Lossy transmission line modeled as a variable neutral density filter (VNDF) represented by channel $\mathcal{A}_T(\rho)$ of (\ref{eq:37}) and an ideal phase shifter (PS) or fiber represented by unitary $U$. When \smash{$T\!=\!0$}, \smash{$\mathcal{A}_{0}(\rho)\!=\!|0\rangle\langle 0|$}, and \smash{$\rho'\!\equiv\! U\mathcal{A}_{0}(\rho)U^{\dag}\!=\!|0\rangle\langle 0|$}, where \smash{$U\equiv e^{i\phi a^{\dag}a}$}, with annihilation operator $a$, and $\phi\equiv\omega\Delta t$ where $\omega$ is the field's angular frequency and $\Delta t\equiv(n_{P}-n_M)\frac{Z}{c}$ is the extra time delay for the field to propagate through the phase shifter of length $Z$ and refractive index $n_P$ beyond the time it would take in the surrounding medium of index $n_M$, where $c$ is the speed of light in vacuum.  Any error channels inserted at any point in this sequence can all be combined with these elements as a single error channel to model a \textit{noisy} transmission line, as well.}
\label{fig:7}
\end{figure}
Note that although we use the standard single-mode model of a phase shifter, they are really two-mode devices, as mentioned earlier, which is partly responsible for the teleportation effect.  See \App{App.C} for a two-mode model of a phase shifter.

The \textit{worst-case} scenario is if the transmission probability is $T=0$, corresponding to placing a perfect absorber before the line, for which the output is
\begin{equation}
\rho ' = U\mathcal{A}_{0}(\rho )U^{\dag}   = \sum\limits_{k = 0}^\infty  {U{A_{0}}_k \rho {A_{0}}_k^{\dag}  } U^{\dag}   \equiv \sum\limits_{k = 0}^\infty  {E_k \rho E_k^{\dag}  },
\label{eq:39}
\end{equation}
where the net Kraus operators are $E_k  \equiv U{A_{0}}_k $ where ${A_{T}}_k  \equiv (I^{(1)}  \otimes \langle k^{(2)} |)B_{\theta (T)} (I^{(1)}  \otimes |0^{(2)} \rangle )$ are Kraus operators of the attenuation channel, so that here where $T=0$, $\rho'=|0\rangle\langle 0|$.  Thus, we see that even the \textit{worst} transmission line can be viewed as one big error channel \smash{$\mathcal{E}(\rho ) = \sum\nolimits_{k = 0}^\infty  {E_k \rho E_k^{\dag}  }$}.  Furthermore, if any \textit{additional} error channels act along the way, they can simply be lumped into the total definition of $\mathcal{E}$.

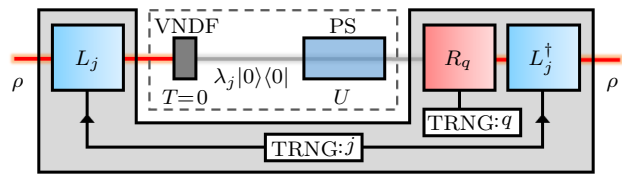
\begin{figure}[H]
\centering
\includegraphics[width=0.99\linewidth]{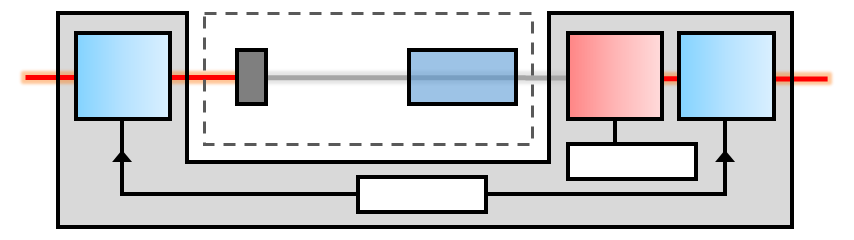}%
\vspace{-12pt}
\setlength{\unitlength}{0.01\linewidth}
\begin{picture}(100,0)
\put(3,14){\footnotesize $\rho$}
\put(34,14.5){\footnotesize $\lambda_{j}|0\rangle\langle 0|$}
\put(94.5,14){\footnotesize $\rho$}
\put(24.88,22.25){\footnotesize VNDF}
\put(51.8,22.25){\footnotesize PS}
\put(26.1,11.40){\footnotesize $T\!\!=\!0$}
\put(52.5,11.40){\footnotesize $U$}
\put(66.9,7.30){\footnotesize TRNG}
\put(77.1,7.85){\footnotesize :}
\put(78.3,8.05){\footnotesize $q$}
\put(42.9,3.30){\footnotesize TRNG}
\put(53.1,3.85){\footnotesize :}
\put(54.3,4.05){\footnotesize $j$}
\put(12.7,17.4){\footnotesize $L_j$}
\put(69.5,17.4){\footnotesize $R_q$}
\put(82.5,17.4){\footnotesize $L_j^{\dag}$}
\end{picture}
\caption[]{(color online) Fully attenuating transmission line (enclosed by dashed line) with limited-direct UQEC applied as a communication protocol to recover the input state. Transmission line definitions match those in \Fig{7}, and the UQEC notation is that of \Fig{4}. The large light-gray region is a sealed box indicating that the results of \textit{which} operators are applied do not leave the box.  Just after the VNDF, a particular result is $\lambda_{j}|0\rangle\langle 0|$, where $\lambda_{j}\equiv\langle e_{j}|\rho|e_{j}\rangle$ is a scalar.  Thus only vacuum exists within the transmission line, but the recovery and decoding \textit{still} recover $\rho$, if true projectors are used.}
\label{fig:8}
\end{figure}

This means that if we use UQEC as in \Fig{8} to encode prior to the transmission line, then  after recovery and decoding, we can get $\rho$ at the end of the line, \textit{even though we assumed total absorption of the input state!}
Thus, using UQEC as a communication protocol, if we use the paradigm of sending many identically prepared systems in the same state, then the party on the receiving side can tomographically measure $\rho$, regardless of how noisy the transmission line is.  

The message sent can be encoded into $\rho$ by some pre-agreed method, such as chopping up its scalars into accurately discernible intervals and assigning letter values to each (see \Sec{V} for more details about this).  The states sent could also each last for a certain time period, and series of states can be sent in a cycle.  The method of message encoding is trivial.  The amazing thing here is that \textit{we may not even need a transmission line at all!}

This last observation comes from the fact that we assumed total absorption prior to the transmission line and were still able to treat it as a correctable error channel.  If true, some of the startling implications of this are:
\begin{itemize}
\item[$1.$]\hypertarget{imp:1}{}No transmission lines are needed, and the encoded state can be into a perfect absorber (see (\ref{eq:36})).
\item[$2.$]\hypertarget{imp:2}{}Messages can be sent as states across \textit{any distance}.
\item[$3.$]\hypertarget{imp:3}{}There is no classical transmission signal.
\item[$4.$]\hypertarget{imp:4}{}Messages can be received faster than a classical signal can send them.  (See \Sec{III.G} for why this does not violate special relativity, but still permits \textit{effective} superluminal communication.)
\item[$5.$]\hypertarget{imp:5}{}Alignment is irrelevant; relative position of sender and receiver is irrelevant because all information about the input state is encoded in subsystem-independent scalars \smash{$\lambda_j$}.
\item[$6.$]\hypertarget{imp:6}{}Messages cannot be classically eavesdropped, since the sender's output is always vacuum.
\item[$7.$]\hypertarget{imp:7}{}Messages cannot be corrupted or faked by interference in transit, since any mid-transit operation is lumped with the total error and corrected.
\end{itemize}

In the next few sections, we discuss the application of the above hypothetical implications, including how to realize them and why we should even consider them, given that they seem to contradict known physics.

As we will discuss in \Sec{IV} and the \hyperlink{Concl}{Conclusions}, all of these seemingly impossible implications depend on the existence of true projection operations.  Thus throughout \Sec{III}, we suppose that true projectors are possible, and then use the extraordinary consequences of that supposition to suggest a ``\hyperlink{NoProjThm}{no-projector theorem}'' in \Sec{IV}.
\subsection{\label{sec:III.B}Design of an Ansible}
The term \textit{ansible} was coined by science fiction author Ursula K. Le Guin to describe any device capable of superluminal communication, an idea dating back to at least Isaac Asimov's ``hyper-wave relay.''  While these were just plot devices, they happen to describe the application of teleporting unassisted QEC (UQEC) as a communication device with enforced total absorption.

\begin{figure}[H]
\centering
\includegraphics[width=0.99\linewidth]{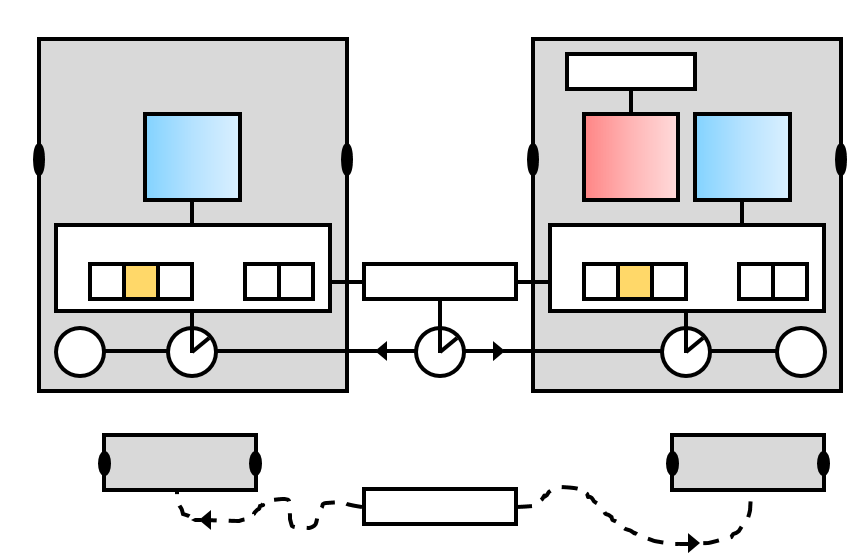}%
\vspace{-12pt}
\setlength{\unitlength}{0.01\linewidth}
\begin{picture}(100,0)
\put(0.7,61){\small (a)}
\put(0.7,13){\small (b)}
\put(17.3,54){\footnotesize encoder}
\put(82.7,56){\footnotesize recovery/}
\put(82.7,52.9){\footnotesize decoder}
\put(13,34.75){\footnotesize index memory}
\put(7.9,30.5){\footnotesize $j${\kern 0.03pt}:}
\put(12,30.4){\footnotesize $3$}
\put(15.8,30.4){\footnotesize $8$}
\put(19.8,30.4){\footnotesize $4$}
\put(23.6,30.4){\footnotesize $\cdots$}
\put(30,30.4){\footnotesize $5$}
\put(34,30.4){\footnotesize $2$}
\put(70.23,34.75){\footnotesize index memory}
\put(65.13,30.5){\footnotesize $j${\kern 0.03pt}:}
\put(69.23,30.4){\footnotesize $3$}
\put(73.03,30.4){\footnotesize $8$}
\put(77.03,30.4){\footnotesize $4$}
\put(81.18,30.4){\footnotesize $\cdots$}
\put(87.23,30.4){\footnotesize $5$}
\put(91.23,30.4){\footnotesize $2$}
\put(43.5,30.2){\footnotesize TRNG}
\put(53.7,30.75){\footnotesize :}
\put(56.5,30.65){\footnotesize $j$}
\put(54.8,30.67){\scriptsize $\{$}
\put(58.0,30.67){\scriptsize $\}$}
\put(66.9,54.50){\footnotesize TRNG}
\put(77.1,55.05){\footnotesize :}
\put(78.3,55.25){\footnotesize $q$}
\put(7.9,22){\footnotesize IT}
\put(91.53,22){\footnotesize IT}
\put(21.85,24.9){\tiny $\blacktriangle$}
\put(23.45,24.1){\tiny \rotatebox[origin=c]{-48.15}{$\blacktriangle$}}
\put(50.60,24.9){\tiny $\blacktriangle$}
\put(52.20,24.1){\tiny \rotatebox[origin=c]{-48.15}{$\blacktriangle$}}
\put(79.14,24.9){\tiny $\blacktriangle$}
\put(80.74,24.1){\tiny \rotatebox[origin=c]{-48.15}{$\blacktriangle$}}
\put(20.6,45){\footnotesize $L_j$}
\put(71.6,45){\footnotesize $R_q$}
\put(84.4,45){\footnotesize $L_{j}^{\dag}$}
%
\put(46.45,4.285){\footnotesize factory}
\put(17.8,9.45){\footnotesize $\{{\kern 0.5pt}\!L_{j}\!{\kern 0.5pt}\}$}
\put(80.1,9.45){\footnotesize $\{{\kern 0.5pt}\!R_{q}\!{\kern 0.5pt}\}\,\!{\kern 0.5pt}\{{\kern 0.5pt}\!L_{j}^{\dag}\!{\kern 0.5pt}\}$}
\end{picture}
\caption[]{(color online) Preparation of a pair of ansible halves. (a) At the factory, a true random number generator (TRNG) generates an exhaustive random number sequence and writes it to an index memory in both halves, and no record of the sequence is retained externally. Internal clocks of the halves are synchronized to the factory clock so that they always point to the same index $j$ in both devices at the same time.  Inertial trackers (IT) within the halves apply relativistic corrections to the clocks to keep them synchronized without need for communication between them. All notation is that of previous figures. (b) After the sequence is loaded into the halves, they are disconnected and travel away from the factory. Thus, a matched pair of ansible halves contains index-linked encoding and decoding operators $L_j$ and $L_{j}^{\dag}$.}
\label{fig:9}
\end{figure}
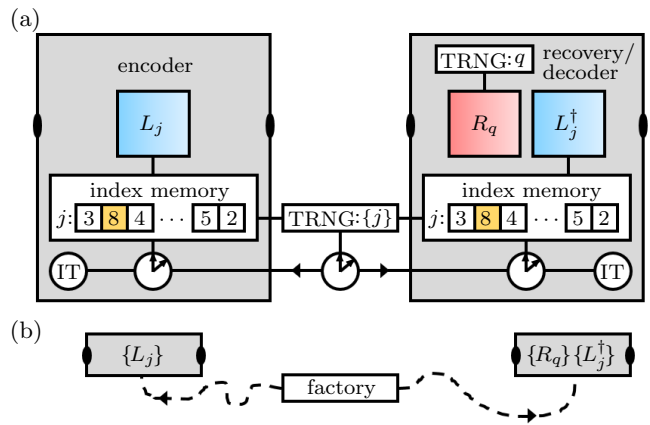

The implementation of an ansible is simple: two separate devices called \textit{ansible halves} (\textit{sender} and \textit{receiver}) are prepared in such a way that they enforce the index-linking of encoders \smash{$\{L_{j}\}$} and decoders \smash{$\{L_{j}^{\dag}\}$} without any communication between them (basically cutting the light-gray-box connection in \Fig{8}).  A general receiver also contains recovery operators \smash{$\{R_{q}\}$}, and both ansible halves must apply their operators with equal probability. The information about \textit{which} operators are applied must be kept inside the ansible halves in the classical sense, meaning that availability of this information is acceptable as long as users do not learn the outcomes.  \Figure{9} illustrates how this might be achieved.  (Later, in \Sec{V} we will see that the recovery operators can be omitted if proper steps are taken, but for now we keep them to highlight the ansible's origin as UQEC.)

The index linking of \smash{$\{L_{j},L_{j}^{\dag}\}$} may be achieved by using a single true random number generator (TRNG) to generate and store the same random string of integers in each ansible half.  This integer string must contain all the integers from each outcome at least once and in equal proportions.  Internal clocks are synchronized to the factory clock, and thereafter are kept running.  The clocks enforce the equal probability by selecting each index for an equal amount of time, while their synchronization enforces the index linking.  The randomness of the index generation ensures that the outside world is ignorant about \textit{which} pair \smash{$\{L_{j},L_{j}^{\dag}\}$} is applied.
After the factory synchronizes the ansible halves, they are disconnected and go off in their separate directions.  Both ansible halves contain inertial trackers (IT) which apply relativistic corrections to ensure that the clocks remain synchronized in spite of time-dilation effects.  The reference frame of the factory is the reference that both devices treat as ``correct.''  For example, if the sender is moving away from the factory, its IT causes its clock to cycle faster to match the factory clock.  If the \textit{receiver} moves away from the factory at a different velocity, \textit{its} IT adjusts \textit{its} clock to run faster as well, again to match the factory clock.  Thus, both ansible clocks remain synchronized without communicating with each other or with the factory.  For this reason, the index-matching is not achieved through classical communication, but rather, through classical \textit{correlation}.

Although the ansible halves \textit{could} be operated with a line-of-sight pathway between them, perhaps by directing the output beam of the sender through a noisy fiber, as depicted in \Fig[a]{10}, we have already seen in (\ref{eq:39}) that the fiber and any noise channel, \textit{including} full absorption can be grouped as one big correctable error channel, and thus, inserting a perfect absorber in the path as in \Fig[b]{10} does not change the output state, resulting in the effect of \textit{beam rebirth}, which we will discuss shortly.  Then, since only vacuum exists in the path, we can remove the fiber altogether and incorporate the absorber in the sender ansible half, as in \Fig[c]{10}.  Thus, for these reasons, we see that \hyperlink{imp:1}{Implications 1}, \hyperlink{imp:2}{2}, and \hyperlink{imp:3}{3} from \Sec{III.A} are each reasonable in the context of such an ansible pair.
\begin{figure}[H]
\centering
\includegraphics[width=0.99\linewidth]{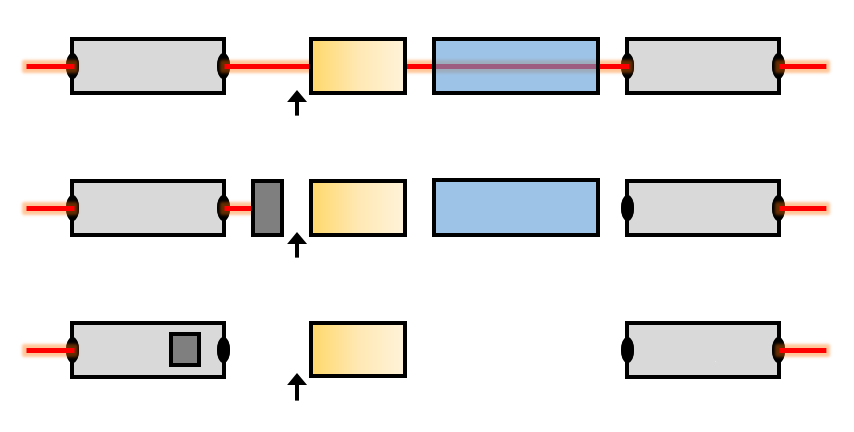}%
\vspace{-12pt}
\setlength{\unitlength}{0.01\linewidth}
\begin{picture}(100,0)
\put(0.7,46.5){\small (a)}
\put(0.7,30){\small (b)}
\put(0.7,13.5){\small (c)}
\put(4,38){\footnotesize $\rho$}
\put(93.5,38){\footnotesize $\rho$}
\put(4,21.5){\footnotesize $\rho$}
\put(93.5,21.5){\footnotesize $\rho$}
\put(4,5){\footnotesize $\rho$}
\put(93.5,5){\footnotesize $\rho$}
\put(26.70,33){\footnotesize $\lambda_{j}|\psi\rangle\langle\psi|$}
\put(27.30,16.5){\footnotesize $\lambda_{j}|0\rangle\langle 0|$}
\put(27.30,0){\footnotesize $\lambda_{j}|0\rangle\langle 0|$}
\put(39,41.1){\footnotesize $\mathcal{E(\cdot)}$}
\put(39,24.6){\footnotesize $\mathcal{E(\cdot)}$}
\put(39,8.1){\footnotesize $\mathcal{E(\cdot)}$}
\put(14.0,41.1){\footnotesize $\{{\kern 0.5pt}\!L_{j}\!{\kern 0.5pt}\}$}
\put(14.0,24.6){\footnotesize $\{{\kern 0.5pt}\!L_{j}\!{\kern 0.5pt}\}$}
\put(11.4,8.1){\footnotesize $\{{\kern 0.5pt}\!L_{j}\!{\kern 0.5pt}\}$}
\put(75.0,41.1){\footnotesize $\{{\kern 0.5pt}\!R_{q}\!{\kern 0.5pt}\}\,\!{\kern 0.5pt}\{{\kern 0.5pt}\!L_{j}^{\dag}\!{\kern 0.5pt}\}$}
\put(75.0,24.6){\footnotesize $\{{\kern 0.5pt}\!R_{q}\!{\kern 0.5pt}\}\,\!{\kern 0.5pt}\{{\kern 0.5pt}\!L_{j}^{\dag}\!{\kern 0.5pt}\}$}
\put(75.0,8.1){\footnotesize $\{{\kern 0.5pt}\!R_{q}\!{\kern 0.5pt}\}\,\!{\kern 0.5pt}\{{\kern 0.5pt}\!L_{j}^{\dag}\!{\kern 0.5pt}\}$}
\put(58,46){\footnotesize PS}
\put(58,29.5){\footnotesize PS}
\end{picture}
\caption[]{(color online) Operation of matched ansible halves. (a) A transmission path with an arbitrary error channel $\mathcal{E}$ and ideal fiber modeled as a phase shifter (PS). (b) Same transmission path with but with perfect absorber inserted (darkest rectangles). (c) Perfect absorber incorporated into the encoding ansible half, and the fiber removed completely.  In each case, there is no classical communication between ansible halves, but rather each pair is classically \textit{correlated}, due to their synchronized initialization at the factory as in \Fig{9}. See \Sec{III.F} for a discussion of the strange prediction of the ``beam rebirth'' in (b) and (c).  Arrows show the $j$th result at that point, where $\lambda_{j}\equiv\langle e_{j}|\rho|e_{j}\rangle$.}
\label{fig:10}
\end{figure}
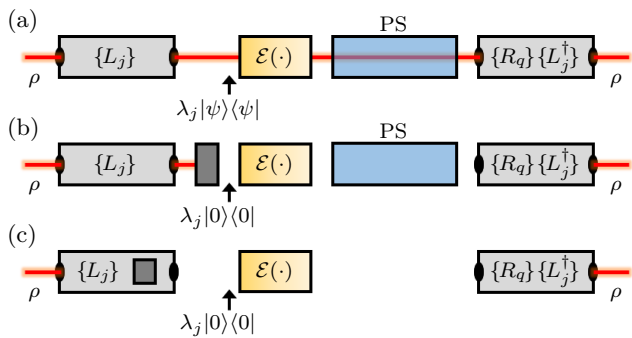
\hyperlink{imp:4}{Implication 4} from \Sec{III.A} is more troubling because it asserts the validity of superluminal communication. In part, this is due to \hyperlink{imp:2}{Implication 2}, which is valid because we could use a fiber/phase-shifter of \textit{any} physical path length, and it would still be treated as part of a correctable error channel. Since the ansible halves can then be moved to any relative distance apart, they could easily be moved to a distance such that the time it takes to tomographically ``read'' the outupt state is much less than the time it would take a classical signal to send the same information.  For example, if the ansible halves are moved $60$ light-minutes apart, but it only takes $1$ minute to tomographically measure a one-qubit message, then the message is learned by users of the ansible receiver $59$ minutes before the same message could reach them by radio.  \Section{III.G} explains exactly how this is possible without violating special relativity.

The main reason this is even remotely plausible is the \textit{correlation} of quantum operators in the ansible halves.  The operators themselves are combinations of unitary gates and projectors, as shown in (\ref{eq:17}), and thus the fact that they are synchronized means that there is \textit{inherent nonlocality} in the ansible pairs.  This need not be entanglement; for example, quantum discord \cite[]{OlZu,HeVe} includes quantum correlations beyond classical correlation that are not necessarily related to entanglement.  Furthermore, the classical idea of \textit{state} is inherently nonlocal, for example the solution to the heat equation has infinite propagation velocity.  On the other hand, entanglement has been shown to have a lower-bound "propagation speed" of $10^{4}c$ \cite[]{VaWo,GiSc,SaBa,CoFa,YiCa}, suggesting instantaneous correlation, as well.  Thus, there are many natural mechanisms that can allow strong nonlocal correlations.

The idea of \textit{beam rebirth} as seen in \Fig{8} and \Fig{10} is plausible because all of the information about the input state is reduced to the nonnegative scalars $\lambda_j$, and therefore it really makes no difference what the reference state $|\psi\rangle$ is between ansible halves.  By including the perfect absorber in the design, we simply force the reference state to be initially vacuum, but any error channel can then change it to another state without affecting the correctability of the net error channel.  Thus, as \hyperlink{imp:3}{Implication 3} asserts, there is no standard concept of transmission signal here, and beam rebirth can be thought of as a kind of teleportation through localization-shifting of the wave function, caused by the built-in classical correlation of quantum operators. \Section{III.F} also explains that the recovery and encoding/decoding operators are generally \textit{active} devices, supplying the energy necessary for beam rebirth from vacuum.  \Section{V} gives a detailed example explicitly showing how the computational basis interfaces to the vacuum.

\hyperlink{imp:5}{Implication 5} that alignment is irrelevant is a powerful and important feature of this device.  The reason for this is that the scalars $\lambda_j\equiv\langle e_{j}|\rho|e_{j}\rangle$, which contain all the information of the input state $\rho$, do not have any local dependence on a particular subsystem.  

For example, suppose a mirror is placed just before the receiver in \Fig[c]{10}.  The mirror acts as a beam splitter with device unitary \smash{$B_\theta \equiv e^{\theta(a_{1}^{\dag}a_{2}-a_{1}a_{2}^{\dag})}$} where $a_1$ and $a_2$ are annihilation operators, with $\theta=\frac{3\pi}{2}$ causing full reflectivity on its primary input so that \smash{$B_{\theta}a_{1}B_{\theta}^{\dag}=a_{2}$} and \smash{$B_{\theta}a_{2}B_{\theta}^{\dag}=-a_{1}$}. For \textit{any state} $\varsigma$ in its auxiliary input, it sends a version of $\varsigma$ to the main output, and redirects the primary input to the auxiliary output in a location \textit{other than} the receiver's input, as seen in \Fig{11}.

Recalling that $L_j  = |\psi \rangle \langle e_j |$ where $\{|e_{j}\rangle\}$ are eigenstates of $\rho$, $R_q  = |\psi \rangle \langle \phi _q |$ where $\{ |\phi _q \rangle \} $ is a complete set of orthonormal states, and $\mathcal{A}_0 (\rho ) = |0\rangle \langle 0|\;\forall \rho$, then the particular results in \Fig{11} just after the sender are
\begin{equation}
\begin{array}{*{20}l}
   { \mathcal{A}_{0}( L_j \rho L_j^{\dag} )  } &\!\! { = \mathcal{A}_0 (|\psi \rangle \langle e_j |\rho |e_j \rangle \langle \psi |)}  \\
   {} &\!\! { = \lambda_j \mathcal{A}_0 (|\psi \rangle \langle \psi |)}  \\
   {} &\!\! {= \lambda_j |0\rangle \langle 0|.}  \\
\end{array}
\label{eq:40}
\end{equation}
Next, apply an arbitrary error channel $\mathcal{E}$ with errors $E_k$, so that for any one of them, the result is
\begin{equation}
\rho _{(k,j)} ' \equiv E_k \mathcal{A}_{0}( L_j \rho L_j^{\dag}  )  E_k^ {\dag}   = \lambda_j E_k |0\rangle \langle 0|E_k^{\dag} . 
\label{eq:41}
\end{equation}
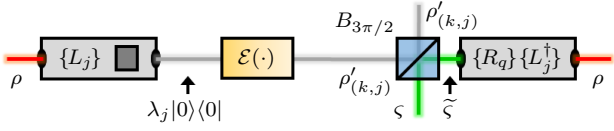
\begin{figure}[H]
\centering
\includegraphics[width=0.99\linewidth]{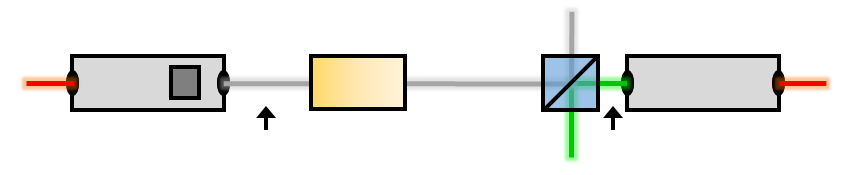}%
\vspace{-12pt}
\setlength{\unitlength}{0.01\linewidth}
\begin{picture}(100,0)
\put(4,6.1){\footnotesize $\rho$}
\put(93.5,6.1){\footnotesize $\rho$}
\put(25.00,1.3){\footnotesize $\lambda_{j}|0\rangle\langle 0|$}
\put(39,9.2){\footnotesize $\mathcal{E(\cdot)}$}
\put(11.4,9.2){\footnotesize $\{{\kern 0.5pt}\!L_{j}\!{\kern 0.5pt}\}$}
\put(75.0,9.2){\footnotesize $\{{\kern 0.5pt}\!R_{q}\!{\kern 0.5pt}\}\,\!{\kern 0.5pt}\{{\kern 0.5pt}\!L_{j}^{\dag}\!{\kern 0.5pt}\}$}
\put(63.0,1.3){\small $\varsigma$}
\put(70.5,1.3){\small $\widetilde{\varsigma}$}
\put(54.5,6.2){\footnotesize $\rho _{(k,j)} '$}
\put(67.9,16.4){\footnotesize $\rho _{(k,j)} '$}
\put(53.9,14.9){\footnotesize $B_{3\pi/2}$}
\end{picture}
\caption[]{(color online) Ansible halves with a mirror modeled as a fully reflective beam splitter $B_{3\pi/2}$ just prior to the receiver.  The notation is the same as in \Fig[c]{10}, except that here a gray line indicates the intermediate ``beam'' starting as the vacuum-proportional result just after the sender.  The mirror sends intermediate result $\rho _{(k,j)} '$ given in (\ref{eq:41}) away from the receiver, and sends some \textit{completely unrelated} state $\widetilde{\varsigma}$ defined in (\ref{eq:42}) into the receiver instead.  As (\ref{eq:47}) shows, the proper state $\rho$ is \textit{still} recovered at the end.  This proves \hyperlink{imp:5}{Implication 5}, that alignment of the ansible halves is irrelevant.}
\label{fig:11}
\end{figure}
{\noindent}Then the mirror acts, modeled as a fully reflecting beam splitter with arbitrary state $\varsigma$ as auxiliary input, yielding
\begin{equation}
B_{{\textstyle{3\pi  \over 2}}} (\rho _{(k,j)} ' \otimes \varsigma)B_{{\textstyle{3\pi  \over 2}}}^ {\dag}   = \widetilde{\varsigma} \otimes \rho _{(k,j)} ',
\label{eq:42}
\end{equation}
where \smash{$\widetilde{\varsigma}  \equiv \sum\nolimits_{y,z = 0,0}^{\infty,\infty}  {  {( - 1)^{y + z} \varsigma _{y,z} |y\rangle \langle z|} } $}, and $\varsigma _{y,z}\!\equiv\!\langle y|\varsigma|z\rangle$.  Meanwhile, the correction procedure still continues on subsystem 1, with a given recovery operator producing
\begin{equation}
R_q \widetilde{\varsigma} R_q ^{\dag}   \otimes \rho _{(k,j)} ' = \langle\phi_{q}|\widetilde{\varsigma} |\phi_{q} \rangle |\psi \rangle \langle \psi | \otimes \rho _{(k,j)} '.
\label{eq:43}
\end{equation}
Then the $j$th decoder acts, and abbreviating the result as $\sigma _{(q,k,j)}  \equiv L_j^{\dag}  R_q \widetilde{\varsigma} R_q ^{\dag}  L_j  \otimes \rho _{(k,j)} '$, yields
\begin{equation}
\begin{array}{*{20}l}
   {\sigma _{(q,k,j)} } &\!\! {\! =\! \langle\phi_{q}|\widetilde{\varsigma} |\phi_{q} \rangle |e_j \rangle \langle \psi |\psi \rangle \langle \psi |\psi \rangle \langle e_j | \otimes \rho _{(k,j)} '}  \\
   {} &\!\! { \!=\! \langle\phi_{q}|\widetilde{\varsigma} |\phi_{q} \rangle |e_j \rangle \langle e_j | \otimes \rho _{(k,j)} '}.  \\
\end{array}
\label{eq:44}
\end{equation}
Putting $\rho _{(k,j)} '$ from (\ref{eq:41}) into (\ref{eq:44}) gives
\begin{equation}
\sigma _{(q,k,j)} \! =\! \lambda_j \langle\phi_{q}|\widetilde{\varsigma} |\phi_{q} \rangle |e_j \rangle \langle e_j | \otimes E_k |0\rangle \langle 0|E_k^{\dag},
\label{eq:45}
\end{equation}
and then, since the user operating the receiver will be performing tomography on subsystem 1, we trace over subsystem 2, which gives
\begin{equation}
\begin{array}{*{20}l}
   {\text{tr}_2 (\sigma _{(q,k,j)} )} &\!\! { \!=\! \lambda_j \langle\phi_{q}|\widetilde{\varsigma} |\phi_{q} \rangle |e_j \rangle\! \langle e_j |\text{tr}(E_k |0\rangle \!\langle 0|E_k^{\dag}  )}  \\
   {} &\!\! {\! =\! \lambda_j \langle\phi_{q}|\widetilde{\varsigma} |\phi_{q} \rangle |e_j \rangle \!\langle e_j |\langle 0|E_k^{\dag}  E_k |0\rangle .}  \\
\end{array}\!\!
\label{eq:46}
\end{equation}
Finally, since the user does not know anything about \textit{which} operators were applied, and since the ansible halves enforce the index linking of their encoder/decoder pairs, then the receiver's output state is
\begin{equation}
{\sum\nolimits_{q,k,j}\! {\text{tr}_2 (\sigma _{(q,k,j)} )} }=\sum\nolimits_j \!{\lambda_j |e_j \rangle \langle e_j |}   = \rho,
\label{eq:47}
\end{equation}
where we used the facts that $\sum\nolimits_{q}  {\langle\phi_{q}|\widetilde{\varsigma} |\phi_{q} \rangle } =\text{tr}(\widetilde{\varsigma}) = 1$ and \smash{$\sum\nolimits_k {E_k^{\dag}  E_k }  = I$}.  

Thus, we see that the receiver user \textit{does} obtain the original state $\rho$, \textit{regardless} of the fact that we inserted a mirror which sent the intermediate state to a different subsystem than that to which the receiver ansible half was applied!  Furthermore, the auxiliary input state $\varsigma$ is completely arbitrary and does not affect the result.  Also, though we used a perfect mirror for a simple example, if it had been the mirror with phase shift such as what happened to $\varsigma$, the result would have been to simply add another error channel, which would get lumped with $\mathcal{E}$. The conclusion is that of \hyperlink{imp:5}{Implication 5}: alignment of the ansible halves is unimportant, and the preceding derivation supports the assertion that the scalar nature of the encoded state is the reason that this is possible.

Note that the preceding proof can be used as an alternative to the NRQT proof in \Sec{II.E}, the difference being that here, the noise was local, and we explicitly used a mirror to prove that alignment does not matter.

Also, this proof treats directly transmitted modes as the same mode to simplify labeling, whereas \Sec{II.E} explicitly gives the sender and receiver their own modes; this is analogous to using one mode to describe a phase shifter instead of two modes as in \App{App.C}.  The reason for using the single-mode transmission here is to highlight the action of the mirror to prove alignment irrelevance.

\begin{figure}[H]
\centering
\includegraphics[width=0.99\linewidth]{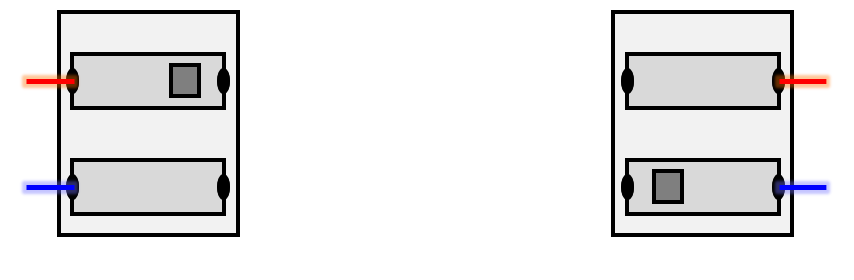}%
\vspace{-12pt}
\setlength{\unitlength}{0.01\linewidth}
\begin{picture}(100,0)
\put(13.5,0.5){\footnotesize Alice}
\put(79,0.5){\footnotesize Bob}
\put(3,19){\footnotesize $\rho$}
\put(93.9,19){\footnotesize $\rho$}
\put(3,6.1){\footnotesize $\rho'$}
\put(93.9,6.1){\footnotesize $\rho'$}
\put(11.5,26.8){\footnotesize sender $1$}
\put(74.8,26.8){\footnotesize receiver $1$}
\put(10.5,14.8){\footnotesize receiver $2$}
\put(75.8,14.8){\footnotesize sender $2$}
\put(11.4,21.7){\footnotesize $\{{\kern 0.5pt}\!L_{j}\!{\kern 0.5pt}\}$}
\put(75.0,21.7){\footnotesize $\{{\kern 0.5pt}\!R_{q}\!{\kern 0.5pt}\}\,\!{\kern 0.5pt}\{{\kern 0.5pt}\!L_{j}^{\dag}\!{\kern 0.5pt}\}$}
\put(10.7,9.2){\footnotesize $\{{\kern 0.5pt}\!L_{l}^{\dag}\!{\kern 0.5pt}\} \,\!{\kern 0.5pt} \{{\kern 0.5pt}\!R_{s}\!{\kern 0.5pt}\}$}
\put(81.9,9.2){\footnotesize $\{{\kern 0.5pt}\!L_{l}\!{\kern 0.5pt}\}$}
\put(2.7,24){\footnotesize $\rightarrow$}
\put(94.5,24){\footnotesize $\rightarrow$}
\put(2.0,11.5){\footnotesize $\leftarrow$}
\put(93.5,11.5){\footnotesize $\leftarrow$}
\end{picture}
\caption[]{(color online) Two-way ansible.  Alice sends a message as $\rho$ into sender $1$, which comes out as $\rho$ from Bob's receiver $1$.  Bob uses tomography and obtains Alice's message from $\rho$ and then sends his response as $\rho'$ into sender $2$, and it comes out as $\rho'$ from Alice's receiver $2$, where she can obtain Bob's message from $\rho'$ through tomography.}
\label{fig:12}
\end{figure}
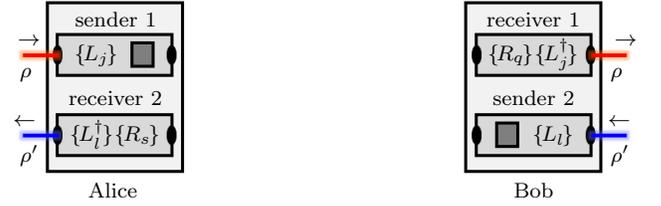

On a practical note, the ``ansible'' we have described thus far is only a one-way communication device.  For full two-way communication, ansibles would need to be manufactured in crossed pairs, so that each two-way anisble half would contain a sending half and a receiving half of two separate one-way ansible pairs, as shown in \Fig{12}.
Now that we have described the realization of an ansible as a possible application of UQEC (supposing the possibility of true projectors), we are ready to examine some of its encryption properties.
\subsection{\label{sec:III.C}Strong Encryption Between Ansible Halves}
As described in \Sec{III.B}, operation of an ansible can be done without need of any broadcast signal or even a transmission line or alignment.  Here, we discuss the encryption properties of this hypothetical technology.

In \Fig[c]{10}, we see that a message can be sent as a state that is proportional to the \textit{vacuum state} in the region between ansible halves.  Therefore the state ``sent out'' by the sender is \textit{always} the vacuum, which means that no information about $\rho$ can be learned by attempting to perform tomography \textit{between} the ansible halves.

The ``sent state'' may not remain vacuum, since any error channel may act along the way, but since those states have nothing to do with the input $\rho$, they do not permit $\rho$ to be learned either.

To test this, imagine that the auxiliary output of the beam splitter in \Fig{11} is intercepted by an eavesdropper.  What state will they get?  The joint-system outcomes at that point are given in (\ref{eq:44}) and (\ref{eq:45}), and so the state available to the eavesdropper is
\begin{equation}
{\sum\nolimits_{q,k,j} {\text{tr}_1 (\sigma _{(q,k,j)} )} } = \sum\nolimits_k {E_k |0\rangle \langle 0|E_k ^{\dag}  } ,
\label{eq:48}
\end{equation}
which is simply error-corrupted vacuum $\mathcal{E}(|0\rangle \langle 0|)$, containing no information about the message $\rho$.

What if the eavesdropper uses recovery and decoding operators?  In that case, the recovery operators are definitely independent because those can be generated instantaneously within each ansible half, so the eavesdropper can at best have index-independent set $\{R_{s}\}$.  Furthermore, since the random-number strings and clock phase are unknown to the eavesdropper, their decoding operators will also be index-independent such as $\{L_{h}\}$.  Thus, defining the eavesdropper's possible outcomes as \smash{$\eta _{(q,k,j,h,s)}  \equiv (I \otimes L_h ^{\dag}  R_s )\sigma _{(q,k,j)} (I \otimes R_s ^{\dag}  L_h )$}, they get
\begin{equation}
\frac{{\sum\nolimits_{q,k,j,h,s} {\text{tr}_1 (\eta _{(q,k,j,h,s)}) } }}{{\text{tr}(\sum\nolimits_{q,k,j,h,s} {\text{tr}_1 (\eta _{(q,k,j,h,s)}) } )}} = {\textstyle{1 \over n}}I,
\label{eq:49}
\end{equation}
which is the maximally mixed state, also containing no information about $\rho$.

However, there is one case in which the eavesdropper could be successful: if they guess or obtain the clock phase and the exact internal random-number string of the ansible pair, \textit{and} maintain inertially tracked corrections from the creation point at the factory, then the eavesdropper's decoding operators would be index-linked with those of the ansible such that their outcomes would be \smash{$\eta _{(q,k,j,j,s)}  \equiv (I\! \otimes L_j ^{\dag}  R_s )\sigma _{(q,k,j)} (I\! \otimes R_s ^{\dag}  L_j )$}, which sum to
\begin{equation}
\sum\nolimits_{q,k,j,s} {\eta _{(q,k,j,j,s)} }  = \sum\nolimits_j {\lambda_j } |e_j \rangle \langle e_j | \otimes |e_j \rangle \langle e_j |,
\label{eq:50}
\end{equation}
and the state obtained by the eavesdropper is
\begin{equation}
{\text{tr}_1 (\sum\nolimits_{q,k,j,s} {\eta _{(q,k,j,j,s)} } )}= \rho,
\label{eq:51}
\end{equation}
which is \textit{exactly} the state obtained by the ansible receiver.  Worse than that, this is done \textit{silently}, since the receiver can detect no effect of this eavesdropping.

While it is highly unlikely that such eavesdropping could be accomplished by guessing or searching due to the high degree of synchronization required, it can be made far less likely by lengthening the random-number strings stored in the ansible pair, such that all numbers still occur an equal number of times.  In that way, the probability of successful eavesdropping by luck or trial-and-error can be made arbitrarily close to zero by increasing the stored random-number string length.

Thus, \hyperlink{imp:6}{Implication 6} from \Sec{III.A} is mainly justified since eavesdropping the vacuum or any error-distorted version of it yields no information about the input state, with the caveat that the unpredictability of the internal random numbers needs to be as strong as possible.

Similarly, \hyperlink{imp:7}{Implication 7} from \Sec{III.A} is justified because \textit{any} operation performed between the ansible halves is merely treated as another error channel, completely corrected away by the receiver.  The caveat here is that the internal workings of the sender and receiver ansible halves must be protected from tampering; the error resistance only holds true \textit{between} the ansible halves.

Of course, placing a spy at the receiver also circumvents the secure connection, but that is true for \textit{all} security protocols.  However, as long as some method of user authentication is employed, then the message is safe.  Note that authentication requires a two-way ansible.

Thus, we can say that the ansible provides a \textit{strongly encrypted pathway for communication, because messages cannot be altered or easily eavesdropped in transit}.
\subsection{\label{sec:III.D}Quantum Broadcasting}
The suggested possibility for successful eavesdropping of an ansible pair in \Sec{III.C} can actually be used intentionally to enable multi-party communication, which we might call \textit{quantum broadcasting}.

This can easily be seen from the joint-system state in (\ref{eq:50}), from which \textit{both parties will perceive the same state} $\rho$ in their respective subsystems.

This broadcasting can be achieved by preparing multiple receiver ansible ``halves'' at the factory, in the same way as described in \Fig{9}, such that all receivers contain the same random number string, and all clocks are synchronized and maintained via internal inertial trackers.  In the case of $N$ subsystems with matched receivers, the joint-system output state is
\begin{equation}
\eta \equiv \sum\nolimits_j {\lambda_j } |e_j^{(1)} \rangle \langle e_j^{(1)} | \otimes  \cdots  \otimes |e_j^{(N)} \rangle \langle e_j^{(N)} |,
\label{eq:52}
\end{equation}
where parenthetical superscripts are subsystem labels. The reduction for each subsystem, tracing over the other $N-1$ subsystems, is always $\rho$.

Thus, we can achieve quantum broadcasting since a single input state $\rho$ can be simultaneously sent to all parties possessing the receivers index-matched to the sender.  An added advantage of this system is that if any one of the receivers breaks, the others keep receiving.
\subsection{\label{sec:III.E}Effective Quantum Cloning}
The quantum broadcasting application of \Sec{III.D} is suggestively similar to quantum cloning.  However, it is \textit{not} quantum cloning because the joint-system state $\eta$ is \textit{not} a tensor product of identical states, even though each \textit{reduction} is the identical state $\rho$. For true quantum cloning, the joint output state would need to be \smash{$\rho^{\otimes N}\equiv\rho^{(1)}\otimes\cdots\otimes\rho^{(N)}$}.  Furthermore, limited-direct UQEC limits the family of states $\rho$ to those sharing the same set of eigenstates, while ideal quantum cloning would have no such limitation.  Similarly, in insulating UQEC, though it can handle and transmit all possible states, the reductions would be in the form of the insulated state $\rho_D$, which is merely isomorphic to $\rho$.

However, although the joint output state is not $\rho^{\otimes N}$, we might nevertheless refer to the scenario where all the output \textit{reductions} are $\rho$ as \textit{effective quantum cloning}, and it may prove to be useful.

Note that the ``no-cloning'' theorem \cite[]{WoZu,Diek} only applies to \textit{unitary} operations, and therefore in the context of more general quantum operations, such as QEC, the no-cloning theorem is generally \textit{not} valid.
\subsection{\label{sec:III.F}Wireless Power Transfer?}
While scenarios such as \Fig[c]{10} make it \textit{appear} as if a beam (and all its energy) is disappearing in one location and reappearing in another location, carrying with it some nonzero mean particle number and thus power, there is actually \textit{no power transfer}.

The reason no power is transfered is that, as seen in (\ref{eq:50}), the input state's ability to be effectively quantum-cloned means that energy is not conserved in the broadcasting process, even in the case of just one receiver.

This nonconservation of energy is perfectly explainable: the recovery/decoding process is an \textit{active} process, meaning that it pumps energy into the system from the power used to operate the ansible receiver.

But how can the recovery/decoding operators be active devices if we showed in \Sec{II.D.3} that they are combinations of projective filters and unitary gates?  The answer is that \textit{unitary operations can be active power sources}.

For example, consider the \textit{unitary} vacuum displacement operator, \smash{$D(\alpha)\equiv e^{\alpha a^{\dag}-\alpha^{*}a}$}, where $a$ is the annihilation operator and $\alpha$ is a complex number, such that
\begin{equation}
|\alpha\rangle=D(\alpha)|0\rangle,
\label{eq:53}
\end{equation}
where $|\alpha\rangle$ is a coherent state of mean particle number $|\alpha|^2$.  Since the input state is \textit{vacuum}, with mean particle number $0$, then $D(\alpha)$ is an \textit{active} unitary operator since it transforms the vacuum to a new state with a generally larger mean particle number.

To see how this relates to the ansible, the possible results after applying the recovery operator are
\begin{equation}
R_q E_k L_j \rho L_j ^{\dag}  E_k ^{\dag}  R_q ^{\dag} = \lambda_j |\langle \phi _q |E_k |\psi \rangle |^2 |\psi \rangle \langle \psi |,
\label{eq:54}
\end{equation}
where everything is as defined in \Sec{II}.

Now suppose that we use vacuum as the reference state, setting $|\psi \rangle  \equiv |0\rangle$.  In that case, each possible result in (\ref{eq:54}) is proportional to the vacuum as
\begin{equation}
R_q E_k L_j \rho L_j ^{\dag}  E_k ^{\dag}  R_q ^{\dag} = \lambda_j |\langle \phi _q |E_k |0 \rangle |^2 |0 \rangle \langle 0 |,
\label{eq:55}
\end{equation}
from which we see that each result has mean particle number $0$, the same value we would get for the normalized sum of all results.

However, applying the decoder to (\ref{eq:55}) yields
\begin{equation}
L_j ^{\dag}  R_q E_k L_j \rho L_j ^{\dag}  E_k ^{\dag}  R_q ^{\dag}  L_j  = \lambda_j |\langle \phi _q |E_k |0 \rangle |^2 |e_j \rangle \langle e_j |,
\label{eq:56}
\end{equation}
and since each state $|e_j \rangle \langle e_j |$ has nonzero mean particle number, and the same for the total state, then the mean particle number has been \textit{increased} by the decoders.

Therefore, in this example, the decoders \textit{raise} the vacuum to various other states of larger mean particle number, which proves that the decoders \smash{$L_j^{\dag}$} are injecting energy into the output system.  Specifically, recalling the operator implementation of \Sec{II.D.3}, since the projective part of \smash{$L_j^{\dag}$} only incurs loss, then it is the \textit{unitary} part of \smash{$L_j^{\dag}$} that is responsible for injecting the power.

Thus, \textit{globally, energy is conserved} in the ansible because the unitary operators used to realize the recovery and decoding operators are generally \textit{active} devices which provide the energy to create the output beams.

That means that although the power outputs of the receivers do indeed \textit{mimic} the input power of the sender, the actual power of each output beam would come from the receiver's local power source, thereby making this device incapable of power transfer.
\subsection{\label{sec:III.G}Effective Superluminal Communication Without Violating Special Relativity}
Strictly speaking, the ideal ansible of \Sec{III.B} is \textit{not} capable of superluminal (faster-than-light) communication, because it would technically take an \textit{infinite} number of measurements to perform perfect tomography, even assuming all equipment is ideal.  Therefore, special relativity is not violated by the ansible, because in theory, it cannot send a \textit{perfect} message in a finite time.

However, for communication purposes, we merely need to measure the state to within some acceptable tolerance, and typically that can be done very well using unbiased estimators for which the mean of the estimator is the quantity sought, and the error can be made acceptably small for reasonable sample times.

Furthermore, the classical correlations involved have theoretically infinite propagation velocity, which is acceptable since they do not carry classical information by themselves.  Thus, the inherent nonlocality of the strongly-correlated ansible halves and the nonlocality of the scalars $\lambda_j$ are what make long-range superluminal \textit{correlation} possible (supposing that true projectors are possible), and the acceptable approximation of finite samples of unbiased estimators in the receiver's tomography is what makes effective superluminal \textit{communication} possible without violating special relativity.  Note that this is \textit{not} the same thing as proposing superluminal communication with entanglement.

Now that we have discussed the ansible as if it were possible, recall that superluminal communication and quantum cloning are familiar indicators of impossible processes.  Therefore we now propose a theorem that explains why the ansible is likely impossible.
\section{\label{sec:IV}Proposal of a ``No-Projector Theorem''}
To this point, all the abnormal effects related to the ansible, such as effective superluminal communication and effective quantum cloning, can be traced back to the supposition that we can implement a single true projector deterministically (see \Sec{II.D.4} and \App{App.B}).

Although \Sec{II.G} proposed a seemingly logical method of realizing true projectors, we will see in \Sec{V} that even that will fail since it does not allow one projector to be applied without the other.  Therefore, we propose the following theorem:

\hypertarget{NoProjThm}{\textbf{No-Projector Theorem:~~}}It is impossible to deterministically implement any operation consisting of only a projector such as $P\equiv|\psi\rangle\langle\psi|$, where $|\psi\rangle$ is a pure state, without postselection.

Notice that if we use $|\psi\rangle\langle\psi|$ to represent a pure state, there is no problem; the \hyperlink{NoProjThm}{no-projector theorem} only observes the impossibility of \textit{operations} in the form of $|\psi\rangle\langle\psi|$.  Furthermore, as we saw in \Sec{II.G}, it is quite possible to realize true projectors in the context of a Kraus-complete quantum operation, but that is different from deterministically producing a lone projector.

\hypertarget{NoSubtractionOpCoro}{\textbf{No-Subtraction-Operator Corollary:~~}}It is impossible to deterministically implement the annihilation or subtraction operator, without postselection.

\hypertarget{NoOpSuperposCoro}{\textbf{No-Operator-Superposition Corollary:~~}}It is impossible to deterministically implement superposition of operators such as $cA+dB$, where $c$ and $d$ are complex scalars and $A$ and $B$ are operators, without postselection.

These corollaries allude to the amazing work on subtraction operators and operator superposition in \cite[]{BelZ}, which demonstrates postselected application of operator superposition such as $aa^{\dag}-e^{i\phi}a^{\dag}a=I$, where $a$ is the annihilation operator.  It is easy to show that if postselection were not necessary, then these techniques could be used to implement true projectors.  Therefore, the postselection is an important part of realizing projectors in these methods, and we will see that it allows us to emulate scalar zero, and that it is also the reason that classical communication must be involved.

Actually, the \hyperlink{NoProjThm}{no-projector theorem} and its corollaries above are all manifestations of a more general theorem; that \textit{it is impossible to implement any nonunitary rank-1 operation without postselection}, where the \textit{rank} of a quantum operation is the minimum number of Kraus operators it can have over all possible decompositions.  However, since the polar decomposition allows \textit{any} operator to be expressed as a product of a nonnegative Hermitian projector $P$ and a unitary operation $U$, then the problem can be simplified to the realization of projectors, though it is not limited to rank-1 projectors (in this case, \textit{rank} is the number of nonzero eigenvalues of $P$).

A more general way to state the \hyperlink{NoProjThm}{no-projector theorem} would be: In an $n$-level quantum system, with complete orthonormal basis $\{|\phi_k \rangle\}$, no projector $P\equiv\sum\nolimits_{k=1}^{k\leq n}r_{k}|\phi_k \rangle\langle\phi_k |$, where $r_{k}$ are real nonnegative scalars, can be implemented as a rank-$1$ quantum operation, without postselection, unless $P=I$ is the identity (in which case it is unitary and not merely a projector).

To \textit{prove} the \hyperlink{NoProjThm}{no-projector theorem}, first we will show that \textit{Kraus completeness is a necessary consequence of the completeness of the orthonormal basis of the environment} for a joint system that yields some physical quantum operation.  Then, using polar decomposition, we will see that \textit{all Kraus completeness statements can be written with projectors only}.  Then, the fact that the smallest nontrivial environment is a qubit will prove that \textit{there must always be at least two projectors} for a nonunitary operation to obey Kraus completeness, which proves that no physical quantum operation can consist of fewer projectors than what is needed for Kraus completeness.

First, following \cite[]{HedD}, start with the identity-identity
\begin{equation}
I^{(S)}  \otimes I^{(E)}  = I,
\label{eq:57}
\end{equation}
for $n_S$-level system $S$ and $n_E$-level environment $E$.  Then, expand \smash{$I^{(E)}$} with any complete orthonormal basis as \smash{$I^{(E)}  = \sum\nolimits_{k = 1}^{n_E } |\phi _k ^{(E)} \rangle \langle \phi _k ^{(E)} |$}, so (\ref{eq:57}) becomes
\begin{equation}
\sum\limits_{k = 1}^{n_E } (I^{(S)}  \otimes |\phi _k ^{(E)} \rangle )(I^{(S)}  \otimes \langle \phi _k ^{(E)} |) = I.
\label{eq:58}
\end{equation}
Then apply joint-system unitary $U^{\dag}$ from the left and $U$ from the right and use \smash{$U^{\dag}U=I$} to get
\begin{equation}
\sum\limits_{k = 1}^{n_E } U^{\dag}  (I^{(S)}  \otimes |\phi _k ^{(E)} \rangle )(I^{(S)}  \otimes \langle \phi _k ^{(E)} |)U = I.
\label{eq:59}
\end{equation}
Then apply \smash{$(I^{(S)}\!  \otimes \langle \psi _0 ^{(E)} |)$} from the left, \smash{$(I^{(S)}\!  \otimes |\psi _0 ^{(E)} \rangle )$} from the right, and use \smash{$(I^{(S)}\!  \otimes \langle \psi _0 ^{(E)} |)I(I^{(S)}  \!\otimes |\psi _0 ^{(E)} \rangle )=$} \smash{$I^{(S)}\!  \otimes \langle \psi _0 ^{(E)} |\psi _0 ^{(E)} \rangle  = I^{(S)} $} on the right to get
\begin{equation}
\sum\limits_{k = 1}^{n_E } E_k^{(S)\dag } E_k^{(S)}  = I^{(S)},
\label{eq:60}
\end{equation}
which is a Kraus-completeness relation where \smash{$E_k^{(S)}  \equiv$} \smash{$(I^{(S)}  \otimes \langle \phi _k ^{(E)} |)U(I^{(S)}  \otimes |\psi _0 ^{(E)} \rangle )$} are the Kraus operators.  Thus, since (\ref{eq:60}) was \textit{derived} from the completeness of the basis spanning $E$, that means Kraus completeness is a direct consequence of the completeness of the basis of $E$.  Although these \smash{$E_k^{(S)}$} imply a joint system starting in \textit{product form} as \smash{$\rho^{(S)}\otimes|\psi _0 ^{(E)} \rangle \langle \psi _0 ^{(E)} |$}, it also applies to a \textit{mixed initial environment}, which just means more Kraus operators, and a \textit{generally correlated initial joint system}, since that can always be converted to product form by inserting a preliminary quantum operation converting a general product of states to the correlated state, resulting in more Kraus operators again.

To see that only projectors really matter, recall that all operators $\Omega$ have a right-polar decomposition $\Omega  = UP$ where $U \equiv WV^{\dag}$ is unitary, \smash{$P \equiv \sqrt {\Omega ^{\dag}  \Omega }  = V\Sigma V^{\dag} $} is nonnegative Hermitian, where $\Omega  = W\Sigma V^{\dag} $ is the singular-value decomposition (SVD) of $\Omega$ where $W$ and $V$ are unitary and $\Sigma$ is the nonnegative diagonal matrix of singular values of $\Omega$. Therefore, all Kraus operators can be decomposed as (omitting $S$ labels on the right),
\begin{equation}
E_k^{(S)}  = U_k P_k ,
\label{eq:61}
\end{equation}
and then putting this into (\ref{eq:60}) gives
\begin{equation}
\sum\limits_{k = 1}^{n_E } (U_k P_k )^{\dag}  U_k P_k  = \sum\limits_{k = 1}^{n_E } P_k ^{\dag}  U_k ^{\dag}  U_k P_k  = \sum\limits_{k = 1}^{n_E } P_k ^{\dag}  P_k  = I^{(S)},
\label{eq:62}
\end{equation}
which shows that the only thing that really \textit{ever} matters about Kraus completeness is the projectors in this ``projective Kraus completeness'' relation in (\ref{eq:62}) (left unsimplified to highlight its relation to Kraus completeness), where the \smash{$P_k =V_{k}\Sigma_{k} V_{k}^{\dag}$} are general weighted projectors of form \smash{$P_{k}=\sum\nolimits_{m=1}^{m\leq n}r_{k|m}|\phi_{k|m} \rangle\langle\phi_{k|m} |$} where \smash{$r_{k|m}\!=\!(\Sigma_{k})_{m,m}\!\geq \! 0$} and \smash{$|\phi_{k|m} \rangle$} is the $m$th column vector of \smash{$V_{k}$}, both from the SVD of \smash{$E_k$}.  Thus, we have proven that \textit{projective Kraus completeness is also a necessary consequence of the completeness of the environment{\kern 2pt}basis}.

Finally, since the smallest nontrivial environment has \smash{$\min(n_{E})=2$} levels, (\ref{eq:62}) shows that \textit{there must always be at least two projectors in any physical nonunitary quantum operation to maintain the necessary projective Kraus completeness}.  Thus we have proven the \hyperlink{NoProjThm}{no-projector theorem} up to postselection, and we get another corollary:

\hypertarget{NoKrausUnderCompleteCoro}{\textbf{No-Kraus-Undercompleteness Corollary:~~}} A quantum operation \smash{$\mathcal{E}(\rho)\equiv\sum\nolimits_{k}E_{k}\rho E_{k}^{\dag}$} is physical iff it is Kraus complete \smash{$\sum\nolimits_{k}E_{k}^{\dag}E_{k}=I$} on the full Hilbert space of \smash{$\mathcal{E}(\rho),\; \forall\rho$}.

The reason that postselection allows ``deterministic'' rank-1 nonunitary operations is that by performing a measurement and looking at the result to do the postselection, our reality becomes defined by a particular result(s) of the mixture, which can have a support that is less-than-complete on the original full space.  This permits a ``Kraus-undercompleteness'' to be considered complete within a particular result of a mixture. The price we pay is the fundamental unpredictability of the postselection events, which causes increased wait times and the need to communicate the postselection results.

For example, consider a detector with a complete set of projective measurement operators \smash{$\Pi _j$} such that \smash{$\Sigma _j \Pi _j ^{\dag}  \Pi _j  = I$}.  Then, the Kraus operators for the full event (evolution by $U$, detection, reduction) are \smash{$E_{k|j}^{(S)}  \equiv$} \smash{$ (I^{(S)}  \otimes \langle \phi _k ^{(E)} |)\Pi _j U(I^{(S)}  \otimes |\psi _0 ^{(E)} \rangle )$} for the $j$th detection result.  Then, supposing the simplest case of measurement operators with form \smash{$\Pi _j  = I^{(S)}  \otimes |\phi _j ^{(E)} \rangle \langle \phi _j ^{(E)} |$}, the Kraus operators are
\begin{equation}
E_{k|j}^{(S)}  = \delta _{j,k} (I^{(S)}  \otimes \langle \phi _j ^{(E)} |)U(I^{(S)}  \otimes |\psi _0 ^{(E)} \rangle ),
\label{eq:63}
\end{equation}
so that \textit{given} that a \textit{particular} detection output happens, the projective Kraus completeness gets truncated to
\begin{equation}
\sum\limits_{k = 1}^{n_E } P_{k|j} ^{\dag}  P_{k|j}  = P_j ^{\dag}  P_j  < I^{(S)} 
\label{eq:64}
\end{equation}
which shows how postselection can cause \textit{apparent} Kraus undercompleteness.  By interacting with the detector, we are ``along for the ride'' on the particluar reality of just one of its measurement operators.  The price we pay for this is that the postselection detection acts like an absorptive projective filter, inducing a mixture with the vacuum, and causing the instant of detection to be fundamentally uncertain.

We now go through a step-by-step example of how an ansible would function, and in the process we show why a true ansible would not work, but also find a way to make a \textit{pseudo-ansible}; a device for connection-free noise-resistant light-speed communication.
\section{\label{sec:V}How To Build a Pseudo-Ansible}
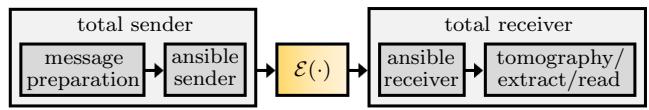
\begin{figure}[H]
\centering
\includegraphics[width=0.99\linewidth]{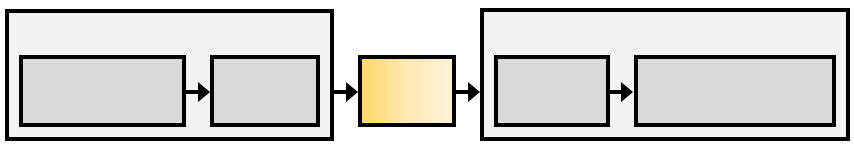}%
\vspace{-12pt}
\setlength{\unitlength}{0.01\linewidth}
\begin{picture}(100,0)
\put(11.5,12.3){\footnotesize total sender}
\put(68,12.3){\footnotesize total receiver}
\put(6.5,7.2){\footnotesize message}
\put(3.7,3.8){\footnotesize preparation}
\put(26,7.3){\footnotesize ansible}
\put(26.4,3.9){\footnotesize sender}
\put(44.9,5.4){\footnotesize $\mathcal{E(\cdot)}$}
\put(59.2,7.2){\footnotesize ansible}
\put(58.7,3.8){\footnotesize receiver}
\put(76.0,7.2){\footnotesize tomography/}
\put(76.2,3.8){\footnotesize extract/read}
\end{picture}
\caption[]{(color online) Overview of general one-way ansible protocol.  The middle stage is general transmission modeled as noise channel $\mathcal{E}$, which can be any noise, even the complete attenuation channel from \Sec{III.A}.}
\label{fig:13}
\end{figure}

Here we begin by first returning to the supposition that true projectors and therefore true ansibles are possible.  We then find a flaw and use it to develop the idea of pseudo-ansibles.  The general setup is shown in \Fig{13}.
In the following sections, we will discuss the details of each of the stages shown in \Fig{13}.  In keeping with previous convention, here we use $\{|\widetilde{1}\rangle,|\widetilde{2}\rangle,\ldots\}$ as the computational basis to distinguish it from the Fock basis.
\subsection{\label{sec:V.A}Message Preparation}
Following \Sec{II.H}, we will use \textit{insulating} UQEC to send our message, allowing us to restrict input to \textit{pure} states, ensuring only \textit{superposition scalars} are involved.  For one qubit, this means we only have two degrees of freedom for information storage.

One way to store this information is to use spherical parameterization, and chop up each parameter  by the number of letters in our alphabet, limited only by resolution accuracy of measurement.

For example, a single-qubit message state has the form
\begin{equation}
|\xi \rangle  \equiv c_\theta  |\widetilde{1}\rangle  + s_\theta  e^{i\phi } |\widetilde{2}\rangle,
\label{eq:65}
\end{equation}
where $c_\theta   \equiv \cos (\theta )$, $s_\theta   \equiv \sin (\theta )$, where $\theta  \in [0,{\textstyle{\pi  \over 2}}]$ and $\phi  \in [0,2\pi )$.  Supposing we could accurately chop these parameters each into, say, $M=41$ pieces, to fit 26 letters, 10 numbers, a period, a question mark, a space ``$\_\,$'', ``$\sim$'', and ``$\ast$'' in that order, where ``$\sim$'' means ``no message'' and ``$\ast$'' means ``start new message,'' then in the range of $[0,1]$, the $m$th character $\chi _m$ of this alphabet would have value $\chi _m  = {\textstyle{{2m - 1} \over {2M}}}$, where $m = \{ 1, \ldots ,M\}$.  Then let $\theta  \equiv {\textstyle{\pi  \over 2}}\chi _m$ for character one, and $\phi  \equiv 2\pi \chi _m $ for character two.  To send a message consisting of any two characters, such as ``GO,'' we simply set $\theta  \equiv {\textstyle{\pi  \over 2}}\chi _7  \approx 0.249[\text{rad}]$ and $\phi  \equiv 2\pi \chi _{15}  \approx 2.222[\text{rad}]$. Thus, the message state is
\begin{equation}
\rho  = |\xi \rangle \langle \xi | \approx \left( {\begin{array}{*{20}l}
   {0.939} & {0.239e^{ - i2.222} }  \\
   {0.239e^{i2.222} } & {0.061}  \\
\end{array}} \right)\!.
\label{eq:66}
\end{equation}
Typically, there is some fiduciary state $|\xi_0\rangle$ naturally produced by our single-photon source (SPS), and we must use $\theta$ and $\phi$ to apply \smash{$U=\epsilon_{|\xi\rangle}\epsilon_{|\xi_{0}\rangle}^{\dag}$} as the particular unitary that transforms $|\xi_0\rangle$ to message $|\xi\rangle$, where $\epsilon_{|A\rangle}$ is the eigenvector matrix of $|A\rangle\langle A|$.
\subsection{\label{sec:V.B}Ansible Sender: Encoding and Vacuum Conversion}
Here we examine two different methods of encoding, both of which use conversion to vacuum for connection-free communication and full noise resistance.
\subsubsection{\label{sec:V.B.1}Time-Bin Projector Method}
As in \Sec{II.G}, if the input photon's momentum qubit has state \smash{$|\widetilde{1}\rangle$}, our joint-system input state is \smash{$\rho\otimes|\widetilde{1}\rangle\langle\widetilde{1}|$}.  Then, recalling the insulating-UQEC encoding operators \smash{$D_j  = P_{|\widetilde{1}\rangle } \epsilon _{|D_j \rangle }^{\dag}$}, only applying the unitary part of this as \smash{$|\xi _j \rangle  \equiv \epsilon _{|D_j \rangle }^{\dag}  |\xi \rangle$} gives \smash{$\rho_{j}\otimes|\widetilde{1}\rangle\langle\widetilde{1}|$}, analogous to the input of \Fig{6}, where \smash{$\rho _j  \equiv |\xi _j \rangle \langle \xi _j |$}.  Then, the application of the time-separated projector channel of \Sec{II.G} yields
\begin{equation}
\rho _j ' \equiv (\rho _j )_{1,1} |1_{\tau _1 } ,0_{\tau _2 } \rangle \langle 1_{\tau _1 } ,0_{\tau _2 } | + (\rho _j )_{2,2} |0_{\tau _1 } ,1_{\tau _2 } \rangle \langle 0_{\tau _1 } ,1_{\tau _2 } |,
\label{eq:67}
\end{equation}
where $\rho _j '\equiv (\rho _j)_{\text{CNOT}}$, $(\rho _j )_{a,b}  \equiv \langle \widetilde{a}|\rho _j |\widetilde{b}\rangle$, and $\tau _k$ labels time-bin $k$.  From (\ref{eq:E.14}), we recognize the scalars in (\ref{eq:67}) as $(\rho _j )_{1,1}  = d_j $ and $(\rho _j )_{2,2}  = 1 - d_j$, so that (\ref{eq:67}) becomes
\begin{equation}
\rho _j ' \equiv d_j |1_{\tau _1 } ,0_{\tau _2 } \rangle \langle 1_{\tau _1 } ,0_{\tau _2 } | + (1-d_j )|0_{\tau _1 } ,1_{\tau _2 } \rangle \langle 0_{\tau _1 } ,1_{\tau _2 } |.
\label{eq:68}
\end{equation}

Now we are ready to expose the flaw of this projector method. First, though it seems like (\ref{eq:68}) is a step-function in time, which would separate its scalars in time, it is actually a true mixture over both time bins.  This means that we only get a photon in one time bin and vacuum in the other, but never get a photon in each time bin or vacuum in both; a step function in time would yield a photon in both time bins, so that is not what (\ref{eq:68}) means.

If we just focus on $\tau_1$, then tracing over $\tau_2$ gives
\begin{equation}
\text{tr}_{\tau _2 } (\rho _j ') = d_j |1_{\tau _1 } \rangle \langle 1_{\tau _1 } | + (1-d_j )|0_{\tau _1 } \rangle \langle 0_{\tau _1 } |,
\label{eq:69}
\end{equation}
which is a mixture with vacuum with \textit{both results in} $\tau_1$. Then, applying a total absorber $\mathcal{A}_0$ gives
\begin{equation}
\mathcal{A}_0 (\text{tr}_{\tau _2 } (\rho _j '))\! =\! d_j |0_{\tau _1 } \rangle \langle 0_{\tau _1 } | \!+\! (1\! -\! d_j )|0_{\tau _1 } \rangle \langle 0_{\tau _1 } |\! =\! |0_{\tau _1 } \rangle \langle 0_{\tau _1 } |,
\label{eq:70}
\end{equation}
so \textit{the information in the scalars is lost}.  Thus, focusing on either time bin does not yield the results we would expect from the application of a lone projector.

If we try to use the whole state, applying $\mathcal{A}_0$ gives
\begin{equation}
\begin{array}{*{20}l}
   {\mathcal{A}_0 (\rho _j ')} &\!\! { = d_j |0_{\tau _1 } ,\! 0_{\tau _2 } \rangle \langle 0_{\tau _1 } ,\! 0_{\tau _2 } | \!+\! (1 - d_j )|0_{\tau _1 } ,\! 0_{\tau _2 } \rangle \langle 0_{\tau _1 } ,\! 0_{\tau _2 } |}  \\
   {} &\!\! { = |0_{\tau _1 } ,\! 0_{\tau _2 } \rangle \langle 0_{\tau _1 } ,\! 0_{\tau _2 } |,}  \\
\end{array}
\label{eq:71}
\end{equation}
so again, the scalar information is lost.

Therefore, we have shown that implementing true projectors as part of a Kraus-complete channel yields the same type of vacuum mixtures as the application of an absorptive projector, thus preventing the ansible from achieving superluminal communication.

The reason it is easy to think that a time-bin projector channel would behave like a step function of projectors over time is because if true \textit{lone} projectors were possible, then if we did implement a step function of them over time, choosing each one at random with equal probability and applying only one every two time bins, then an observer who chose a sample time at least two-time bins long but was ignorant of which projector was applied, would report the output to be in the state shown in (\ref{eq:67}).  But applying a lone projector at a given instant in time is impossible because then we would be free to stop before applying the other, and the operation would not be Kraus complete and therefore not physical.

However, we can still get connection-free light-speed communication if we use \textit{postselection} in the form of a detector that tells us \textit{which} result of the mixture is happening, and therefore \textit{which} scalar is active at that instant.  This will require us to classically transmit the results of the postselection, but we can now abandon this time-bin projector method and just use detectors instead. This will lead to a more robust version of light-speed limited communication as will soon see.
\subsubsection{\label{sec:V.B.2}Postselective Projector Method}
Here, we bypass the time-bin projector method completely and return to a prepared state $\rho$ from (\ref{eq:66}) and (\ref{eq:65}).  We still apply \smash{$\epsilon _{|D_j \rangle }^{\dag}$} the unitary part of the encoders to get \smash{$\rho_{j}$} as in \Sec{V.A}, but now we simply apply a \textit{polarizer} as the projector \smash{$P_{|\widetilde{1}\rangle }$} for that part of the encoding.

Since a polarizer is an absorptive projective filter (APF), its action on results \smash{$\rho_{j}$} yields (as in \App{App.B})
\begin{equation}
P_{|\widetilde{1}\rangle }^{(\text{APF})} (\rho _j ) = d_j |1\rangle \langle 1| + (1 - d_j )|0\rangle \langle 0|,
\label{eq:72}
\end{equation}
where $d_j \! =\! (\rho _j )_{1,1}$, and we have converted to the Fock basis and suppressed polarization.

Here we note that while total absorption would again conceal the scalars as in \Sec{V.B.1} and (\ref{eq:30}), we can \textit{emulate} a true projector by using \textit{postselection}, effectively annihilating the vacuum and replacing it with a scalar zero without actually doing so.

Since we need an absorber anyway (to force vacuum, so we can eliminate the need for a quantum connection), we instead place a \textit{detector} in the ``output beam,'' and we only recognize instances of a detector click as being part of the process, resulting in
\begin{equation}
\Pi _{\overline{0}} P_{|\widetilde{1}\rangle }^{(\text{APF})} (\rho _j )\Pi _{\overline{0}} ^{\dag} = d_j |1\rangle \langle 1|,
\label{eq:73}
\end{equation}
where $\Pi _{\overline{0}}  \equiv \sum\nolimits_{k = 1}^\infty  {|k\rangle \langle k|}$ is the nonvacuum measurement operator of an ideal on-off single-photon detector.  \textit{Then}, accounting for the fact that the detector absorbs the photon in the act of detecting it we apply $\mathcal{A}_0$ to get
\begin{equation}
(D_{j}\rho D_{j}^{\dag})_{\text{PS}}^{(\text{APF})}\equiv\mathcal{A}_0(\Pi _{\overline{0}} P_{|\widetilde{1}\rangle }^{(\text{APF})} (\rho _j )\Pi _{\overline{0}} ^{\dag}  ) = d_j |0\rangle \langle 0|,
\label{eq:74}
\end{equation}
which are the desired results, where PS means ``postselected,'' since we looked at the measurement result.  While it may seem like cheating to involve  the detector's projectors, we pay for it by interacting with the system, discarding unwanted results, and then having to communicate those results to the receiver.  Thus, realizing a true projector via postselection imposes a light-speed limit.

In practice, since polarizers are not perfect, it may be better to use a polarizing beam splitter (PBS) and send one output beam to a detector and the other beam to a beam dump (BD), as in \Fig[b]{14} (BD not shown).

Also, there is no need to elongate the trial time as in the temporal purification methods mentioned in \App{App.B}.  Here, we actually \textit{want} the form of (\ref{eq:72}), and therefore we keep the trial time constant from input to output.
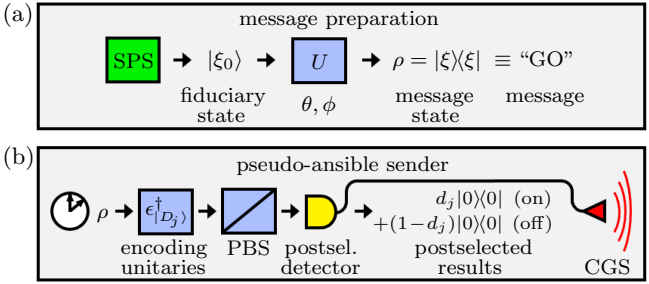
\begin{figure}[H]
\centering
\includegraphics[width=0.99\linewidth]{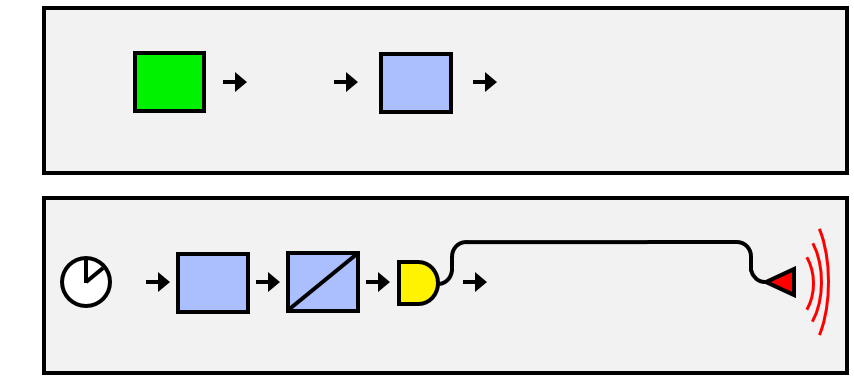}%
\vspace{-12pt}
\setlength{\unitlength}{0.01\linewidth}
\begin{picture}(100,0)
\put(0.0,40.2){\small (a)}
\put(0.0,18.2){\small (b)}
%
%
\put(36.4,39.6){\footnotesize message preparation}
\put(17.10,33.0){\footnotesize SPS}
\put(31.5,33.3){\footnotesize $|\xi_{0}\rangle$}
\put(47.5,32.8){\footnotesize $U$}
\put(46,26.3){\footnotesize $\theta,\phi$}
\put(60,33.3){\footnotesize $\rho = |\xi\rangle\!\langle\xi|\;\equiv\; $``GO''}
\put(27.4,27.7){\footnotesize fiduciary}
\put(30,24.6){\footnotesize state}
\put(60.4,27.7){\footnotesize message}
\put(63,24.6){\footnotesize state}
\put(77.4,27.7){\footnotesize message}
%
%
\put(36,17.0){\footnotesize pseudo-ansible sender}
\put(14.5,9.8){\footnotesize $\rho$}
\put(21.95,10.3){\footnotesize $\epsilon ^{\dag}$}
\put(22.25,9.9){\scriptsize ${\;}_{|D_{\!j} \rangle}$}
\put(18.2,4.3){\footnotesize encoding}
\put(18.2,1.2){\footnotesize unitaries}
\put(34.5,4.1){\footnotesize PBS}
\put(43.4,4.3){\footnotesize postsel.}
\put(42.8,1.2){\footnotesize detector}
\put(63.4,4.3){\footnotesize postselected}
\put(67,1.2){\footnotesize results}
\put(66.6,12){\scriptsize $d_{j}|0\rangle\!\langle 0|$}
\put(57,8.2){\scriptsize $+(1\!-\! d_{j})|0\rangle\!\langle 0|$}
\put(78.6,12){\scriptsize (on)}
\put(78.4,8.2){\scriptsize (off)}
\put(89.5,1.2){\footnotesize CGS}
\put(9.55,12.5){\tiny $\blacktriangle$}
\put(11.15,11.7){\tiny \rotatebox[origin=c]{-48.15}{$\blacktriangle$}}
\end{picture}
\caption[]{(color online) Schematic of total sender of a pseudo-ansible. (a) Message preparation. (b) Ansible sender, emitting a classical grouping signal (CGS) to communicate the postselection results. Details are given in the text.  (Stored random-number strings are not shown.)}
\label{fig:14}
\end{figure}
\subsection{\label{sec:V.C}Ansible Receiver: Conversion to Single-Photon Basis and Decoding}
An interesting property of the recovery operators $R_{q}\equiv |\psi\rangle\langle\phi_{q}|$ is that they form a ``trace-and-replace'' channel, meaning that whatever the input state is, it is traced-away and replaced by the reference state $|\psi\rangle$.  

In terms of the receiver operations, this means that a single-photon source (SPS) effectively realizes the recovery and the projective part of the decoding operators.

The joint state of sender and receiver after encoding and postselection is
\begin{equation}
(D_{j}\rho D_{j}^{\dag})_{\text{PS}}^{(\text{APF})}\otimes |1\rangle \langle 1|=d_j |0\rangle \langle 0| \otimes |1\rangle \langle 1|,
\label{eq:75}
\end{equation}
where $|1\rangle \langle 1|$ is the initial state of the receiver's SPS output.  If noise $\mathcal{E}$ acts on the sender's output, we get
\begin{equation}
\eta_{j}=d_j \mathcal{E}(|0\rangle \langle 0|) \otimes |1\rangle \langle 1|,
\label{eq:76}
\end{equation}
where \smash{$\eta_{j}\equiv \mathcal{E}((D_{j}\rho D_{j}^{\dag})_{\text{PS}}^{(\text{APF})})\otimes |1\rangle \langle 1|$}.  The receiver's photon state is found by tracing over the sender as
\begin{equation}
\text{tr}_1 (\eta_{j})=d_j |1\rangle \langle 1|.
\label{eq:77}
\end{equation}
Since this is already the pass-state for the true projective part of the decoding operator, we can omit that stage and directly apply the unitary part of the decoders as
\begin{equation}
\epsilon _{|D_j \rangle } \text{tr}_1 (\eta_{j} )\epsilon _{|D_j \rangle } ^{\dag}   = d_j |D_j \rangle \langle D_j |,
\label{eq:78}
\end{equation}
where the matched indices are enforced by the factory synchronization of these operators with their counterparts in the sender \textit{and} ``discarding'' trials in which the ``no click'' is reported by the classical grouping signal (we will discuss that more below).

Then, since the receiver does not keep track of \textit{which} decoding unitaries are applied, tomography will find the system to be in the state (where $m_n \equiv n(2n-1)=6$)
\begin{equation}
\frac{\sum\nolimits_{j=1}^{m_n}\epsilon _{|D_j \rangle } \text{tr}_1 (\eta_{j} )\epsilon _{|D_j \rangle } ^{\dag}}{\text{tr}(\sum\nolimits_{j=1}^{m_n}\epsilon _{|D_j \rangle } \text{tr}_1 (\eta_{j} )\epsilon _{|D_j \rangle } ^{\dag}  )}= {\textstyle{1 \over {2n - 1}}}\sum\limits_{j = 1}^{m_n} {d_j |D_j \rangle \langle D_j |}  = \rho _D ,
\label{eq:79}
\end{equation}
which is the insulated version of message $\rho$.

The classical grouping signal (CGS) is needed to determine \textit{which} trials contain $d_j$ and which trials contain $1-d_j$.  But we do not need to discard trials with $1-d_j$; we simply interpret their message state as a bit-flipped version of the intended message.

If both SPSs in sender and receiver are \textit{deterministic}, meaning that photon emission can be controlled by classical pulses, then they can be synchronized by the inertial trackers, allowing for the time of CGS travel.

Thus, the tomography initially collects data during every trial, only marking results by local receiver time.  Then, when the CGS arrives, the data, previously stored in classical memory, is then sorted into two groups corresponding to scalars $d_j$ and $1-d_j$.  Amazingly, this means that the \textit{message transmission} happens \textit{instantaneously}, regardless of distance, but that no sense can be made out of the data until the CGS arrives.

Note that deterministic photon production has been achieved \cite[]{WeiH}, and since all operations we need can be kept within the same cryogenically cooled region, it will retain its ideal properties.

If the SPSs are not deterministic, the CGS information may need to either be transmitted as a very narrow pulse in time, or include information about the exact times the postselecting detector fired.  The benefit of deterministic SPSs is that the CGS can just be a rectangular pulse train with a peak or trough filling each entire trial window.  In either case, the tomographic clicks accepted as data must be coincidences with the preordained time of photon production or with a heralding photon from a nondeterministc SPS, and an inertial history of the receiver will be needed to match the events properly.
\begin{figure}[H]
\centering
\includegraphics[width=0.99\linewidth]{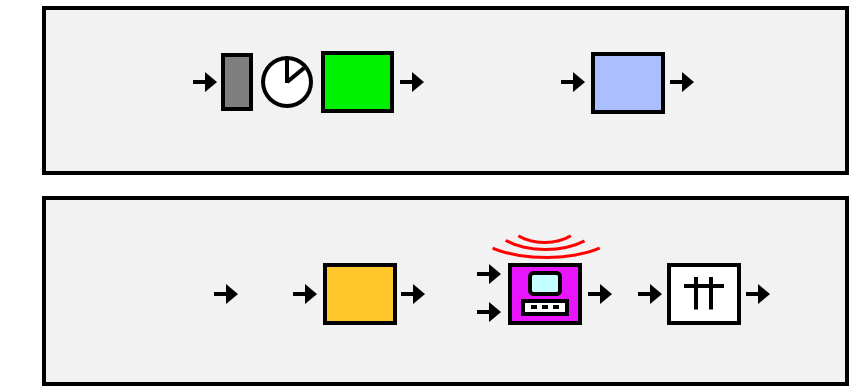}%
\vspace{-12pt}
\setlength{\unitlength}{0.01\linewidth}
\begin{picture}(100,0)
\put(0.0,41.3){\small (a)}
\put(0.0,19.2){\small (b)}
%
%
\put(36,40.5){\footnotesize pseudo-ansible receiver}
\put(13.7,36.8){\scriptsize \scalebox{0.9}{$d_{{\kern -0.5pt}j}|0\rangle\!\langle 0|$}}
\put(6.4,32.6){\scriptsize \scalebox{0.9}{$+({\kern -1pt}1{\kern -2pt}-{\kern -1.5pt} d_{{\kern -0.5pt}j}{\kern -1.5pt}){\kern -0.5pt}|0\rangle\!\langle 0|$}}
\put(21.4,28.8){\footnotesize isolation}
\put(21.4,25.7){\footnotesize absorber}
\put(38.8,34.35){\footnotesize SPS}
\put(56.2,36.8){\scriptsize \scalebox{0.9}{$d_{{\kern -0.5pt}j}|1\rangle\!\langle 1|$}}
\put(48.9,32.6){\scriptsize \scalebox{0.9}{$+({\kern -1pt}1{\kern -2pt}-{\kern -1.5pt} d_{{\kern -0.5pt}j}{\kern -1.5pt}){\kern -0.5pt}|1\rangle\!\langle 1|$}}
\put(70.15,35.0){\footnotesize $\epsilon$}
\put(70.45,34.6){\scriptsize ${\;}_{|D_{\!j} \rangle}$}
\put(85.8,36.8){\scriptsize \scalebox{0.9}{$d_{{\kern -0.5pt}j}|{\kern -0.5pt}D_{{\kern -1.0pt}j}{\kern -1.0pt}\rangle\!\langle{\kern -1.0pt} D_{{\kern -1.0pt}j}{\kern -0.5pt}|$}}
\put(78.5,32.6){\scriptsize \scalebox{0.9}{$+({\kern -1pt}1{\kern -2pt}-{\kern -1.5pt} d_{{\kern -0.5pt}j}{\kern -1.5pt}){\kern -0.5pt}|{\kern -0.5pt}D_{{\kern -1.0pt}j}{\kern -1.0pt}\rangle\!\langle{\kern -1.0pt} D_{{\kern -1.0pt}j}{\kern -0.5pt}|$}}
\put(66.5,28.8){\footnotesize decoding}
\put(66.6,25.7){\footnotesize unitaries}
%
%
\put(28,18.5){\footnotesize tomography/extraction/reading}
\put(13.7,12.3){\scriptsize \scalebox{0.9}{$d_{{\kern -0.5pt}j}|{\kern -0.5pt}D_{{\kern -1.0pt}j}{\kern -1.0pt}\rangle\!\langle{\kern -1.0pt} D_{{\kern -1.0pt}j}{\kern -0.5pt}|$}}
\put(6.4,8.0){\scriptsize \scalebox{0.9}{$+({\kern -1pt}1{\kern -2pt}-{\kern -1.5pt} d_{{\kern -0.5pt}j}{\kern -1.5pt}){\kern -0.5pt}|{\kern -0.5pt}D_{{\kern -1.0pt}j}{\kern -1.0pt}\rangle\!\langle{\kern -1.0pt} D_{{\kern -1.0pt}j}{\kern -0.5pt}|$}}
\put(30.3,12.3){\scriptsize \scalebox{0.9}{$\rho_D$}}
\put(28,8.0){\scriptsize \scalebox{0.9}{$+\rho_{D}'$}}
\put(38.6,9.90){\scriptsize tomo}
\put(51.3,12.3){\scriptsize \scalebox{0.9}{$\rho_D$}}
\put(49,8.0){\scriptsize \scalebox{0.9}{$+\rho_{D}'$}}
\put(49.8,15.4){\footnotesize CGS}
\put(72,10){\footnotesize $\rho$}
\put(89.45,9.7){\footnotesize ``GO''}
\put(19.3,4.3){\footnotesize sum over}
\put(21.0,1.3){\footnotesize results}
\put(54,4.3){\footnotesize CGS sorting}
\put(53.9,1.3){\footnotesize \&\! extraction}
\put(77,4.3){\footnotesize lookup}
\put(78.2,1.3){\footnotesize table}
\put(32.87,36.96){\tiny $\blacktriangle$}
\put(34.47,36.17){\tiny \rotatebox[origin=c]{-48.15}{$\blacktriangle$}}
\end{picture}
\caption[]{(color{\kern 2.0pt}online){\kern 2.0pt}Schematic{\kern 2.0pt}of{\kern 2.0pt}total{\kern 2.0pt}receiver.{\kern 2.0pt}(a){\kern 2.0pt}Pseudo-ansible{\kern 2.0pt}receiver.{\kern 2.0pt}(b){\kern 2.0pt}Tomography,{\kern 2.0pt}extraction,{\kern 2.0pt}reading.  See the text for explanations.}
\label{fig:15}
\end{figure}
\subsection{\label{sec:V.D}Tomography, Extraction, and Message Reading}
Here, we get an estimate of $\rho$ via state tomography on the receiver output, where only detections \textit{coincident} to the time-frame-adjusted CGS are treated as data.

The tomography yields a numerical representation of $\rho_D$ from (\ref{eq:79}).  Then, we perform extraction by computing
\begin{equation}
\rho  = 3\rho _D  - I.
\label{eq:80}
\end{equation}
Next, we read the data from the extracted state $\rho$.  In the scheme of \Sec{V.A}, the character-label integer is \smash{$m = {\textstyle{1 \over 2}}(2M\chi _m  + 1)$}, and since the superposition character is \smash{$\chi _m  = {\textstyle{2 \over \pi }}\cos ^{ - 1} (\sqrt {\rho _{1,1} } ) \pm {\textstyle{1 \over {2M}}}$}, and the phase character is \smash{$\chi_m  = {\textstyle{1 \over {2\pi}}}(\arg (\rho _{2,1} ) + 2\pi \delta _{\text{sgn}(\arg (\rho _{2,1})),-1})\pm {\textstyle{1 \over {2M}}}$} (which remaps the phase to $[0,2\pi)$ if $\text{arg}$ outputs on $[-\pi,\pi)$ as for $\text{atan}2(y,x)$), then the character-label integers are given from the extracted state $\rho$ by
\begin{equation}
\begin{array}{*{20}l}
   {m_\theta  } &\!\! { = \text{round}({\textstyle{{2M} \over \pi }}\cos ^{ - 1} (\sqrt {\rho _{1,1} } ) + {\textstyle{1 \over 2}})}  \\
   {m_\phi  } &\!\! { = \text{round}({\textstyle{M \over {2\pi }}}\arg (\rho _{2,1} ) + M\delta_{\text{sgn}(\arg (\rho _{2,1})),-1}  + {\textstyle{1 \over 2}})}.  \\
\end{array}
\label{eq:81}
\end{equation}
Treating (\ref{eq:66}) as the extracted $\rho$ (since that presentation was rounded, we can use it to simulate measurement error here), then putting it into (\ref{eq:81}) yields
\begin{equation}
\left. {\begin{array}{*{20}l}
   {m_\theta  } &\!\! { = \text{round}(7.014)} &\!\! { = 7} &\!\! { \to \text{``G''}}  \\
   {m_\phi  } &\!\! { = \text{round}(14.999)} &\!\! { = 15} &\!\! { \to \text{``O''}}  \\
\end{array}} \right\} \to \text{``GO''},
\label{eq:82}
\end{equation}
where a look-up table of the character meanings was used to translate the message, successfully completing the one-way pseudo-ansible protocol, as depicted in \Fig[b]{15}.

Note that message preparation can be built into the sender, and tomography/extraction/message-reading can be built into the receiver, so that all their users have to do is type their messages in at one end, and read the message from a display screen at the other end.
\section{\label{sec:VI}Conclusions}\raisebox{43pt}[0pt][0pt]{\hypertarget{Concl}{}}%
This paper presented a theoretically justified proposition for noise-resistant quantum teleportation (NRQT) -- a method of transporting a state between spatial modes while protecting it from any quantum noise channel that acts in the region between source and destination.

\Section{II} gave an overview of \textit{unassisted quantum error correction} (UQEC), which is the mechanism that enables protection of a quantum state against all noise channels, where the term \textit{unassisted} means that no ancilla systems are necessary. In particular, UQEC does not need encoding or measurement ancillas.

It must be emphasized that UQEC is \textit{not} practical for QEC.  As explained in \App{App.A}, the term ``unassisted'' only applies in a theoretical sense because a true UQEC channel requires tracing over a larger system with an error-free subsystem, and though applying the UQEC channel classically in the local system is technically ``unassisted,'' its tomographic nature and the nonexistence of lone projectors from \Sec{IV} makes it impractical for quantum computation.  Nevertheless, UQEC proved to be a valuable tool for investigating the existence of true projectors.

The fundamental principle of UQEC is that it reduces the input state to \textit{scalars} which are not bound to any particular subsystem since scalars pass through Kronecker products.  The index-synchronization of encoding and decoding operators allows the state to be rebuilt perfectly, even when some arbitrary error channel acts between them.  When the rebuilding (decoding) happens in a different spatial mode than the encoding, we get a teleportation effect combined with the UQEC.

\Section{II.D} showed how to implement the UQEC quantum operations in a state-free manner, and used a \textit{classical} realization of the channel that utilizes random numbers to enforce ignorance on the outside world, and to enable index-linking of the encoding and decoding operations over great distances without a connection.  Since the communication protocol is tomographical at the receiving end, this classical realization does not affect the estimation of the state.

Then, \Sec{II.D.3} showed that the Kraus operators needed require nonunitary operators such as projectors, and \Sec{II.D.4} showed that absorptive projective filters (APFs) such as polarizers are \textit{not} actually faithful implementations of true projectors since they send the ``no-pass'' polarization to vacuum instead of scalar zero.

\Section{II.E} proved that teleportation can coexist with UQEC, while the UQEC does not reverse the teleportation.  Furthermore, it showed that we can even assume a \textit{nonlocal error channel} over both sender and receiver, and the input state is still perfectly teleported.

\Section{II.F} then showed that the projector problem identified in \Sec{II.D.3} leads to a need to represent successful projection as an \textit{event}.  This means that if we use APFs to realize our projectors, then we would need to maintain a lossless connection and accept the limitation of light-speed delay for the nonvacuum \textit{events} to reach the receiver in a communication protocol.  Later in \Sec{V}, we showed that by involving a postselection detector, we can achieve connection-free communication, though still limited to a light-speed delay.

The remainder of \Sec{II} focused on issues that might improve the possibility of the ideal communication protocol, including a method for realizing true projectors as part of a Kraus-complete channel, using \textit{insulating} UQEC to enable pure-state input to carry information and guarantee true \textit{superposition scalars}, and the theoretical possibility of converting the reference state to vacuum, if true projectors could be realized.

Then, armed with all of the necessary ingredients for ideal NQRT, \Sec{III} supposed that true projectors were possible, and proposed a plan to build an \textit{ansible}, a long-range communication device without any classical broadcast signal, and seemingly capable of superluminal communication.  We found that the supposition of true projectors did indeed permit effective superluminal communication (\Sec{III.G}), as well as quantum broadcasting (\Sec{III.D}) and effective quantum cloning (\Sec{III.E}).

While superluminality and quantum cloning are not sufficient to preclude the possibility of realizing true projectors, the fact that the no-cloning theorem is tied to the impossibility of superluminal signals suggests that their theoretical possibility here stems from the one supposition that we made; that true projectors are possible to implement.  Thus, \Sec{IV} proposed a \hyperlink{NoProjThm}{no-projector theorem} to explain why we should not actually be able to build a true ansible.

Then, \Sec{V} gave a detailed example of how to build a \textit{pseudo-ansible}, a device for \textit{connection-free} communication, limited to light-speed delay.  Along the way, we revisited the time-bin projector idea from \Sec{II.G}, and showed that even that will cause the ansible to fail because it does not deterministically implement a \textit{lone} projector, even though its projectors seem to act during different time bins.  Finally, \textit{postselection} is shown to enable the desired communication, at the price of the need to communicate the postselection results to the receiver, which is where the light-speed delay enters the protocol, since a classical radio signal would be the fastest means of sending that information.

Interestingly, the postselection method that enables pseudo-ansibles does not prevent the quantum broadcasting and effective quantum cloning applications of \Sec{III.D} and \Sec{III.E}.  This means that we \textit{can} achieve effective quantum cloning, and nearly instantaneously if done within a local region.  In limited-direct UQEC, all reductions of the multipartite output would be the same unknown state as the input, but the input would be limited to a family of constant eigenstates.  While such input has the same number of degrees of freedom as a classical state, it would nevertheless enable cloning of complete sets of maximally entangled states, such as cases where the Bell states form the family of constant eigenstates.  In the case of \textit{insulating} UQEC, there are \textit{no limitations on input}, but the effectively cloned states are all \textit{insulated versions} of the input, isomorphic to it, but not the same state.  However, in some cases this may be just as useful as a true quantum cloner, and is an intriguing area for further research.

The \textit{shocking} thing is that if pseudo-ansibles are truly possible, then they exhibit a \textit{spookiness without entanglement}, since it is possible for a distant receiver to have finished collecting all the tomographic data for a particular message \textit{before} the classical grouping signal for that message arrives.  Then, all of the data about the message is \textit{already sitting in classical memory at the receiver} when the classical signal arrives, and it merely tells us how to sort the data to read the message.  Thus, here the ``spookiness'' comes from the inherently nonlocal properties of wave functions and their overlaps, which is the origin of our ``superposition scalars.''  Since wave functions of free particles are infinite in extent, this is part of the nonlocality, which can be harnessed with synchronization and classically-transmitted postselection.

However, the classical grouping signal (CGS) \textit{does} transmit something about the message; it sends the message in the form of \textit{estimators} of measurement populations, scrambled to a ``one-time pad,'' meaning the internal string of random numbers shared by sender and receiver.  This is because, for each encoding/decoding pair, the total number of ``clicks'' reported by the CGS, divided by the total number of trials, forms an estimator for the nonvacuum probability of the mixture induced by the postselection projector.  It is this set of estimators that is sent by the CGS at light speed.  But there the superposition scalars $d_j$ and $1-d_j$ \textit{also} transfer to the receiver, and do so instantaneously and without estimation.  The CGS simply tells us \textit{which} detected events in the receiver correspond to $d_j$ and which belong to $1-d_j$.

In closing, we have proposed a method of noise-resistant quantum teleportation and its application as a pseudo-ansible for long-distance connection-free light-speed communication, and developed a \hyperlink{NoProjThm}{no-projector theorem} to explain why the theoretically possible superluminal ansible is not possible in reality.  While the light-speed limitation of the \textit{pseudo}-ansible may make it no more useful than classical communication, its error-protection properties may have many applications of their own.  Thus, the pseudo-ansible can at least serve as an interesting test of our understanding of quantum mechanics and how far it can bend classical limitations.  Most importantly, the \hyperlink{NoProjThm}{no-projector theorem} provides an interesting sequel to the no-clone theorem, and highlights the importance of Kraus-completeness to physicality.
\begin{acknowledgments}
This work was funded by the Johns Hopkins University Applied Physics Laboratory's Postdoctoral Fellowship in Quantum Information Science.  The author is grateful to Ting Yu, B.~D.~Clader, Michael Fitch, Jeffrey Barnes, and Dennis Lucarelli for helpful discussions.
\end{acknowledgments}
\begin{appendix}
\section{\label{sec:App.A}Simple Derivation of Limited-Direct Unassisted Quantum Error Correction}
The unassisted quantum error correction (UQEC) derivation shown here uses the idea of a \textit{virtual environment}, at first seemingly unrelated to the main system but which then enables us to derive the UQEC procedure entirely in one system.

It should be noted that UQEC is \textit{not} a practical means for implementing QEC for quantum computing because the existence of the UQEC operations implies assistance from the larger quantum system needed to cause it.  In that perspective, it essentially assumes the existence of an error-free ancilla, and then its encoding swaps the state to be protected into the error-free space, while then letting the error act only on the original input's subsystem, followed by swapping back.  If we had such an error-free subsystem, we would just use that exclusively!  

However, the value of UQEC (in its various forms) is that it provides the interesting hypothetical situation that allowed us to show the connection between true projectors and violations of special relativity, ultimately yielding the \hyperlink{NoProjThm}{no-projector theorem}.  Yet, the fact that the UQEC channels can still be implemented in the classical sense using absorptive projective filters means (as explained in this paper) that we can achieve freedom from assistance, but at the price of tomographic behavior and the need to postselect, which prohibits efficient computational scaling (repeated computations cause exponential increase in minimum sample time at the output).
 
To \textit{derive} UQEC, we use the idea of a \textit{scalar channel}, meaning a channel where the Kraus operators are scalars.  To construct this scalar channel, consider an $n\times n$ bipartite system starting in a state $\rho\otimes |\psi\rangle\langle\psi|$, subject to some error channel acting only on the second subsystem, $I\otimes\mathcal{E}(\sigma)\equiv$ \smash{$I\otimes\sum\nolimits_{k}E_{k}\sigma E_{k}^{\dag}$}, with Kraus completeness \smash{$\sum\nolimits_{k}E_{k}^{\dag} E_{k}=I$}. The scalar channel is then
\begin{equation}
\begin{array}{*{20}l}
   {\mathcal{S}(\rho  \otimes |\psi \rangle \langle \psi |)} &\!\! { = \text{tr}_E (\rho  \otimes \mathcal{E}(|\psi \rangle \langle \psi |))}  \\
   {} &\!\! { = \rho \sum\limits_{q = 1}^n {} \langle \phi_{q}|\sum\limits_{k} {} E_k |\psi \rangle \langle \psi |E_k ^{\dag}  |\phi_{q}\rangle, }  \\
\end{array}
\label{eq:A.1}
\end{equation}
where $\{|\phi_{q}\rangle\}$ is any complete basis for an $n$-level system.  Then, treating $|\psi\rangle$ as an intrinsic property of $\mathcal{S}$, we can simply relabel as $\mathcal{S}(\rho ) \equiv \mathcal{S}(\rho  \otimes |\psi \rangle \langle \psi |)$, and using the commutation of the scalars in (\ref{eq:A.1}), we obtain
\begin{equation}
\mathcal{S}(\rho ) = \sum\limits_{q = 1}^n {}\! \sum\limits_{k} {}\! \langle \phi_{q}|E_k |\psi \rangle \rho \langle \psi |E_k ^{\dag}  |\phi_{q}\rangle  = \sum\limits_{q = 1}^n {}\! \sum\limits_{k} {} S_{q,k} \rho S_{q,k}^{\dag}, 
\label{eq:A.2}
\end{equation}
which is a scalar channel since it has \textit{scalar Kraus operators} $S_{q,k} \! \equiv\! \langle \phi_{q}|E_k |\psi \rangle$.  (Although popular notation would use ``$\langle k|U|\psi \rangle$'' to represent ``$(I\otimes\langle k|)U (I\otimes|\psi \rangle)$,'' which is an \textit{operator}, here $|\phi_{q}\rangle$, $E_k $, and $|\psi \rangle$ all belong to the same subsystem, so $S_{q,k} \! \equiv\! \langle \phi_{q}|E_k |\psi \rangle$ really are scalars.)

$\mathcal{S}(\rho)$ is also an \textit{identity channel}, meaning $\mathcal{S}(\rho)=\rho$, as seen by shuffling the scalars around,
\begin{equation}
\begin{array}{*{20}l}
   {\mathcal{S}(\rho )} &\!\! { = \rho \langle \psi |\sum\limits_{k} {} E_k ^{\dag}  \sum\limits_{q = 1}^n {} |\phi_{q}\rangle \langle \phi_{q}|E_k |\psi \rangle }  \\
   {} &\!\! { = \rho \langle \psi |\sum\limits_{k} {} E_k ^{\dag}  E_k |\psi \rangle }  \\
   {} &\!\! { = \rho. }  \\
\end{array}
\label{eq:A.3}
\end{equation}
Thus, in place of $\mathcal{S}(\rho)$, we can just write $\rho$.

At this point, $\mathcal{S}(\rho)$ contains an arbitrary error channel of the same \textit{dimension} as $\rho$, and it \textit{outputs} $\rho$ in all cases.  Yet, the Kraus operators are scalars.  Thus, we need a way to \textit{elevate the scalar channel} to the dimension of $\rho$.

To achieve this, recall the \textit{spectral decomposition} of $\rho$, given by $\rho \equiv \Lambda(\rho) = \sum\nolimits_{j = 1}^n {|e_j \rangle \langle e_j |\rho |e_j \rangle \langle e_j |} $, where the $|e_j\rangle$ and $\lambda_j$ are eigenstates and eigenvalues of $\rho$.  

$\Lambda(\rho)$ can be viewed as a quantum operation with Kraus-diagonalized operators of the ``bowtie'' form $|A\rangle\langle B|$, in between which we could insert scalars of the form $\langle Y|E_{k}|Z\rangle$.  Therefore, inserting $\Lambda(\rho)$ into $\mathcal{S}$ allows us to elevate our scalar channel to the dimension of $\rho$ as
\begin{equation}
\begin{array}{*{20}l}
\rho &\!\! ={\mathcal{S}(\rho)= \mathcal{S}(\Lambda(\rho) )} \\
{} &\!\! = { \sum\limits_{q = 1}^n {} \sum\limits_{k } {} \langle \phi_{q}|E_k |\psi \rangle \sum\limits_{j = 1}^n {}|e_j \rangle \langle e_j |\rho  |e_j \rangle \langle e_j | \langle \psi |E_k ^{\dag}  |\phi_{q}\rangle}  \\
{}  &\!\! { = \sum\limits_{j = 1}^n {} \sum\limits_{q = 1}^n {} \sum\limits_{k } {} |e_j \rangle \langle \phi_{q}|E_k |\psi \rangle \langle e_j |\rho |e_j \rangle \langle \psi |E_k ^{\dag}  |\phi_{q}\rangle \langle e_j |}  \\
   {} &\!\! { = \sum\limits_{j = 1}^n {} \sum\limits_{q = 1}^n {} \sum\limits_{k} {} L_j ^{\dag}  R_q E_k L_j \rho L_j ^{\dag}  E_k ^{\dag}  R_q ^{\dag}  L_j, }  \\
\end{array}
\label{eq:A.4}
\end{equation}
where we inserted one as $|e_j \rangle \langle \phi_{q}| = |e_j \rangle \langle \psi |\psi \rangle \langle \phi_{q}|$ and $|\phi_{q}\rangle \langle e_j | = |\phi_{q}\rangle \langle \psi |\psi \rangle \langle e_j |$, so that $L_j  \equiv |\psi \rangle \langle e_j |$ and $R_q  \equiv |\psi \rangle \langle \phi_{q}|$.

While this version of UQEC is limited to the family of constant eigenstates that define its encoding and decoding operators, it nevertheless protects this family against all possible noise channels.  

A more general method of UQEC with \textit{no limits} on input is given in \App{App.E}, however that only outputs a state that is \textit{isomorphic} to the input state.  A variation of UQEC that corrects \textit{all errors on all input states, protecting the input perfectly} is \textit{direct} UQEC, which we merely summarize here as
\begin{equation}
\begin{array}{*{20}l}
   \rho  &\!\! { = \sum\limits_{a,b = 1,1}^{n,n} {} \sum\limits_{q = 1}^n {} \sum\limits_k^{} {} |\chi _a \rangle \langle \phi_{q}|E_k |\psi \rangle \langle \chi _a |\rho |\chi _b \rangle \langle \psi |E_k ^{\dag}  |\phi_{q}\rangle \langle \chi _b |}  \\
   {} &\!\! { = \sum\limits_{a,b = 1,1}^{n,n} {} \sum\limits_{q = 1}^n {} \sum\limits_k^{} {} C_a ^{\dag} R_q E_k C_a \rho C_b ^{\dag}  E_k ^{\dag}  R_q ^ {\dag}  C_b ,}  \\
\end{array}
\label{eq:A.5}
\end{equation}
where $C_a  \equiv |\psi \rangle \langle \chi _a |$, $C_b  \equiv |\psi \rangle \langle \chi _b |$, $R_q  \equiv |\psi \rangle \langle \phi_{q}|$, where $\{|\chi _a \rangle\}$ and $\{|\phi_{q}\rangle\}$ are complete orthonormal basis sets for $n$-level systems, and where we inserted unity as $|\chi _a \rangle \langle \phi_{q}| = |\chi _a \rangle \langle \psi |\psi \rangle \langle \phi_{q}|$ and $|\phi_{q}\rangle \langle \chi _b | = |\phi_{q}\rangle \langle \psi |\psi \rangle \langle \chi _b |$.  

The difference in (\ref{eq:A.5}) is that since we used \smash{$I\rho I$} to help elevate the scalar channel \smash{$\mathcal{S}$} instead of the spectral decomposition, the operation is not Kraus diagonal, meaning that the left-side encoding pairs \smash{$\{C_a, C_a^{\dag} \}$} have different indices than those on the right.

Thus, although direct UQEC is ideal, it requires \textit{operator superposition} of index-linked pairs of operators, making it impractical for theoretical communication schemes, and its theoretical reliance on error-free virtual ancillas makes it impractical for quantum computing as well.
\subsection{\label{sec:App.A.1}Proof that Error Constant $C$ is Unity for All Channels}
To prove the claim in \Sec{II.B} that $C=1$ always, where $C$ is the error constant defined in (\ref{eq:12}), recall that the standard definition of Kraus operators, given a system $S$ and environment $E$ and joint unitary $U^{(S,E)}$, is \smash{$E_k  \equiv (I^{(S)}  \otimes \langle \tau _k ^{(E)} |)U^{(S,E)} (I^{(S)}  \otimes |\psi _0^{(E)} \rangle ) \in S$}, where \smash{$\{|\tau _k ^{(E)}\rangle\}$} is a complete orthonormal basis for $E$ and \smash{$|\psi _0^{(E)} \rangle$} is any initial pure state of $E$.  Then, if \smash{$\{|\phi _q ^{(S)}\rangle\}$} is a complete orthonormal basis for $S$, the matrix elements of $E_{k}$ in basis \smash{$\{|\phi _q ^{(S)}\rangle\}$} are \smash{$(E_k )_{q,r}  \equiv \langle \phi _q ^{(S)} |E_k |\phi _r ^{(S)} \rangle$}, producing
\begin{equation}
(E_k )_{q,r}  = (\langle \phi _q ^{(S)} | \otimes \langle \tau _k ^{(E)} |)U^{(S,E)} (|\phi _r ^{(S)} \rangle  \otimes |\psi _0^{(E)} \rangle ).
\label{eq:A.6}
\end{equation}
The square magnitudes of these elements are
\begin{equation}
|(E_k )_{q,r} |^2 \! =\! \langle \phi _r | \otimes \langle\! \psi _0 |U^{\dag}  |\phi _q \rangle \langle \phi _q | \otimes |\tau _k \rangle \langle \tau _k |U|\phi _r \rangle\!  \otimes |\psi _0 \rangle,
\label{eq:A.7}
\end{equation}
where the subsystem labels have been suppressed.  Then, summing over all Kraus operators and all rows,
\begin{equation}
\begin{array}{*{20}l}
   {\sum\limits_{q = 1}^{n } {\sum\limits_{k} {|(E_k )_{q,r} |^2 } } } &\!\! { = \langle \phi _r | \otimes \langle \psi _0 |U^{\dag}  U|\phi _r \rangle  \otimes |\psi _0 \rangle }  \\
   {} &\!\! { = \langle \phi _r |\phi _r \rangle  \otimes \langle \psi _0 |\psi _0 \rangle }  \\
   {} &\!\! { = 1,}  \\
\end{array}
\label{eq:A.8}
\end{equation}
where $n$ is the number of levels in $S$. Note that since (\ref{eq:A.8}) was defined for any arbitrary basis \smash{$\{|\phi _q ^{(S)}\rangle\}$} and equals $1$ always, then it is true regardless of the basis that we use to define the recovery operators, since this derivation was chosen so that the indices $q$ align with those contributed by $R_q$ in (\ref{eq:12}) when using the computational basis.

Therefore, since (\ref{eq:A.8}) is true for any particular value of index $r$, then we can use its basis independence to set \smash{$\{|\phi _q ^{(S)}\rangle\}\equiv \{|q\rangle\}$} to coincide with the form used in (\ref{eq:12}), and then setting $r=1$ in (\ref{eq:A.8}) proves that $C=1$, by the definition in (\ref{eq:12}).
\subsection{\label{sec:App.A.2}Proof that Channels with Index-Linked Kraus Operators Can be Physical}
The strange thing about (\ref{eq:2}) is its index-linked operators, the $j$-indexed pairs in \smash{$L_{j}^{\dag}R_{q}E_{k}L_{j}$}.  Thus, here we prove that these index-linked operations can be physical.

First, we define \textit{interleaved channels} as quantum operations containing one or more sets of nonconsecutive index-linked operators.  A simple example would be
\begin{equation}
\Omega (\rho ) \equiv \sum\nolimits_{j,k} {F_j E_k D_j \rho D_j^{\dag}  E_k^{\dag}  F_j^{\dag}  } ,
\label{eq:A.9}
\end{equation}
where the physicality condition is Kraus completeness,
\begin{equation}
\sum\nolimits_{j,k} {(F_j E_k D_j )^{\dag}  F_j E_k D_j }  = I.
\label{eq:A.10}
\end{equation}

To prove that interleaved channels exist, it is sufficient to find one example that obeys the physicality condition.  Therefore, suppose that we have two physical channels,
\begin{equation}
\mathcal{E}(\rho ) \equiv \sum\nolimits_k {E_k \rho E_k^{\dag}  } \;\;\;\text{and}\;\;\;\mathcal{G}(\rho ) \equiv \sum\nolimits_j {G_j \rho G_j^{\dag} },
\label{eq:A.11}
\end{equation}
each with Kraus completeness
\begin{equation}
\sum\nolimits_k {E_k^ {\dag}  E_k }  = I\;\;\;\text{and}\;\;\;\sum\nolimits_j {G_j^ {\dag}  G_j }  = I.
\label{eq:A.12}
\end{equation}
Next, since every matrix has a polar decomposition, then
\begin{equation}
G_j  = U_j P_j ,
\label{eq:A.13}
\end{equation}
where \smash{$U_j \equiv W_j V_j^{\dag}$} is unitary, \smash{$P_j \equiv V_j S_j V_j^{\dag} $} is nonnegative Hermitian, and where \smash{$G_j  = W_j S_j V_j^{\dag} $} is the singular value decomposition of $G_j$, where $W_j$ and $V_j$ are unitary, and $S_j$ is the nonnegative diagonal matrix of singular values of $G_j$.  Then, putting (\ref{eq:A.13}) into (\ref{eq:A.12}) shows that
\begin{equation}
\sum\nolimits_j {G_j^{\dag}  G_j }  = \sum\nolimits_j {P_j ^{\dag}  U_j ^{\dag}  U_j P_j }  = \sum\nolimits_j {P_j ^{\dag}  P_j }  = I,
\label{eq:A.14}
\end{equation}
thus \smash{$\{P_{j}\}$} is \textit{also} a Kraus-complete set.  Then, if $\mathcal{E}(\rho )$ could act so that its operators came between the polar parts of $G_j =U_j P_j$, the resulting channel would have operators $\Omega_{j,k}\equiv U_j E_k P_j $.  Thus, to see if $\{\Omega_{j,k}\}$ constitutes a physical channel, test it for Kraus completeness:
\begin{equation}
\begin{array}{*{20}l}
   {\sum\nolimits_{j,k} {(U_j E_k P_j )^{\dag}  U_j E_k P_j }} &\!\! { = \sum\nolimits_{j,k} {P_j ^{\dag}  E_k ^{\dag}  U_j ^{\dag}  U_j E_k P_j } }  \\
   {} &\!\! { = \sum\nolimits_j {P_j ^ {\dag}  \sum\nolimits_k {E_k ^{\dag}  E_k } P_j } }  \\
   {} &\!\! { = \sum\nolimits_j {P_j ^{\dag}  P_j } }  \\
   {} &\!\! { = I,}  \\
\end{array}
\label{eq:A.15}
\end{equation}
where we used \smash{$U_j ^{\dag}U_j =I$} and the Kraus completeness of $\{E_k\}$ and $\{P_j\}$ from (\ref{eq:A.12}) and (\ref{eq:A.14}).  Therefore, (\ref{eq:A.15}) proves that the interleaved channel with Kraus operators $\Omega_{j,k}\equiv U_j E_k P_j$ is a \textit{physical} channel, and so this example proves the existence of physical interleaved channels.
\section{\label{sec:App.B}Absorptive Projective Filters are Not True Projectors}
Absorptive projective filters (APFs) act by selectively replacing certain states with the vacuum.  For example, in optics, a linear polarizer is an APF, and can be modeled as follows.  First, start with a general polarization state,
\begin{equation}
|\psi \rangle  = d|1_H ,0_V \rangle  + e|0_H ,1_V \rangle,
\label{eq:B.1}
\end{equation}
where $H$ and $V$ mean horizontal or vertical polarization, and we use the Fock basis, and commas within kets denote internal degrees of freedom.  Then, suppose we use a polarizing beam splitter to convert to a dual-rail basis,
\begin{equation}
|\psi '\rangle  = d|1_H ,0_V \rangle  \otimes |0_H ,0_V \rangle  + e|0_H ,0_V \rangle  \otimes |0_H ,1_V \rangle,
\label{eq:B.2}
\end{equation}
the outer product of which is $\rho ' \equiv |\psi '\rangle \langle \psi '|$, so
\begin{equation}
\rho ' =\! \left( \begin{array}{l}
 |d|^2 |1_H ,0_V ^{(1)} \rangle \langle 1_H ,0_V ^{(1)} | \otimes |0_H ,0_V ^{(2)} \rangle \langle 0_H ,0_V ^{(2)} | \\ 
  + de^* |1_H ,0_V ^{(1)} \rangle \langle 0_H ,0_V ^{(1)} | \otimes |0_H ,0_V ^{(2)} \rangle \langle 0_H ,1_V ^{(2)} | \\ 
  + ed^* |0_H ,0_V ^{(1)} \rangle \langle 1_H ,0_V ^{(1)} | \otimes |0_H ,1_V ^{(2)} \rangle \langle 0_H ,0_V ^{(2)} | \\ 
  + |e|^2 |0_H ,0_V ^{(1)} \rangle \langle 0_H ,0_V ^{(1)} | \otimes |0_H ,1_V ^{(2)} \rangle \langle 0_H ,1_V ^{(2)} | \\ 
 \end{array} \right)\!.
\label{eq:B.3}
\end{equation}
Then, suppose there exists an \textit{absorption channel} $\mathcal{A}_{0}$, defined in more detail in \Sec{III.A}, for which $\mathcal{A}_{0}(\rho)=|0\rangle\langle 0|\;\forall\rho$.  If we place a perfect absorber in the path of the second beam, where only the vertically polarized photon appears, letting $\rho''\equiv (I \otimes A_0 ( \cdot ))\rho '$, the result is
\begin{equation}
\rho'' = \left( \begin{array}{l}
 |d|^2 |1_H ,0_V \rangle \langle 1_H ,0_V | \\ 
  + |e|^2 |0_H ,0_V \rangle \langle 0_H ,0_V | \\ 
 \end{array} \right) \otimes |0_H ,0_V \rangle \langle 0_H ,0_V |,
\label{eq:B.4}
\end{equation}
where \smash{$(I \otimes \mathcal{A}_{0}())$} is a superoperator, and the ``off-diagonal'' terms containing \smash{$|0_H ,0_V ^{(2)} \rangle \langle 0_H ,1_V ^{(2)} |$} and \smash{$|0_H ,1_V ^{(2)} \rangle \langle 0_H ,0_V ^{(2)} |$} vanish because $A_0 ( \cdot )$ is defined in \Sec{III.A} as the application of a partial trace after a swap with vacuum, meaning that these orthogonal outer products are directly traced, resulting in zero.  Then, letting \smash{$\rho ''' \equiv \text{tr}_2 (\rho '')$}, the APF output can be written as
\begin{equation}
\rho '''\! =\! \left\{\!\! {\begin{array}{*{20}l}
   {|d|^2 |1_H ,0_V \rangle \langle 1_H ,0_V | \! +\! |e|^2 |0_H ,0_V \rangle \langle 0_H ,0_V |;} &\!\! {e \ne 1}  \\
   {{\kern 55.5pt}|0_H ,0_V \rangle \langle 0_H ,0_V |;} &\!\! {e = 1.}  \\
\end{array}} \right.\!\!
\label{eq:B.5}
\end{equation}
However, applying a \textit{true} projector \smash{$P_H  \equiv P_{|1_H ,0_V \rangle \langle 1_H ,0_V |}$} \smash{$  \equiv |1_H ,0_V \rangle \langle 1_H ,0_V |$} to (\ref{eq:B.1}) gives, before normalization,
\begin{equation}
P_H \rho P_H  =\! \left\{\! {\begin{array}{*{20}l}
   {|d|^2 |1_H ,0_V \rangle \langle 1_H ,0_V |;} & {e \ne 1}  \\
   {\kern 48.0pt}0; & {e = 1.}  \\
\end{array}} \right. 
\label{eq:B.6}
\end{equation}
Thus, a \textit{true projector} produces \textit{scalar zero} in the case of orthogonal input, while an APF yields the \textit{vacuum} state for orthogonal input, and leaves the unblocked polarization in a mixture with the vacuum, where the probability of the unblocked polarization is the same as its probability in the pure-state input.  Furthermore, the scalar coefficient $|d|^2$ of the pass-state is only achieved in (\ref{eq:B.5}) by focusing on nonvacuum results, which leads to the need for \textit{postselection}, discussed in the main text.

So how can a state like that in (\ref{eq:B.5}) be considered pure in situations such as the Stern-Gerlach experiment \cite[]{StGe1,StGe2,StGe3} that use APFs?  The answer is that we can ``purify'' the state by increasing the output trial time, a result provable with a time-bin basis and number operators, but which is not needed here.  In any case, such purification away from vacuum effectively eliminates the coefficient of $|d|^2$, making that process more of a state-preparation scheme, and not usefull in the present application.

Thus, although absorptive projective filters can be useful approximations to true projectors, \textit{they are not technically valid implementations of true projectors}.
\section{\label{sec:App.C}Two-Mode Model of Phase Shifters and the Teleportation Property of Free Space}
A phase shifter is typically modeled with a single spatial mode, so that its action on Fock state $|n\rangle$ is
\begin{equation}
P_\phi  |n\rangle  = e^{ - i\phi a^{\dag}  a} |n\rangle  = e^{ - in\phi } |n\rangle,
\label{eq:C.1}
\end{equation}
where $a$ is the annihilation operator for the spatial mode, $\phi\equiv\omega\Delta t$ is the phase-shift angle where $\Delta t =(n_{P}-n_{M})\frac{Z}{c}$, $n_P$ and $n_M$ are the indices of refraction of the phase shifter and the surrounding medium, $Z$ is the thickness of the phase shifter in the direction of field propagation, and $c$ is the speed of light in vacuum.  Note that this model treats the device as a single harmonic oscillator with Hamiltonian $H=\hbar\omega(a^{\dag}a+\frac{1}{2})$, so that technically, \smash{$P_\phi =e^{-i\frac{1}{\hbar}H\Delta t}=e^{-i\phi a^{\dag}a}e^{-i\frac{\phi}{2}}$}, but it is common practice to leave off the global phase factor \smash{$e^{-i\frac{\phi}{2}}$}, as in (\ref{eq:C.1}). 

However, a more physically accurate model for a phase shifter has \textit{two} spatial modes, one for the field at its input port and one for its output port, separated in space by the physical thickness $Z$ of the phase shifter.

To achieve a linear two-mode model of a phase shifter, recall that the general beam-splitter Hamiltonian is
\begin{equation}
H = \hbar g(e^{i\varphi } a_1 ^{\dag}  a_2  + e^{ - i\varphi } a_1 a_2 ^{\dag}  ),
\label{eq:C.2}
\end{equation}
where $a_1$ and $a_2$ are annihilation operators for modes 1 and 2, $\varphi$ is a phase-shift angle, and $g$ is a coupling strength related to the balance of transmission between the two output modes.  The device unitary is
\begin{equation}
B_{(\theta ,\varphi )}  = e^{ - i\theta (e^{i\varphi } a_1 ^{\dag}  a_2  + e^{ - i\varphi } a_1 a_2 ^{\dag}  )},
\label{eq:C.3}
\end{equation}
where $\theta  = g\Delta t_B$, which is related to the mode-1 transmittivity $T$ as \smash{$\cos(\theta)\equiv\sqrt{\,\!T^{\,^{\,}}\!\!}$}, and $\Delta t_B$ is the optical delay between mode 1 and mode 2.

A simpler form is obtained by setting $\varphi  \equiv {\textstyle{\pi  \over 2}} + \phi$, where $\phi$ is the phase-shift angle defined in (\ref{eq:C.1}), so
\begin{equation}
B_{(\theta ,{\textstyle{\pi  \over 2}} + \phi )}  = e^{\theta (e^{i\phi } a_1 ^{\dag}  a_2  - e^{ - i\phi } a_1 a_2 ^{\dag}  )}.
\label{eq:C.4}
\end{equation}
Then, using the Baker-Campbell-Hausdorff formula,
\begin{equation}
e^{sK} Ae^{ - sK}  = \sum\limits_{n = 0}^\infty  {{\textstyle{{s^n } \over {n!}}}C_n } ,\;\;\;C_n  \equiv [K,C_{n - 1} ],\;\;\;C_0  \equiv A,
\label{eq:C.5}
\end{equation}
where $[A,B]\equiv AB-BA$, the output modes of (\ref{eq:C.4}) are
\begin{equation}
\begin{array}{*{20}l}
   {a_1 '^{\dag}  } &\!\! { \equiv B_{(\theta ,{\textstyle{\pi  \over 2}} + \phi )} a_1 ^{\dag}  B_{(\theta ,{\textstyle{\pi  \over 2}} + \phi )} ^{\dag}  } &\!\! { = c_\theta  a_1 ^{\dag}   - e^{ - i\phi } s_\theta  a_2 ^{\dag}  }  \\
   {a_2 '^{\dag}  } &\!\! { \equiv B_{(\theta ,{\textstyle{\pi  \over 2}} + \phi )} a_2 ^{\dag}  B_{(\theta ,{\textstyle{\pi  \over 2}} + \phi )} ^{\dag}  } &\!\! { = e^{i\phi } s_\theta  a_1 ^{\dag}   + c_\theta  a_2 ^{\dag}  },  \\
\end{array}
\label{eq:C.6}
\end{equation}
where $c_{\theta}\equiv \cos(\theta)$ and $s_{\theta}\equiv \sin(\theta)$.  

Then, to model the teleportation action of the phase shifter, we set $\theta  = {\textstyle{{3\pi } \over 2}}$, since that flips the modes while getting rid of the minus sign in the mode-1 output, as
\begin{equation}
\begin{array}{*{20}l}
   {a_1 '^{\dag}  } &\!\! { \equiv B_{({\textstyle{{3\pi } \over 2}},{\textstyle{\pi  \over 2}} + \phi )} a_1 ^{\dag}  B_{({\textstyle{{3\pi } \over 2}},{\textstyle{\pi  \over 2}} + \phi )} ^{\dag}  } &\!\! { = e^{ - i\phi } a_2 ^ {\dag}  }  \\
   {a_2 '^{\dag}  } &\!\! { \equiv B_{({\textstyle{{3\pi } \over 2}},{\textstyle{\pi  \over 2}} + \phi )} a_2 ^{\dag} B_{({\textstyle{{3\pi } \over 2}},{\textstyle{\pi  \over 2}} + \phi )} ^{\dag}  } &\!\! { =  - e^{i\phi } a_1 ^{\dag}  }.  \\
\end{array}
\label{eq:C.7}
\end{equation}
Thus, by setting
\begin{equation}
P_\phi   \equiv B_{({\textstyle{{3\pi } \over 2}},{\textstyle{\pi  \over 2}} + \phi )}  = e^{{\textstyle{{3\pi } \over 2}}(e^{i\phi } a_1 ^{\dag}  a_2  - e^{ - i\phi } a_1 a_2 ^{\dag}  )},
\label{eq:C.8}
\end{equation}
the output modes of (\ref{eq:C.7}) cause
\begin{equation}
P_\phi  |n\rangle  \otimes |0\rangle  = e^{ - in\phi } |0\rangle  \otimes |n\rangle,
\label{eq:C.9}
\end{equation}
yielding the same phase factor as the single-mode model in (\ref{eq:C.1}), but \textit{also} accounting for the teleportation of $|n\rangle$.

In general, we can consider free space to be made up of a chain of infinitesimally thin phase shifters all back-to-back in each direction.  Light entering free space at any point is then continually teleported forward to new spatial modes. Since free space is the vacuum, then the phase shift incurred is zero at all times, and we find that free space is the ideal quantum teleporter.
\section{\label{sec:App.D}Example of Implementation of True Projectors}
In \Sec{II.G}, we proposed using a CNOT gate along with time-bin adjustments to create a channel of orthogonal projectors where the projectors are effectively applied during different time bins to allow us to properly synchronize two sides of a communicator.  Here we give the details of the proposed example for achieving this.

In \cite[]{FiWo}, a single-photon deterministic CNOT gate was demonstrated, utilizing a photon's momentum and polarization qubits.  If the polarization qubit contains our message, then preparing the momentum qubit in its first basis state and applying such a CNOT gives
\begin{equation}
\rho ' = \sum\limits_{r = 1}^2 {P_{|\widetilde{r}\rangle } \rho P_{|\widetilde{r}\rangle } } 
\label{eq:D.1}
\end{equation}
in the polarization system.  Then, to convert from polarization to a time-bin basis, first convert to dual-rail with a polarizing beam splitter (PBS).  Then send one rail through a delay line longer than the photon's coherence time, send the second rail through a half-wave plate (HWP) to make both polarizations the first computational basis state, and use a time-controlled switch to convert the time-bin qubit to single rail (this is the dual-rail/time-bin converter from \cite[]{BSBL}), resulting in
\begin{equation}
\rho_{\text{CNOT}} = \rho _{1,1} |\widetilde{1}_{\tau _1 } ,0_{\tau _2 } \rangle \langle \widetilde{1}_{\tau _1 } ,0_{\tau _2 } | + \rho _{2,2} |0_{\tau _1 } ,\widetilde{1}_{\tau _2 } \rangle \langle 0_{\tau _1 } ,\widetilde{1}_{\tau _2 } |,
\label{eq:D.2}
\end{equation}
where $\tau_k$ means time-bin $k$, and $\rho _{a,b}\equiv \langle\widetilde{a} |\rho|\widetilde{b}\rangle$.  

Since the time-bin conversion is \textit{after} the projector channel, the net Kraus operators are not time-bin projectors, so the input $\rho$ does not need to be in a time-bin product with vacuum to survive the projectors; a constant input over both time bins is enough to yield (\ref{eq:D.2}).

Rather than ignoring the time-bin-2 output, since only its scalar is different, and we can just treat the time-bin 2 output as a bit-flipped version of that in time-bin 1.

The index-linking of the operators is then entirely realized by synchronizing unitary operators on both sides of the communicator, and since those are deterministic (channels having only one Kraus operator), we can control their application precisely, as discussed in \Sec{III.B}.

Thus it would \textit{seem} as if we have designed a practical quantum operation for achieving \textit{true} projectors separated in time, enabling us to know exactly when each projector is acting.  However, \Sec{V.B.1} exposes the flaw of this thinking and explains why it seems reasonable in spite of its failure.
\section{\label{sec:App.E}Insulating UQEC}
In \Sec{II.H} we saw that it might be advantageous to use the method of \textit{insulating} UQEC in an NRQT scheme since it would permit locally generated pure-state input to perform communication with \textit{superposition scalars}, thereby avoiding any problems with event-based statistical scalars.  So here we review insulating UQEC.

Insulating UQEC has two stages; transmission and extraction.  To transmit $n$-level state $\rho$ through noise channel $\mathcal{E}$ with errors $E_k$, the procedure is
\begin{equation}
\rho _D = {\textstyle{1 \over {2n - 1}}}\sum\limits_{j = 1}^{m_n}\sum\limits_{q = 1}^n\sum\limits_{k} {D_j^{\dag} {R_q  {E_k D_j \rho D_j^{\dag}  E_k^{\dag}  R_q^{\dag} D_j } } },
\label{eq:E.1}
\end{equation}
where $m_n  \equiv n(2n - 1)$, \smash{$R_q \equiv |\psi\rangle\langle \phi_q|$} are the familiar recovery operators, \smash{$D_j$} and \smash{$D_j^{\dag}$} are matched encoding and decoding operators (defined below), and \smash{$\rho_D$} is an \textit{insulated} form of $\rho$, where
\begin{equation}
\begin{array}{*{20}l}
   {\rho _D  \equiv \mathcal{D}(\rho )} &\!\! { \equiv {\textstyle{1 \over {2n - 1}}}\sum\limits_{j = 1}^{m_n } {|D_j \rangle \langle D_j |\rho |D_j \rangle \langle D_j |} }  \\
   {} &\!\! { = {\textstyle{1 \over {2n - 1}}}(I + (n - 2)\Delta (\rho ) + \rho ),}  \\
\end{array}
\label{eq:E.2}
\end{equation}
where \smash{$\Delta (\sigma ) \equiv \sum\nolimits_{u = 1}^n {|u\rangle \langle u|} \sigma |u\rangle \langle u|$} is the maximal dephasing channel where \smash{$\{ |u\rangle \}$} are computational basis kets generically labeled starting on $1$ as \smash{$\{ |1\rangle , \ldots ,|n\rangle \}$}, and \smash{$\{|D_j \rangle \}$} is defined in (\ref{eq:E.5}).  Then, since \smash{$\rho_D$} is isomorphic to $\rho$, the fully recovered input $\rho$ is given by the \textit{extraction},
\begin{equation}
\rho  = (2n - 1)\rho _D  - {\textstyle{{(2n - 1)(n - 2)} \over {n - 1}}}\Delta (\rho _D ) - {\textstyle{1 \over {n - 1}}}I.
\label{eq:E.3}
\end{equation}
Although (\ref{eq:E.3}) is not a standard quantum operation on $\rho_D$, it is simply a mathematical way to convert the tomographical estimation of $\rho_D$ to $\rho$, and thus the message.

The encoding operators $D_j$ are
\begin{equation}
D_j \equiv |\psi\rangle\langle D_j|,
\label{eq:E.4}
\end{equation}
where $|\psi\rangle$ is the reference state, and $\{|D_j \rangle \}$ is an overcomplete basis, one possible set of which is
\begin{equation}
\{|D_j \rangle \}  \equiv \left\{ {{\textstyle{1 \over {\sqrt 2 }}}(|a\rangle  \pm |b\rangle ),{\textstyle{1 \over {\sqrt 2 }}}(|a\rangle  \pm i|b\rangle ),|a\rangle ,|b\rangle } \right\},
\label{eq:E.5}
\end{equation}
generating $n$ diagonal states in total, and four nondiagonal states for each of the $(n^2 -n)/2$ index-pairs $\{a,b\}$ such that $1 \le a \le n - 1$ and $a + 1 \le b \le n$, where $|a\rangle$ and $|b\rangle$ are the orthonormal computational basis states.  Note that (\ref{eq:E.2}) is specifically for use with (\ref{eq:E.5}), as are the pre-factor ${\textstyle{1 \over {2n - 1}}}$ and upper bound $m_n$ in (\ref{eq:E.1}), which stem from the overcompleteness \smash{${\textstyle{1 \over {2n - 1}}}\sum\nolimits_{j = 1}^{m_n } {|D_j \rangle \langle D_j |}  = I$}.

The advantage of insulating UQEC is that it allows pure-state inputs to carry information (since it admits \textit{any} states), and that lets us use a locally generated pure state to carry information in its superposition scalars.

The best way to get a feel for insulating UQEC is to see an example.  Here we will use a general one-qubit ($n=2$) mixed state $\rho$ as input.  First, choosing $|\psi \rangle \! =\! |1\rangle$ and \smash{$\{|\phi_q\rangle\}=\{|1\rangle,|2\rangle\}$}, where $|1\rangle\!\equiv\!\binom{1}{0}$ and $|2\rangle\!\equiv\!\binom{0}{1}$, the recovery operators are
\begin{equation}
\begin{array}{*{20}c}
   {R_1 = \left(\! {\begin{array}{*{20}c}
   1 & 0  \\
   0 & 0  \\
\end{array}}\! \right)} & {R_2 = \left(\! {\begin{array}{*{20}c}
   0 & 1  \\
   0 & 0  \\
\end{array}}\! \right)\!,}  \\
\end{array}
\label{eq:E.6}
\end{equation}
and (\ref{eq:E.4}) gives the encoding operators,
\begin{equation}
\begin{array}{*{20}c}
   {\begin{array}{*{20}l}
   {D_1 } &\!\! { = {\textstyle{1 \over {\sqrt 2 }}}\left(\! {\begin{array}{*{20}c}
   1 & 1  \\
   0 & 0  \\
\end{array}}\! \right)}  \\
   {D_2 } &\!\! { = {\textstyle{1 \over {\sqrt 2 }}}\left(\! {\begin{array}{*{20}c}
   1 & { - 1}  \\
   0 & 0  \\
\end{array}}\! \right)}  \\
\end{array}} &\!\!\! {\begin{array}{*{20}l}
   {D_3 } &\!\! { = {\textstyle{1 \over {\sqrt 2 }}}\left(\! {\begin{array}{*{20}c}
   1 & { - i}  \\
   0 & 0  \\
\end{array}}\! \right)}  \\
   {D_4 } &\!\! { = {\textstyle{1 \over {\sqrt 2 }}}\left(\! {\begin{array}{*{20}c}
   1 & i  \\
   0 & 0  \\
\end{array}}\! \right)}  \\
\end{array}} &\!\!\! {\begin{array}{*{20}l}
   {D_5 } &\!\! { =\! \left(\! {\begin{array}{*{20}c}
   1 & 0  \\
   0 & 0  \\
\end{array}}\! \right)}  \\
   {D_6 } &\!\! { =\! \left(\! {\begin{array}{*{20}c}
   0 & 1  \\
   0 & 0  \\
\end{array}}\! \right)\!,}  \\
\end{array}}\!\!\!  \\
\end{array}
\label{eq:E.7}
\end{equation}
where \Sec{II.D.3} explains how to ``realize'' (\ref{eq:E.6}) and (\ref{eq:E.7}).  

Now, the first step in the correction procedure is to randomly choose and apply one of the $D_j$ to $\rho$, keeping in mind that later, we will also have to apply its adjoint $D_j^{\dag}$.  The results of the encoding step are
\begin{equation}
D_j \rho D_j^{\dag} = {d_j}\!\left(\! {\begin{array}{*{20}c}
   1 & 0  \\
   0 & 0  \\
\end{array}}\! \right),
\label{eq:E.8}
\end{equation}
where the coefficients $d_{j}\equiv\langle D_{j}|\rho|D_{j}\rangle$ are given by
\begin{equation}
\begin{array}{*{20}l}
   {d_1 } &\!\! { = {\textstyle{1 \over 2}} + \text{Re}(\rho _{2,1} )} &\; {d_3 } &\!\! { = {\textstyle{1 \over 2}} + \text{Im}(\rho _{2,1} )} &\; {d_5 } &\!\! { = \rho _{1,1} }  \\
   {d_2 } &\!\! { = {\textstyle{1 \over 2}} - \text{Re}(\rho _{2,1} )} &\; {d_4 } &\!\! { = {\textstyle{1 \over 2}} - \text{Im}(\rho _{2,1} )} &\; {d_6 } &\!\! { = \rho _{2,2}, }  \\
\end{array}
\label{eq:E.9}
\end{equation}
where $\rho _{a,b}\equiv\langle a|\rho|b\rangle$.  The matrix elements in (\ref{eq:E.9}) lead to the results in (\ref{eq:34}) for pure input.  

To check this with the reference-state projector method of \Sec{II.D.3}, the encoding operators are
\begin{equation}
D_j  \equiv |1\rangle \langle D_j | = |1\rangle \langle 1|\epsilon _{|D_j \rangle } ^{\dag}, 
\label{eq:E.10}
\end{equation}
where the descending-order eigenvector matrices are
\begin{equation}
\begin{array}{*{20}l}
   {\epsilon _{|D_1 \rangle } } &\!\! { = {\textstyle{1 \over {\sqrt 2 }}}\left( {\begin{array}{*{20}c}
   1 & { - 1}  \\
   1 & 1  \\
\end{array}} \right)\!,} & {\epsilon _{|D_2 \rangle } } &\!\! { = {\textstyle{1 \over {\sqrt 2 }}}\left( {\begin{array}{*{20}c}
   { - 1} & { - 1}  \\
   1 & { - 1}  \\
\end{array}} \right)\!,}  \\
   {\epsilon _{|D_3 \rangle } } &\!\! { = {\textstyle{1 \over {\sqrt 2 }}}\left( {\begin{array}{*{20}c}
   { - i} & { - i}  \\
   1 & { - 1}  \\
\end{array}} \right)\!,} & {\epsilon _{|D_4 \rangle } } &\!\! { = {\textstyle{1 \over {\sqrt 2 }}}\left( {\begin{array}{*{20}c}
   i & i  \\
   1 & { - 1}  \\
\end{array}} \right)\!,}  \\
   {\epsilon _{|D_5 \rangle } } &\!\! { = \left( {\begin{array}{*{20}c}
   1 & 0  \\
   0 & 1  \\
\end{array}} \right)\!,} & {\epsilon _{|D_6 \rangle } } &\!\! { = \left( {\begin{array}{*{20}c}
   0 & 1  \\
   1 & 0  \\
\end{array}} \right)\!.}  \\
\end{array}
\label{eq:E.11}
\end{equation}
Then, application of the \smash{$\epsilon _{|D_j \rangle }^{\dag}$} to $\rho$ gives
\begin{equation}
\begin{array}{l}
 \rho _1  = \left( {\begin{array}{*{20}c}
   {{\textstyle{{1 + 2\text{Re}(\rho _{2,1} )} \over 2}}} & {{\textstyle{{1 - 2\rho _{1,1}  - i2\text{Im}(\rho _{2,1} )} \over 2}}}  \\
   {{\textstyle{{1 - 2\rho _{1,1}  + i2\text{Im}(\rho _{2,1} )} \over 2}}} & {{\textstyle{{1 - 2\text{Re}(\rho _{2,1} )} \over 2}}}  \\
\end{array}} \right) \\ 
 \rho _2  = \left( {\begin{array}{*{20}c}
   {{\textstyle{{1 - 2\text{Re}(\rho _{2,1} )} \over 2}}} & {{\textstyle{{2\rho _{1,1}  - 1 - i2\text{Im}(\rho _{2,1} )} \over 2}}}  \\
   {{\textstyle{{2\rho _{1,1}  - 1 + i2\text{Im}(\rho _{2,1} )} \over 2}}} & {{\textstyle{{1 + 2\text{Re}(\rho _{2,1} )} \over 2}}}  \\
\end{array}} \right) \\ 
 \rho _3  = \left( {\begin{array}{*{20}c}
   {{\textstyle{{1 + 2\text{Im}(\rho _{2,1} )} \over 2}}} & {{\textstyle{{2\rho _{1,1}  - 1 - i2\text{Re}(\rho _{2,1} )} \over 2}}}  \\
   {{\textstyle{{2\rho _{1,1}  - 1 + i2\text{Re}(\rho _{2,1} )} \over 2}}} & {{\textstyle{{1 - 2\text{Im}(\rho _{2,1} )} \over 2}}}  \\
\end{array}} \right) \\ 
 \rho _4  = \left( {\begin{array}{*{20}c}
   {{\textstyle{{1 - 2\text{Im}(\rho _{2,1} )} \over 2}}} & {{\textstyle{{2\rho _{1,1}  - 1 + i2\text{Re}(\rho _{2,1} )} \over 2}}}  \\
   {{\textstyle{{2\rho _{1,1}  - 1 - i2\text{Re}(\rho _{2,1} )} \over 2}}} & {{\textstyle{{1 + 2\text{Im}(\rho _{2,1} )} \over 2}}}  \\
\end{array}} \right) \\ 
 \rho _5  = \left( {\begin{array}{*{20}c}
   {\rho _{1,1} } & {\rho _{1,2} }  \\
   {\rho _{2,1} } & {\rho _{2,2} }  \\
\end{array}} \right),\;\;\text{and}\;\;\rho _6  = \left( {\begin{array}{*{20}c}
   {\rho _{2,2} } & {\rho _{2,1} }  \\
   {\rho _{1,2} } & {\rho _{1,1} }  \\
\end{array}} \right), \\ 
 \end{array}
\label{eq:E.12}
\end{equation}
where \smash{$\rho _j  \equiv \epsilon _{|D_j \rangle } ^{\dag}  \rho \epsilon _{|D_j \rangle } $}.  Then, since
\begin{equation}
D_j \rho D_j ^{\dag}   = |1\rangle \langle 1|\epsilon _{|D_j \rangle } ^{\dag}  \rho \epsilon _{|D_j \rangle } |1\rangle \langle 1| = P_{|1\rangle}\rho_{j}P_{|1\rangle}=d_{j}|1\rangle \langle 1|,
\label{eq:E.13}
\end{equation}
applying \smash{$P_{|1\rangle}  \equiv |1\rangle \langle 1|$} to the $\rho_j$ yields $d_{j}|1\rangle \langle 1|$, where
\begin{equation}
d_j  = \langle 1|\epsilon _{|D_j \rangle } ^{\dag}  \rho \epsilon _{|D_j \rangle } |1\rangle = \langle 1|\rho_j |1\rangle =(\rho_{j})_{1,1},
\label{eq:E.14}
\end{equation}
which gives the values in (\ref{eq:E.9}), which can be seen by putting the $\rho_j$ from (\ref{eq:E.12}) into (\ref{eq:E.14}).  The purpose of showing this decomposition is to illustrate that if we use insulating UQEC instead of limited-direct UQEC, we can still use the first computational basis state as the reference state, so we only need to realize one true projector.

Then one of the $E_k$ acts, producing
\begin{equation}
E_k D_j \rho D_j^{\dag}  E_k^{\dag}   = d_j \left(\! {\begin{array}{*{20}c}
   {|(E_k )_{1,1} |^2 } & {(E_k )_{1,1} (E_k )_{2,1} ^* }  \\
   {(E_k )_{2,1} (E_{k})_{1,1} ^* } & {|(E_k )_{2,1} |^2 }  \\
\end{array}}\! \right)\!,
\label{eq:E.15}
\end{equation}
where, $(E_k)_{a,b}\equiv\langle a|E_{k}|b\rangle$.  Next we randomly apply one of the recovery operators, producing
\begin{equation}
R_q E_k D_j \rho D_j^{\dag}  E_k^{\dag}  R_q^{\dag} = d_j |(E_k )_{q,1} |^2 \left( {\begin{array}{*{20}c}
   1 & 0  \\
   0 & 0  \\
\end{array}} \right)\!,
\label{eq:E.16}
\end{equation}
where $q$ is $1$ or $2$.  Then, applying the corresponding decoder $D_j^{\dag}=|D_{j}\rangle\langle 1|$ yields
\begin{equation}
D_j^{\dag}  R_q E_k D_j \rho D_j^ {\dag}  E_k^{\dag}  R_q^{\dag}  D_j  = d_j |(E_k )_{q,1} |^2 |D_j \rangle \langle D_j |.
\label{eq:E.17}
\end{equation}

Now, since we don't know \textit{which} $E_k$ happened, or \textit{which} $R_q$ was applied, or \textit{which pair} $\{D_j ,D_j^{\dag}\}$ was used, then before we look at the result, the unnormalized state is the sum over all possible results, as
\begin{equation}
\sum\limits_{j = 1}^6 \sum\limits_{q = 1}^2 {\sum\limits_k {D_j^{\dag}  R_q E_k D_j \rho D_j^ {\dag}  E_k^{\dag}  R_q^{\dag}  D_j } }  = C\sum\limits_{j = 1}^6 d_j |D_j \rangle \langle D_j |,
\label{eq:E.18}
\end{equation}
where $C$ is an ``error-dependent'' scalar, defined as
\begin{equation}
C \equiv \sum\limits_{q = 1}^n {\sum\limits_k {|(E_k )_{q,1} |^2 } }. 
\label{eq:E.19}
\end{equation}
In \App{App.A.1}, we showed that $C=1$ for all channels, but here we keep it.  Now we see that the matrix to the right of $C$ in (\ref{eq:E.18}) only contains information about $\rho$, and using (\ref{eq:E.5}) and (\ref{eq:E.9}), we find that
\begin{equation}
\sum\limits_{j = 1}^6 {d_j |D_j \rangle \langle D_j |}  = I + \rho.
\label{eq:E.20}
\end{equation}
Finally, using (\ref{eq:E.20}) in (\ref{eq:E.18}) and normalizing, we get
\begin{equation}
\frac{\sum\limits_{j = 1}^6 {\sum\limits_{q = 1}^2 {\sum\limits_k \!{D_j^{\dag}  R_q E_k D_j \rho D_j^{\dag}  E_k^{\dag}  R_q^{\dag}  D_j } } }}{\text{tr}(C(I + \rho ))}= {\textstyle{1 \over 3}}(I \! +\! \rho ) =\rho_D,
\label{eq:E.21}
\end{equation}
from which $\rho$ can be extracted as
\begin{equation}
\rho  = 3\rho _D  - I.
\label{eq:E.22}
\end{equation}

Thus we see that insulating UQEC proceeds analogously to limited-direct UQEC, with the advantage that it works on any input states, which we utilize in \Sec{II.H} to use only pure-state inputs to ensure that the scalars are directly inherited from superposition scalars.
\end{appendix}
%
\end{document}